\documentclass[prb, aps,twocolumn]{revtex4}
\usepackage{amssymb}
\usepackage{amsmath}
\usepackage{graphicx}
\usepackage{subfigure}

\begin{document}

\title{Quantum kinetic approach for studying thermal transport in the presence of electron-electron interactions and disorder }
\author{Karen~Michaeli$^1$ and Alexander~M.~Finkel'stein$^{1,2}$}
\affiliation{$^1$ Department of Condensed Matter Physics, The Weizmann Institute of Science,
Rehovot 76100, Israel \\ $^2$ Department of Physics, Texas A\&M University, College Station, TX $77843-4242$, USA}

\begin{abstract}
A user friendly scheme based on the quantum kinetic equation is developed for studying thermal transport phenomena in the presence of interactions and disorder. We demonstrate that this scheme is suitable for both a systematic perturbative calculation as well as a general analysis. We believe that we present an adequate alternative to the Kubo formula, which for the thermal transport is rather cumbersome.\newline
\end{abstract}

\maketitle

Measurements of thermal and thermoelectric transport in electron systems of various kinds attracted considerable  attention in recent years.~\cite{Hvar} Among such experiments one can find the measurement of the critical behavior of the thermopower near the metal insulator transition,~\cite{Pudalov2001} the Nernst effect in high-$T_c$ and conventional superconductors,~\cite{Ong2000,Ong2005,Aubin2006,Aubin2007} thermal transport in quantum Hall systems~\cite{Eisenstein2009} as well as thermopower in quantum dots.~\cite{Molenkamp1990} To benefit from these experimental efforts, theoretical studies are required. In this paper, we develop a new theoretical apparatus for analyzing thermal and thermoelectric transport in interacting electron systems in the presence of disorder. The strength of our scheme is in its  generality that allow us to apply it for different kinds of interactions. We believe that this scheme can be adopted as an alternative to the Kubo formula.

The validity of the Kubo formula for thermal and thermoelectric conductivities was proved by Luttinger.~\cite{Luttinger1964} A main ingredient in the Kubo formula is the quantum mechanical expression for the current operators entering the correlation functions. The correlation functions describing the thermoelectric or the thermal current involve the heat current operator which was derived by  Luttinger~\cite{Luttinger1964} for a system of interacting electrons. In the absence of the electron-electron interactions, the expression for the heat current operator is
\begin{align}\label{eq:CurrentNonInt}
\mathbf{j}_h(\mathbf{q}=0,\tau)=\sum_{\mathbf{p}}&\frac{\partial\varepsilon_{\mathbf{p}}}{\partial\mathbf{p}}\left(\varepsilon_{\mathbf{p}}-\mu\right)c_{\mathbf{p}}^{\dag}(\tau)c_{\mathbf{p}}(\tau)\\\nonumber
&+\sum_{\mathbf{p},\mathbf{p}'}\frac{\partial\varepsilon_{\mathbf{p}}}{\partial\mathbf{p}}V_{imp}(\mathbf{p},\mathbf{p}')c_{\mathbf{p}}^{\dag}(\tau)c_{\mathbf{p}'}(\tau),
\end{align}
where $c_{\mathbf{p}}^{\dag}(\tau)$ ($c_{\mathbf{p}}(\tau)$) is the creation (annihilation) operator of an electron in a state with energy $\varepsilon_{\mathbf{p}}$. Here $\mu$ is the chemical potential, $V_{imp}(\mathbf{p},\mathbf{p}')$ is the potential created by the disorder and $\tau$ is the imaginary time. With the help of the equations of motion~\cite{comment} and after transforming to the Matsubara frequencies, the current operator can be written as:
\begin{align}\label{eq:CurrentNonIntFrequency}
\mathbf{j}_h(\mathbf{q}=0,\omega_n)=\sum_{\mathbf{p},\epsilon_n}\frac{\partial\varepsilon_{\mathbf{p}}}{\partial\mathbf{p}}\frac{2i\epsilon_n-i\omega_n}{2}c_{\mathbf{p}}^{\dag}(\epsilon_n)c_{\mathbf{p}}(\epsilon_n-\omega_n).
\end{align}
When interactions between the electrons are included, the expression for $\mathbf{j}_h$ becomes more complicated. In general, even when the equations of motion are employed, the resulting expression for the current is not just the frequency multiplied by the velocity but rather contains additional terms.~\cite{Langer1962}  Unfortunately, the heat current operator of free electrons is frequently used for interacting systems even when there is no justification for it. We demonstrate in Appendix~\ref{App:FermiLiquidKubo} that the simplified form of the Kubo formula fails to reproduce the phenomenological thermal conductivity of a Fermi-liquid system, i.e., the result does not satisfy the Wiedemann-Franz law.  The problems induced by the simplified Kubo formula do not necessarily imply that  the use of the Kubo formula for thermal transport is generally wrong. The weak point is in replacing Luttinger's expression for the heat current by the simplified form. The problem with using the full expression for the heat current is that is that it is too complex.~\cite{Reizer1994}

The difficulties related to the Kubo formula lead us to turn to the quantum kinetic equation.  Although derivations of the transport coefficients using the kinetic equation already exist, see for example Refs.~\onlinecite{Mahan1984,Castellani1991,Livanov1991,Raimondi2004,Aleiner2005}, our method differs in few aspects. We developed a simple scheme that, as we demonstrate in this paper, can be applied for various electron systems. The scheme is well suitable for both a general analysis and a systematic perturbation expansion. The novelty of our method is that we were able to derive both the kinetic equation and the currents directly from the action and the corresponding conservation laws. The systematic formulation of the quantum kinetic equation in the presence of a temperature gradient has been achieved  using Luttinger's idea of introducing a gravitational field.~\cite{Luttinger1964} As a result, we  found the expressions for the currents from which all four components of the conductivity tensor can be extracted:
\begin{equation}
\left(\begin{array}{c}
\mathbf{j}_{e} \\
\mathbf{j}_{h}\end{array}\right)=
\left(
\begin{array}{cc}
\hat{\sigma} & \hat{\alpha} \\
\hat{\tilde{\alpha}} & \hat{\kappa} \\
\end{array}\right)\left(\begin{array}{c}
\mathbf{E} \\
-\boldsymbol{\nabla}T
\end{array}\right).
\label{eq:ConductivityTensor}
\end{equation}%
Moreover, we obtained that all the currents share a common simple structure:
\begin{align}\label{eq:current}
\mathbf{j}_{e,h}(\mathbf{r})=\int\frac{d\epsilon}{2\pi}d\mathbf{r}'\chi_{e,h}(\epsilon)\left[\hat{\mathbf{v}}(\mathbf{r},\mathbf{r}',\epsilon)\hat{G}(\mathbf{r}',\mathbf{r},\epsilon)\right]^{<}.
\end{align}
Here we use the notation $\chi_{e}(\epsilon)=-e$ for the electric current and $\chi_{h}(\epsilon)=\epsilon$ for the heat current. Both the velocity and the quasiparticle Green's function in the above equation are fully renormalized by the interaction. The currents are related to the lesser component in the Keldysh space of the product of the renormalized velocity and Green's function. Equation~\ref{eq:current} is one of the central results of this paper; for the first time it is shown that the flow of energy occurs with the renormalized velocity. This structure of the heat current guarantees that the Wiedemann-Franz law is satisfied in the framework of the Fermi-liquid theory. Finally, let us remark that although in this paper we consider electrons interacting via the Coulomb interaction, the generality of the method allows to account for different interactions with a minimal effort. In particular, this scheme was highly useful in the analysis of the Nernst effect in disordered films in the presence of superconducting fluctuations.~\cite{KM2008}

The paper is organized as follows: In Sec.~\ref{sec:ElectricCond} we calculate the electric conductivity. We deliberately choose the well elaborated example of the electric conductivity in order to illustrate the main steps of our approach. We consider two scattering mechanisms; the Coulomb interaction between the electrons and the elastic scattering by impurities.  In Sec.~\ref{sec:KQEGradT} we repeat the scheme introduced in Sec.~\ref{sec:ElectricCond} in order to obtain the quantum kinetic equation for a system of electrons in the presence of a temperature gradient. In Sec.~\ref{sec:HeatCurrent} we derive the expressions for the different currents induced by the temperature gradient.  For completeness, we present the expression for the heat current generated by an electric field. Sections~\ref{sec:KQEGradT} and~\ref{sec:HeatCurrent} constitute the core of the paper. In the rest of the paper we demonstrate how to apply this scheme for various calculations and check that this method reproduces some known results. In particular, in Sec.~\ref{sec:FermiLiquidQKE} we show that the Wiedemann-Franz law is satisfied for a Fermi-liquid system. Then, in Sec.~\ref{sec:WF} we discuss the fate of the Wiedemann-Franz law when diffusive corrections arising due to the interplay of the electron-electron interaction and disorder are considered. In Sec.~\ref{sec:Onsager} we examine Onsager's relations. The scheme developed in this paper allows us to calculate the two thermoelectric currents separately. As an additional test, we demonstrate in Sec.~\ref{sec:Onsager} that our expressions for $\alpha_{xx}$ and $\tilde{\alpha}_{xx}$ satisfy the Onsager relation, $\tilde{\alpha}_{xx}=T\alpha_{xx}$. Appendix~\ref{App:ElecricCond} contains additional technical details of the derivation of the electric conductivity. In Appendix~\ref{App:FermiLiquidKubo} we present the calculation of the thermal conductivity for a Fermi-liquid system using the simplified version of the Kubo formula. We show that the simplified formula leads to an erroneous result. In Appendix~\ref{App:Drag} we concentrate on the contribution of the Coulomb drag to the different transport coefficients. We emphasize the difference between the role of the Coulomb drag in electric conductivity and that in thermal conductivity. Finally, in Appendix~\ref{sec:ElectricCond-SCF} we briefly describe the generalization of the scheme developed in this paper for superconducting fluctuations.

\section{The quantum kinetic scheme for analyzing the electric conductivity}\label{sec:ElectricCond}

In this paper we limit ourselves to the well known example of electrons interacting through the density channel (e.g. the Coulomb interaction) in order to demonstrate the logic of our scheme for studying transport phenomena. For the electric conductivity we start with the following action:
\begin{align}\label{eq:EC-S}\nonumber
\mathcal{S}&=\int{d\mathbf{r}dt}\left\{\phantom{\frac{1}{1}}\hspace{-2mm}\psi^{\dag}(\mathbf{r},t)i\partial_t\psi(\mathbf{r},t)
-\frac{(\boldsymbol{\nabla}\psi^{\dag}(\mathbf{r},t))(\boldsymbol{\nabla}\psi(\mathbf{r},t))}{2m}\right.\\
&\left.-\psi^{\dag}(\mathbf{r},t)\left[e\mathbf{r}\cdot\mathbf{E}+V_{imp}(\mathbf{r})\right]\psi(\mathbf{r},t)\right.\\\nonumber
&\left.-\frac{1}{2}\int{d\mathbf{r}'}\psi^{\dag}(\mathbf{r},t)\psi^{\dag}(\mathbf{r}',t)U(\mathbf{r}-\mathbf{r}')\psi(\mathbf{r}',t)\psi(\mathbf{r},t)\right\}.
\end{align}
Here $\psi(\mathbf{r})$ and $\psi^{\dag}(\mathbf{r})$ are the Grasmannian fields describing the quasiparticles, $\psi^{\dag}(\mathbf{r})=\sum_{\mathbf{p}}c_{\mathbf{p}}^{\dag}e^{-i\mathbf{pr}}$. The electric field is assumed to be constant, and we choose to work in the gauge $\mathbf{E}=-\boldsymbol{\nabla}\varphi(\mathbf{r})$. In general, the field operators include a spin index that is summed over in the above action. Here and in the following  we do not indicate (whenever it is possible) the spin indices because we do not consider any scattering mechanisms that flip the spins, or the Zeeman splitting. For convenience, we introduce the Hubbard-Stratonovich field $\phi(\mathbf{r})$ into the action:
\begin{align}\label{eq:EC-S-HS}\nonumber
\mathcal{S}&=\int{d\mathbf{r}dt}\left\{\phantom{\frac{1}{1}}\hspace{-2mm}\psi^{\dag}(\mathbf{r},t)i\partial_t\psi(\mathbf{r},t)
-\frac{(\boldsymbol{\nabla}\psi^{\dag}(\mathbf{r},t))(\boldsymbol{\nabla}\psi(\mathbf{r},t))}{2m}\right.\\\nonumber
&\left.-\psi^{\dag}(\mathbf{r},t)\left[e\mathbf{r}\cdot\mathbf{E}+V_{imp}(\mathbf{r})\right]\psi(\mathbf{r},t)
-\phi(\mathbf{r},t)\psi^{\dag}(\mathbf{r},t)\psi(\mathbf{r},t)\right.\\
&\left.+\frac{1}{2}\int{d\mathbf{r}'}\phi(\mathbf{r},t)U^{-1}(\mathbf{r}-\mathbf{r}')\phi(\mathbf{r}',t)\right\}.
\end{align}
After this transformation the system is described by two propagators: the quasiparticle Green's function,  $G(\mathbf{r},t;\mathbf{r}',t')$, and the propagator of the $\phi$ fields, $V(\mathbf{r},t;\mathbf{r}',t')$.
[We use the term propagators when referring to both these functions, while separately we name $\hat{G}(\mathbf{r},t;\mathbf{r}',t')$ the quasiparticle Green's function, and
$\hat{V}(\mathbf{r},t;\mathbf{r}',t')$ the propagator of the interactions.]

The general formulation of the quantum kinetic equation for transport phenomena assumes that at time $t=-\infty$ the system is at equilibrium. Then, an external field is adiabatically switched on generating currents in the system. We follow the Keldysh-Schwinger approach and use the matrix form of the propagators.~\cite{Keldysh1964,Rammer1986,Haug} We find it convenient to work in the basis of the retarded, advanced and Keldysh propagators. At equilibrium, the three components of the matrices are not independent,  $G^{K}(\epsilon)=(1-2n_F(\epsilon))(G^{R}(\epsilon)-G^{A}(\epsilon))$ and $V^{K}(\omega)=(1+2n_P(\omega))(V^{R}(\omega)-V^{A}(\omega))$; here $n_F(\epsilon)$ is the Fermi distribution function and $n_P(\omega)$ is the Bose distribution function. The equilibrium propagators give the properties of the system at $t=-\infty$. In order to describe the system in the non-equilibrium state caused by the external field, one should derive the system of equations for the propagators, i.e., the quantum kinetic equation. The starting point is the Dyson equations corresponding to the action which includes the external field. The Dyson equation for the electric field dependent Green's function of the quasiparticles is:
\begin{align}\label{eq:EC-DEG}
&\left[i\frac{\partial}{\partial{t}}+\frac{\boldsymbol{\nabla}^2}{2m}-V_{imp}(\mathbf{r})-e\mathbf{rE}+\mu
\right]\hat{G}(\mathbf{r},t;\mathbf{r}',t')\\\nonumber
&=\delta(\mathbf{r}-\mathbf{r}')\delta(t-t')\hspace{-0.5mm}+\hspace{-1.5mm}\int\hspace{-1mm}{dt_{\scriptscriptstyle1}}d\mathbf{r}_{\scriptscriptstyle1}\hat{\Sigma}(\mathbf{r},t;\mathbf{r}_{\scriptscriptstyle1},t_{\scriptscriptstyle1})\hat{G}(\mathbf{r}_{\scriptscriptstyle1},t_{\scriptscriptstyle1};\mathbf{r}',t').
\end{align}
The chemical potential appears because we work in the grand canonical ensemble. Note that  the specific time dependence of the retarded, advanced and Keldysh propagators is encoded in the definition of the different components of the matrix propagators. Correspondingly, when the matrix propagator is Fourier transformed, each of its components acquires the proper analytic structure in the complex frequency plane.

\begin{figure}[pt]
\begin{flushright}\begin{minipage}{0.5\textwidth}  \centering
        \includegraphics[width=0.85\textwidth]{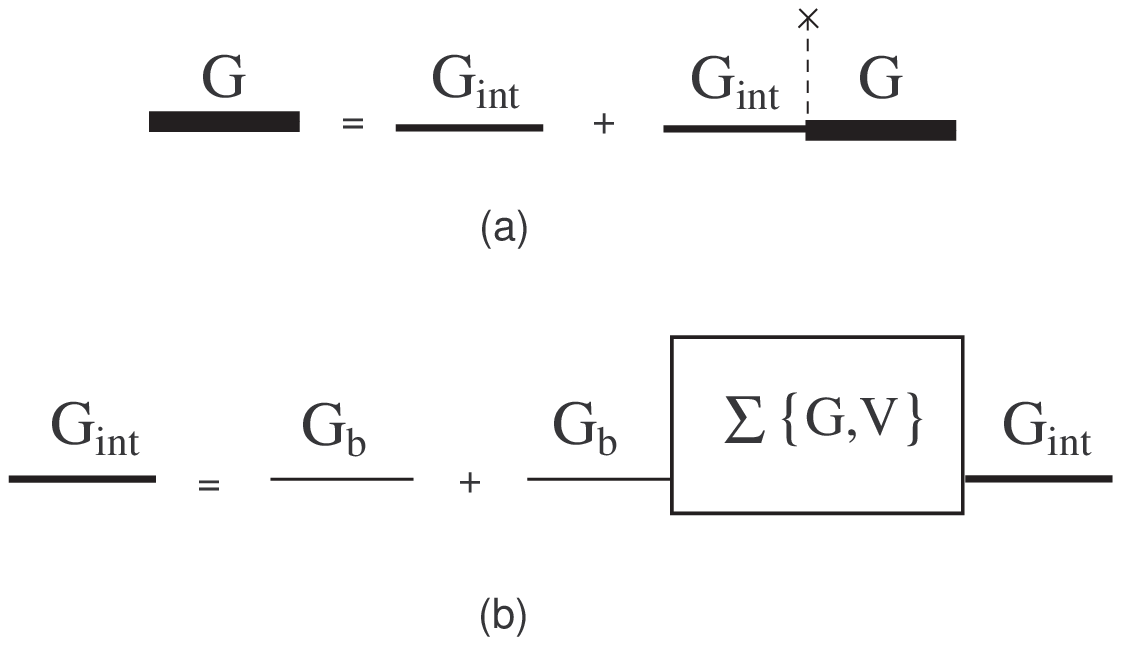}
                 \caption[0.4\textwidth]{\small (a) Illustration of Eq.~\ref{eq:EC-OpenDisorderG1} for the full Green's function $\hat{G}$. (b) The Dyson equation for $\hat{G}_{int}$ (see Eq.~\ref{eq:EC-OpenDisorderG2}). Note that $\hat{G}_{int}$ includes scattering by impurities only through $\hat{\Sigma}\{G,V\}$, which is a functional of the propagators $\hat{G}$ and $\hat{V}$. The bare Green's function, denoted by $\hat{G}_{b}$, is free from the interactions and the scattering by impurities.} \label{fig:Green'sFunction}
\end{minipage}\end{flushright}
\end{figure}

We choose to postpone the averaging over impurities until the last stage of the derivation of the current. Therefore,  the Green's function of
the quasiparticles contains open impurity lines as illustrated in the two coupled equations presented in Fig.~\ref{fig:Green'sFunction}:
\begin{subequations}\label{eq:EC-OpenDisorderG}
\begin{align}\label{eq:EC-OpenDisorderG1}
&\hat{G}(\mathbf{r},t;\mathbf{r}',t')=\hat{G}_{int}(\mathbf{r},t;\mathbf{r}',t')\\\nonumber
&\hspace{5mm}+\int{d\mathbf{r}_{\scriptscriptstyle1}dt_{\scriptscriptstyle1}}\hat{G}_{int}(\mathbf{r},t;\mathbf{r}_{\scriptscriptstyle1},t_{\scriptscriptstyle1})V_{imp}(\mathbf{r}_{\scriptscriptstyle1})\hat{G}(\mathbf{r}_{\scriptscriptstyle1},t_{\scriptscriptstyle1};\mathbf{r}',t');
\end{align}
\begin{align}\label{eq:EC-OpenDisorderG2}\nonumber
&\hat{G}_{int}(\mathbf{r},t;\mathbf{r}',t')=\hat{G}_b(\mathbf{r},t;\mathbf{r}',t')+\hspace{-1.5mm}\int\hspace{-1.5mm}{d\mathbf{r}_{\scriptscriptstyle1}dt_{\scriptscriptstyle1}}{d\mathbf{r}_{\scriptscriptstyle2}dt_{\scriptscriptstyle2}}\hat{G}_b(\mathbf{r},t;\mathbf{r}_{\scriptscriptstyle1},t_{\scriptscriptstyle1})\\
&\times\hat{\Sigma}(\mathbf{r}_{\scriptscriptstyle1},t_{\scriptscriptstyle1};\mathbf{r}_{\scriptscriptstyle2},t_{\scriptscriptstyle2})\hat{G}_{int}(\mathbf{r}_{\scriptscriptstyle2},t_{\scriptscriptstyle2};\mathbf{r}',t').
\end{align}
\end{subequations}
Here, $\hat{G}_{int}(\mathbf{r},t;\mathbf{r}',t')$ is the Green's function of interacting electrons, while $\hat{G}_{b}(\mathbf{r},t;\mathbf{r}',t')$ is free from both interactions and scattering by impurities. Note, that $\hat{G}_{int}(\mathbf{r},t;\mathbf{r}',t')$ includes partially the scattering by impurities. We will rewrite the Green's function and the kinetic equation in terms of the center of mass coordinates, $\mathbf{R}=(\mathbf{r}+\mathbf{r}')/2$, $\mathcal{T}=(t+t')/2$, and the relative coordinates, $\boldsymbol{\rho}=\mathbf{r}-\mathbf{r}'$, $\tau=t-t'$. There are two sources for the dependence of the Green's function on the center of mass coordinate, the electric field and the scattering by the impurities. To separate these two dependencies, we use the notation $\hat{G}(\mathbf{R},\mathcal{T};\boldsymbol{\rho},\tau;imp)$, where the explicit dependence on $\mathbf{R}$ is caused by the electric field, while the one arising due to the impurity potential is incorporated into $imp$. Next, we Fourier transform the Green's function with respect to the relative time coordinate. Simultaneously,  we wish to write the quantum kinetic equation in the gauge invariant form. Owing to the fact that the gauge invariant time derivative includes the scalar potential, one can modify the Fourier transform in the following way:~\cite{Haug, Mahan1983}
\begin{align}\label{eq:EC-Wigner}
\hat{\underline{G}}(\mathbf{R},\mathcal{T};\boldsymbol{\rho},\epsilon;imp)=\int{d\tau}e^{i(\epsilon+e\mathbf{ER})\tau}\hat{G}(\mathbf{R},\mathcal{T};\boldsymbol{\rho},\tau;imp).
\end{align}
The underscore is used to mark gauge invariant quantities.

After linearizing the quantum kinetic equation for $\hat{\underline{G}}$ with respect to the electric field, one obtains:
\begin{align}\label{eq:EC-QKE-E}\nonumber
&\left\{\hspace{-1mm}\left(\epsilon+\mu+\frac{\boldsymbol{\nabla}^2}{2m}\right)\hspace{-1mm}-\frac{e\mathbf{E}}{2m}\frac{\partial}{\partial\epsilon}\boldsymbol{\nabla}-V_{imp}(\mathbf{R}+\boldsymbol{\rho}/2)
-\frac{e\boldsymbol{\rho}\mathbf{E}}{2}\right\}\\&\times\hat{\underline{G}}(\boldsymbol{\rho},\epsilon;imp)
=\delta(\boldsymbol{\rho})
+\int{d\mathbf{r}_{\scriptscriptstyle1}}\hat{\underline{\Sigma}}\left(\boldsymbol{\rho}-\mathbf{r}_{\scriptscriptstyle1},\epsilon;imp\right)\\\nonumber
&\times\left\{1-\frac{e\mathbf{E}}{2}
\left[\mathbf{r}_{\scriptscriptstyle1}\frac{\overleftarrow{\partial}}{\partial\epsilon}-(\boldsymbol{\rho}-\mathbf{r}_{\scriptscriptstyle1})\frac{\overrightarrow{\partial}}{\partial\epsilon}\right]\right\}
\hat{\underline{G}}\left(\mathbf{r}_{\scriptscriptstyle1},\epsilon;imp\right).
\end{align}
A detailed derivation is presented in Appendix~\ref{App:ElecricCond}. Here,  $\boldsymbol{\nabla}=\frac{1}{2}\boldsymbol{\nabla}_{R}+\boldsymbol{\nabla}_{\rho}$. Clearly, for a constant electric field there is no explicit dependence of the gauge invariant Green's function on the center of mass coordinate, and the entire dependence of $\hat{\underline{G}}(\boldsymbol{\rho},\epsilon;imp)$ on $\mathbf{R}$ is incorporated into $imp$. To find $\hat{\underline{G}}$, we write the Green's function as a sum of two terms; the equilibrium Green's function $\hat{g}_{eq}(\boldsymbol{\rho},\epsilon;imp)$  and the $\mathbf{E}$-dependent part $\hat{G}_{\mathbf{E}}(\boldsymbol{\rho},\epsilon;imp)$. The equation for the equilibrium Green's function is:
\begin{align}\label{eq:EC-Geq}
&\left[\epsilon+\frac{\boldsymbol{\nabla}^2}{2m}+\mu-V_{imp}\right]
\hat{g}_{eq}(\boldsymbol{\rho},\epsilon;imp)=\delta(\boldsymbol{\rho})\\\nonumber
&-\int{d\mathbf{r}_{\scriptscriptstyle1}}\hat{\sigma}_{eq}(\boldsymbol{\rho}-\mathbf{r}_{\scriptscriptstyle1},\epsilon;imp)\hat{g}_{eq}(\mathbf{r}_{\scriptscriptstyle1},\epsilon;imp).
\end{align}
Notice that the Laplacian in the above equation includes the derivatives with respect to the center of mass and relative coordinates. [The dependence on the center of mass coordinate is caused by $V_{imp}$.] At equilibrium, the components of the matrix Green's function  are:
\begin{align}\label{eq:GeqMatrix}
g_{eq}^{R,A}&(\boldsymbol{\rho},\epsilon;imp)\\\nonumber
&=\left[\epsilon+\boldsymbol{\nabla}^2/2m+\mu-V_{imp}-\sigma_{eq}^{R,A}(\boldsymbol{\rho},\epsilon;imp)\right]^{-1}\\ \nonumber &\hspace{-7mm}g_{eq}^{K}(\boldsymbol{\rho},\epsilon;imp)\\\nonumber
&= (1-2n_F(\epsilon))[g_{eq}^{R}(\boldsymbol{\rho},\epsilon;imp)-g_{eq}^{A}(\boldsymbol{\rho},\epsilon;imp)].
\end{align}
Here $g_{eq}^{R(A)}$ is analytic on the upper (lower) half of the complex frequency plane $\epsilon$.

The detailed calculation of $\hat{G}_{\mathbf{E}}$ presented in Appendix~\ref{App:ElecricCond} yields:
\begin{align}\label{eq:EC-GE}
&\hat{G}_{\mathbf{E}}(\boldsymbol{\rho},\epsilon)=\hat{g}_{eq}\left(\epsilon\right)
\hat{\Sigma}_{\mathbf{E}}\left(\epsilon\right)\hat{g}_{eq}\left(\epsilon\right)\\\nonumber
&-\frac{ie\mathbf{E}}{2}\left[\frac{\partial\hat{g}_{eq}\left(\epsilon\right)}{\partial\epsilon}\hat{\mathbf{v}}_{eq}(\epsilon)\hat{g}_{eq}\left(\epsilon\right)
-\hat{g}_{eq}\left(\epsilon\right)\hat{\mathbf{v}}_{eq}(\epsilon)\frac{\partial\hat{g}_{eq}\left(\epsilon\right)}{\partial\epsilon}
\right].
\end{align}
From now on, whenever the dependence on the coordinates in the product of matrices is not specified, it  should be understood as a convolution in real space.  The matrix $\hat{\mathbf{v}}_{eq}$ is the velocity of the quasiparticles renormalized by the interactions at equilibrium:
\begin{align}\label{eq:EC-velocityQP}
\hat{\mathbf{v}}_{eq}&(\mathbf{r},\mathbf{r}',\epsilon)=-i\lim_{\mathbf{r}'\rightarrow\mathbf{r}}\frac{\boldsymbol{\nabla}-\boldsymbol{\nabla}'}{2m}
-i(\mathbf{r}-\mathbf{r}')\hat{\sigma}_{eq}(\mathbf{r},\mathbf{r}',\epsilon).
\end{align}
The equation for $\hat{G}_{\mathbf{E}}$ contains the electric field dependent self-energy, $\hat{\Sigma}_{\mathbf{E}}$,  which by itself is a function of the $\mathbf{E}$-dependent propagators. Thus, in order to find $\hat{G}_{\mathbf{E}}$, one has to determine the structure of the self-energy. Once the form of the self-energy is fixed, one should take into consideration that each of the propagators in $\hat{\Sigma}_{\mathbf{E}}$ may depend on the electric field.

To find the response to the applied electric field, we need also to derive the quantum kinetic equation for the propagator of the interaction. The Dyson equation for $\hat{V}$ is:
\begin{align}\label{eq:EC-DEV}\nonumber
\int{d\mathbf{r}_{\scriptscriptstyle1}dt_{\scriptscriptstyle1}}&U^{-1}(\mathbf{r-r}_{\scriptscriptstyle1})\delta(t-t_{\scriptscriptstyle1})\hat{V}(\mathbf{r}_{\scriptscriptstyle1},t_{\scriptscriptstyle1};\mathbf{r}',t')=\delta(\mathbf{r}-\mathbf{r}')\\
&-\int{d\mathbf{r}_{\scriptscriptstyle1}dt_{\scriptscriptstyle1}}\hat{\Pi}(\mathbf{r},t;\mathbf{r}_{\scriptscriptstyle1},t_{\scriptscriptstyle1})\hat{V}(\mathbf{r}_{\scriptscriptstyle1},t_{\scriptscriptstyle1};\mathbf{r}',t'),
\end{align}
where $\hat{\Pi}$ is the self-energy (polarization operator) for the interaction field $\phi(\mathbf{r},t)$. Unlike the kinetic equation for the  quasiparticle Green's function, the electric field does not appear explicitly in the equation for $\hat{V}$. Indeed, since the interaction field $\phi$ is neutral, the electric field can only enter through the self-energy term $\hat{\Pi}$ that contains the Green's functions of the charged quasiparticles $\hat{G}$. For the same reason, the Fourier transform of the relative time in $\hat{V}(\mathbf{r},t;\mathbf{r}',t')$ should be performed without the gauge factor.

The calculation of the electric conductivity requires the expression for the electric current in terms of the  propagators depending on the electric field. We derive the electric current through the continuity equation. We start from the density which can be related to the lesser Green's function in the following way:
\begin{align}\label{eq:EC-ChargeDensity}
n(\mathbf{r},t)&=2\lim_{
                           \begin{array}{c}
                             \scriptstyle{ \mathbf{r}'\rightarrow\mathbf{r}} \\
                             \scriptstyle{t'\rightarrow{t}^{+}} \\
                           \end{array}}
\left\langle{\psi^{\dag}(\mathbf{r}',t')\psi(\mathbf{r},t)}\right\rangle\\\nonumber
&\equiv{-2i}\lim_{
                           \begin{array}{c}
                             \scriptstyle{ \mathbf{r}'\rightarrow\mathbf{r}} \\
                             \scriptstyle{t'\rightarrow{t}} \\
                           \end{array}}G^{<}(\mathbf{r},t;\mathbf{r}',t').
\end{align}
We use the notation $t'\rightarrow{t}^{+}$ to indicate that the limit should be taken in such a way that $t$ is on the upper branch of the Keldysh contour, while $t'$ is on the lower branch.~\cite{Rammer1986} The summation over the spin projection results in a factor of $2$.  The lesser component of the Green's function can be written in terms of the retarded, advanced and Keldysh Green's functions through the relation $G^{<}=(G^{K}-G^{R}+G^{A})/2$. Here $\langle{A}\rangle$ denotes the quantum mechanical averaging with the action given in Eq.~\ref{eq:EC-S-HS}. Therefore, the Green's function is fully dressed by the interactions and depends on the impurity potential. In addition, $\hat{G}$  is a function of the electric field. Since we find the current by extracting it from the continuity equation, we assume that the electric field has  some spatial modulations that will be set to zero at the end of the procedure.

The continuity equation for the charge density, $-e\dot{n}(\mathbf{r},t)+\boldsymbol{\nabla}\mathbf{j}_e(\mathbf{r},t)=0$, can be be written as:
\begin{align}\label{eq:EC-ContinuityEq}
&\boldsymbol{\nabla}\mathbf{j}_e(\mathbf{r},t)=2e\lim_{\begin{array}{c}\scriptstyle{ \mathbf{r}'\rightarrow\mathbf{r}} \\ \scriptstyle{t'\rightarrow{t}^{+}} \\
\end{array}}\left(\frac{\partial}{\partial{t}}+\frac{\partial}{\partial{t}'}\right)\left\langle{\psi^{\dag}(\mathbf{r}',t')\psi(\mathbf{r},t)}\right\rangle.
\end{align}
Under the average, the equations of motion for the fields $\psi$ and $\psi^{\dag}$  allow us to rewrite the continuity equation as follows:
\begin{align}\label{eq:EC-Current}\nonumber
\boldsymbol{\nabla}\mathbf{j}_e&=2ie\lim_{\begin{array}{c}\scriptstyle{ \mathbf{r}'\rightarrow\mathbf{r}} \\ \scriptstyle{t'\rightarrow{t}^{+}} \\
\end{array}}\left\langle{\left[\frac{\boldsymbol{\nabla}^2-\boldsymbol{\nabla}'^2}{2m}-e(\mathbf{r-r'})\mathbf{E}-V_{imp}(\mathbf{r})\right.}\right.\\
&\left.\left.+V_{imp}(\mathbf{r}')-\phi(\mathbf{r},t)+\phi(\mathbf{r}',t')\phantom{\frac{\boldsymbol{\nabla}}{m}}\hspace{-4mm}\right]\psi^{\dag}(\mathbf{r}',t')\psi(\mathbf{r},t)\right\rangle.
\end{align}
We may express the RHS of the above equation in terms of the Green's functions and self-energies:
\begin{align}\label{eq:EC-Current2}
\boldsymbol{\nabla}\mathbf{j}_e&=2e\lim_{\mathbf{r}'\rightarrow\mathbf{r}}\left[\frac{\boldsymbol{\nabla}^2-\boldsymbol{\nabla}'^2}{2m}\hat{G}(\mathbf{r},t;\mathbf{r}',t)\right.\\\nonumber
&\left.-\int{d\mathbf{r}_{\scriptscriptstyle1}}dt_{\scriptscriptstyle1}\hat{\Sigma}(\mathbf{r},t;\mathbf{r}_{\scriptscriptstyle1},t_{\scriptscriptstyle1})\hat{G}(\mathbf{r}_{\scriptscriptstyle1},t_{\scriptscriptstyle1};\mathbf{r}',t)\right.\\\nonumber
&\left.+\int{d\mathbf{r}_{\scriptscriptstyle1}}dt_{\scriptscriptstyle1}\hat{G}(\mathbf{r},t;\mathbf{r}_{\scriptscriptstyle1},t_{\scriptscriptstyle1})\hat{\Sigma}(\mathbf{r}_{\scriptscriptstyle1},t_{\scriptscriptstyle1};\mathbf{r}',t)
\right]^{<}.
\end{align}
We use the notation $\big[...\big]^{<}$ to indicate that the expression inside the square brackets is a matrix and the current corresponds to the lesser component of this matrix. The explicit dependence on $\mathbf{E}$ in Eq.~\ref{eq:EC-Current} dropped out as a result of taking the limit $\mathbf{r}'\rightarrow\mathbf{r}$. However, the Green's functions as well as the self-energies in the expression for the electric current depend on the electric field.

To resolve the expression for the electric current, we have to represent the RHS  of Eq.~\ref{eq:EC-Current2} as a gradient of some function. With this in mind, we write the convolutions of the Green's functions and self-energies in terms of the center of mass and relative coordinates:
\begin{align}\label{eq:EC-Expanssion}\nonumber
&\left[\hat{G}\hat{\Sigma}-\hat{\Sigma}\hat{G}\right]^{<}\\\nonumber
&=\lim_{\boldsymbol{\rho}\rightarrow0}\int{d\mathbf{r}_{\scriptscriptstyle1}}dt_{\scriptscriptstyle1}\left[
\hat{G}\left(\mathbf{R}+\frac{\mathbf{r}_{\scriptscriptstyle1}+\boldsymbol{\rho}/{2}}{2};\frac{\boldsymbol{\rho}}{2}-\mathbf{r}_{\scriptscriptstyle1},-t_{\scriptscriptstyle1};imp\right)\right.\\\nonumber
&\left.\times\hat{\Sigma}\left(\mathbf{R}+\frac{\mathbf{r}_{\scriptscriptstyle1}-\boldsymbol{\rho}/2}{2};\frac{\boldsymbol{\rho}}{2}+\mathbf{r}_{\scriptscriptstyle1},t_{\scriptscriptstyle1};imp\right)\right.\\\nonumber
&\left.-\hat{\Sigma}\left(\mathbf{R}+\frac{\mathbf{r}_{\scriptscriptstyle1}+\boldsymbol{\rho}/{2}}{2};\frac{\boldsymbol{\rho}}{2}-\mathbf{r}_{\scriptscriptstyle1},-t_{\scriptscriptstyle1};imp\right)\right.\\
&\left.\times\hat{G}\left(\mathbf{R}+\frac{\mathbf{r}_{\scriptscriptstyle1}-\boldsymbol{\rho}/2}{2};\frac{\boldsymbol{\rho}}{2}+\mathbf{r}_{\scriptscriptstyle1},t_{\scriptscriptstyle1};imp\right)\right]^{<}
\end{align}
To extract the current out of the continuity equation we expand $[\hat{\Sigma}\hat{G}-\hat{G}\hat{\Sigma}]^{<}$ with respect to the deviation from the center of mass coordinate. [This step resembles the gradient expansion discussed, for example, in Refs.~\onlinecite{Rammer1986,Haug}]  As a consequence of the structure of Eq.~\ref{eq:EC-Expanssion},  one may immediately see that all even terms in the expansion vanish. Generally speaking, the expansion includes both the explicit dependence on $\mathbf{R}$ arising due to the external field and the one entering through the impurity potential. Recall that ultimately we are interested in  the current averaged over space. Owing to the fact that any correlation function between the impurity centers depends only on their relative distance, the derivatives of $V_{imp}$ with respect to the center of mass coordinate vanish upon averaging. Therefore,  in the regime of linear response,  it is enough to keep only the first non-vanishing term in the expansion:
\begin{align}\label{eq:EC-Expanssion2}
&\left[\hat{G}\hat{\Sigma}-\hat{\Sigma}\hat{G}\right]^{<}\approx
\int{d\mathbf{r}_{1}}dt_{1}\boldsymbol{\nabla}_{\mathbf{R}}\left[\hat{G}\left(\mathbf{R};-\mathbf{r}_{\scriptscriptstyle1},-t_{\scriptscriptstyle1};imp\right)\right.\\\nonumber
&\left.\times\frac{\mathbf{r}_{\scriptscriptstyle1}}{2}\hat{\Sigma}\left(\mathbf{R};\mathbf{r}_{\scriptscriptstyle1},t_{\scriptscriptstyle1};imp\right)\right]^{<}+h.c.
\end{align}
After performing the gauge invariant Fourier transform on the current, one come to the very compact expression:
\begin{align}\label{eq:EC-AverageCurrent}
\mathbf{j}_e&=ie\int\frac{d\epsilon}{2\pi}{d\mathbf{r}'}\left[\hat{\underline{\mathbf{v}}}(\mathbf{r},\mathbf{r}',\epsilon)\hat{\underline{G}}(\mathbf{r}',\mathbf{r},\epsilon)
\right]^{<}+h.c.
\end{align}
We used the fact that $\hat{G}$ and $\hat{\mathbf{v}}$ in the expression for the current have the same center of mass coordinate, while their relative time coordinates differ by a sign. Therefore, when we  Fourier transform both the velocity and the Green's function according to Eq.~\ref{eq:EC-Wigner}, the two gauge factors cancel each other. As a result, $\mathbf{j}_e$ becomes a simple convolution of two gauge invariant quantities. The expression for the current contains the renormalized velocity as defined in Eq.~\ref{eq:EC-velocityQP} with the only difference that the self-energy may depend on the electric field, $\hat{\mathbf{\underline{v}}}=\hat{\mathbf{v}}_{eq}+\hat{\mathbf{v}_{\mathbf{E}}}$, where $\hat{\mathbf{v}}_{\mathbf{E}}(\mathbf{r},\mathbf{r}',\epsilon)=-i(\mathbf{r-r}')\hat{\Sigma}_{\mathbf{E}}(\mathbf{r},\mathbf{r}',\epsilon)$.

\begin{widetext}
The final step in the derivation of the electric conductivity is to insert the expressions for $\hat{G}_{\mathbf{E}}$ and $\hat{\mathbf{v}}_{\mathbf{E}}$ into Eq.~\ref{eq:EC-AverageCurrent}, and complete linearizing it with respect to the electric field (for more detailed see the end of Appendix~\ref{App:ElecricCond}). Using the known relations between the components of the Green's function at equilibrium, one gets:
\begin{align}\label{eq:EC-JeFinal}
j_{e}^{i}&=-\frac{e^2E_{j}}{2}\int\frac{d\epsilon}{2\pi}\frac{\partial{n_F(\epsilon)}}{\partial\epsilon}\left[v_{i}^R(\epsilon)g_{eq}^{R}(\epsilon)v_{j}^A(\epsilon)g_{eq}^A(\epsilon)
+v_{i}^R(\epsilon)g_{eq}^{R}(\epsilon)v_{j}^R(\epsilon)g_{eq}^A(\epsilon)-
v_{i}^R(\epsilon)g_{eq}^{R}(\epsilon)v_{j}^R(\epsilon)g_{eq}^R(\epsilon)\right.\\\nonumber&\left.
-g_{eq}^{R}(\epsilon)v_{j}^R(\epsilon)g_{eq}^R(\epsilon)v_{i}^A(\epsilon)
\right]-{e^2E_j}\int\frac{d\epsilon}{2\pi}n_F(\epsilon)\left[v_{i}^R(\epsilon)\frac{\partial{g}_{eq}^R(\epsilon)}{\partial\epsilon}v_{j}^R(\epsilon)g_{eq}^R(\epsilon)-
v_{i}^R(\epsilon){g}_{eq}^R(\epsilon)v_{j}^R(\epsilon)\frac{\partial{g}_{eq}^R(\epsilon)}{\partial\epsilon}\right]\\\nonumber
&-{ie}\int\frac{d\epsilon}{2\pi}v_{i}^R(\epsilon)g_{eq}^R(\epsilon)\left[\Sigma_{\mathbf{E}}^{<}(\epsilon)(1-n_F(\epsilon))+\Sigma_{\mathbf{E}}^{>}(\epsilon)n_F(\epsilon)\right](g_{eq}^{R}(\epsilon)-g_{eq}^A(\epsilon))
+c.c.
\end{align}
\end{widetext}
The second term in the above expression is zero for the longitudinal current, but it becomes important when the transverse current in the presence of a magnetic field is considered.

The obtained result for the electric current is expressed in terms of the renormalized Green's functions and velocities, i.e., it holds to all order with respect to the electron-electron interaction. In practice, to get a quantitative result, one has to specify the form of the self-energy and to average over the disorder. In the case of the \textit{electric conductivity}, the terms in Eq.~\ref{eq:EC-JeFinal} are equivalent to those given by the Kubo formula after performing in the latter the analytic continuation to the real frequency. The derivatives with respect to the frequency are the same derivatives that one gets after expanding the Kubo formula with respect to the external frequency. While the derivatives with respect to the frequency in the first two terms of Eq.~\ref{eq:EC-JeFinal} appear explicitly, in the last term they reveal themselves only after linearizing the self-energy with respect to $\mathbf{E}$.  If interested, one can generate the perturbative expansion order by order in a systematic fashion, and give a diagrammatic interpretation for each of the terms. This technique automatically determines the analytic structure, the way the distribution functions enter as well as the numerical factors of all the diagrams. [We checked that for the electric conductivity the set of diagrams obtained in the quantum kinetic approach coincides with the one given by the Kubo formula. We reproduced all the diagrams contributing to the Altshuler-Aronov corrections to the conductivity.~\cite{Altshuler1985,Raimondi2003} In order to get these known corrections, one should properly perform the averaging over the disorder in Eq.~\ref{eq:EC-JeFinal}.]

To summarize, we demonstrated the main steps in the derivation of currents in response to an external field using  the electric conductivity as an instructive example. In particular, we showed how to find from the continuity equation the current in terms of the renormalized (dressed) quantities. In the subsequent sections we shall follow the same scheme in order to find the heat and electric currents as a response to a temperature gradient. In addition, in Appendix~\ref{sec:ElectricCond-SCF} we describe how this scheme can be applied for an interaction in the Cooper channel.

\section{Derivation of the kinetic equation in the presence of a temperature gradient}\label{sec:KQEGradT}

Similar to the calculation of the electric conductivity described in the previous section, the derivation of the electric and heat currents as a response to a temperature gradient consists of two steps. One has to derive the kinetic equation and to find the expressions for the currents from the continuity equations.

We start with the quantum kinetic equation for the matrix Green's function. Since the temperature gradient is not a mechanical force, one cannot obtain the response to $\boldsymbol{\nabla}T$ just following the route we elaborated for the electric field. To overcome this obstacle, we (following Luttinger~\cite{Luttinger1964}) introduce an auxiliary gravitational field that enters the action. We will show how to establish a direct connection between the response to the gravitational field and the response to the temperature gradient.

\begin{figure}[pt]
\begin{flushright}\begin{minipage}{0.5\textwidth}  \centering
        \includegraphics[width=0.7\textwidth]{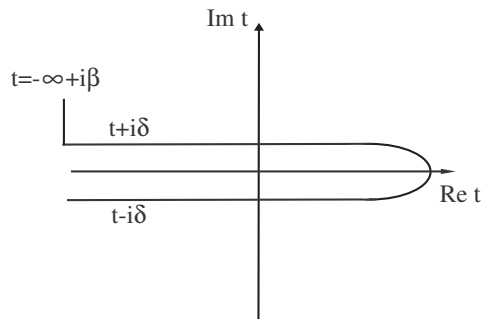}
                 \caption[0.4\textwidth]{\small The Keldysh contour.
                 The first, vertical, part of the contour is parallel to the imaginary axis starting at $t=-\infty+i\beta$ and ending at $t=-\infty+i\delta$. The second, horizontal, part of the contour is parallel to the real time axis. Here $\beta$ is the inverse temperature.} \label{fig:KeldyshContour}
\end{minipage}\end{flushright}
\end{figure}

In the Keldysh-Schwinger approach,~\cite{Keldysh1964,Rammer1986,Haug} the Green's functions are defined using the time ordering operator $T_C$ along the Keldysh contour $C$ (see Fig.~\ref{fig:KeldyshContour}):
\begin{align}\label{eq:QKET-KeldyshGF}
\hat{G}&(\mathbf{r},t;\mathbf{r}',t')\\\nonumber&=-i\left\langle{T_C\left\{e^{-i\int_C{d\tau}(\mathcal{H}(\tau)-\mu{N})}\psi(\mathbf{r},t)\psi^{\dag}(\mathbf{r}',t')\right\}}\right\rangle.
\end{align}
The integration in the exponent contains two parts.  The first part of the contour is parallel to the imaginary axis starting at $t=-\infty+i\beta$ and ending at $t=-\infty+i\delta$. The second, which is parallel to the real time axis, gives the evolution of the system in time. The Green's function along the second part of the contour is described by the kinetic equation. In the derivation of transport properties we usually assume that the driving force is switched on adiabatically, starting at $t=-\infty+i\delta$. Thus, the integration of the Hamiltonian in the exponent along the first part of the contour yields the thermal distribution of a system in the unperturbed state. As a result, at $t=-\infty+i\delta$ when the external perturbation starts to act, the system is at thermal equilibrium.

In contrast to the electric field described in the preceding section, a space dependent temperature, $T(\mathbf{r})$, influences the first part of the integration. In principle, one may try to generalize the integration along the first part of the Keldysh contour in the following way:
\begin{align}\label{eq:QKET-CountourTGradient}
\hat{\rho}=\exp\left\{-i\int{d\mathbf{r}}\int_{-\infty+i\beta(\mathbf{r})}^{-\infty+i\delta}{dt}\left(h(\mathbf{r})-{\mu}n(\mathbf{r})\right)\right\},
\end{align}
where $h(\mathbf{r})$ is the Hamiltonian density. Clearly, the density matrix obtained as a result of this integration corresponds to a non-uniform state. On the other hand, the integration along the second part of the Keldysh contour (parallel to the real axis) is independent of the temperature and, therefore, the quantum kinetic equation does not include any external perturbation. In other words, we face the problem of finding highly non-trivial initial state before  we even start to study its time evolution.  To avoid this complicated task, we shall reformulate the problem in such a way that the initial state of the system is uniform in space.

A similar problem of treating a spatial varying initial state appears when the response to a gradient of the density is studied. According to Einstein's construction, in the case of a density gradient the stationary state  can be obtained by adding a scalar potential at time $t=-\infty$ in such a way that the electro-chemical potential is kept constant, $\zeta(\mathbf{r})=\mu(\mathbf{r})-e\varphi(\mathbf{r})=const$, and the initial state is uniform. Then, the response to the gradient of the chemical potential (density) can be derived by adiabatically switching off the scalar potential, $\varphi(\mathbf{r})$. [This is one way to interpret the Einstein relation.] In a similar fashion, Luttinger~\cite{Luttinger1964} introduced a gravitational field, $\gamma(\mathbf{r})$, as a counterpart of the non-uniform temperature. This auxiliary field allows deriving the equivalent of the Einstein relation for studying thermal transport.

We wish to introduce Luttinger's gravitational field $\gamma(\mathbf{r})$ into the quantum kinetic approach. Since the gravitational field can be considered as a spatial dependent measure of the time coordinate, we shall study the following action:
\begin{align}\label{eq:QKET-S}
&\mathcal{S}=\int{d\mathbf{r}dt}\gamma(\mathbf{r})\left\{\phantom{\frac{1}{1}}\hspace{-2mm}
\psi^{\dag}(\mathbf{r},t)\frac{i\partial_t}{\gamma(\mathbf{r})}\psi(\mathbf{r},t)\right.\\\nonumber
&\left.
-\frac{(\boldsymbol{\nabla}\psi^{\dag}(\mathbf{r},t))(\boldsymbol{\nabla}\psi(\mathbf{r},t))}{2m}
-\left[V_{imp}(\mathbf{r})-\mu\right]\psi^{\dag}(\mathbf{r},t)\psi(\mathbf{r},t)
\right.\\\nonumber
&\left.-\phi(\mathbf{r},t)\psi^{\dag}(\mathbf{r},t)\psi(\mathbf{r},t)
\hspace{-0.5mm}+\hspace{-0.5mm}\frac{1}{2}\int\hspace{-1mm}{d\mathbf{r}'}\phi(\mathbf{r},t)U^{-1}(\mathbf{r}-\mathbf{r}')\phi(\mathbf{r}',t)\right\}.
\end{align}
Let us first fix $\gamma(\mathbf{r}$) in such a way that under the combined effect of the spatially dependent temperature, $T(\mathbf{r})=T+\delta{T}(\mathbf{r})$, and the gravitational field the system remains at equilibrium with a uniform effective temperature $\gamma_0T$.  In other words, the distribution function that describes the state of the system  does not evolve in time, and it is equal to $n_F(\epsilon)=[e^{\epsilon/(\gamma_{0}T)}+1]^{-1}$. One may notice that the gravitational field enters Eq.~\ref{eq:QKET-S} only through  the product $t\gamma(\mathbf{r})$. Correspondingly, after Fourier transforming the time coordinate, this field is coupled to the frequency as $\epsilon/\gamma(\mathbf{r})$. Therefore, under the combined effect of the temperature and the gravitational field, the equilibrium distribution function becomes
\begin{align}\label{eq:CountourGrad+Grav}
n_{F}(\epsilon)&=\left[\exp\left(\frac{\epsilon}{\gamma(\mathbf{r})T(\mathbf{r})}\right)+1\right]^{-1},
\end{align}
where,  the gravitational field has to be chosen so that the product $\gamma(\mathbf{r})\cdot{T}(\mathbf{r})=\gamma_0T$ is constant in space. [We keep $\gamma_0$ unspecified until the end of the derivation of the thermal and thermoelectric currents when it is set to be $1$. We shall see that our choice to leave $\gamma_0$  simplifies the analysis.]

Next, in order to find the response of the system to the temperature gradient, we adiabatically switch off the gravitational field starting at time $t=-\infty+i\delta$ (or equivalently switch on the same field with an opposite sign). Now, the situation is similar to the one we encountered in studying electric conductivity. While for the first part of the integration, parallel to the imaginary axis, the combined effect of the temperature and the gravitational field maintains the system at equilibrium, in the second part of the contour integration the change in the gravitational field perturbs the system.

In the following we shall study the kinetic equation in the presence of the gravitational field as described by the  action in Eq.~\ref{eq:QKET-S}. Since we are specifically interested in switching \textit{off} the field $\gamma(\mathbf{r})=\gamma_{0}T/T(\mathbf{r})$, in the end of the derivation we will set $\delta\gamma(\mathbf{r})=\delta{T}(\mathbf{r})/T$. [The last relation holds for the linear response. Pay attention to the sign in this relation. It is a consequence of the fact that  switching off the field $\delta\gamma(\mathbf{r})$ is equivalent to switching on the field $-\delta\gamma(\mathbf{r})$.] According to Eq.~\ref{eq:QKET-S}, the Dyson equation for the Green's function in the presence of the gravitational field is:
\begin{align}\label{eq:QKET-DEG}\nonumber
&\left[i\frac{\partial}{\partial{t}}+\frac{\boldsymbol{\nabla}(\gamma(\mathbf{r})\boldsymbol{\nabla})}{2m}-\gamma(\mathbf{r})\left(V_{imp}(\mathbf{r})-\mu\right)
\right]\hat{G}(\mathbf{r},t;\mathbf{r}',t')\\
&=\delta(\mathbf{r}-\mathbf{r}')\delta(t-t')\\\nonumber
&+\gamma(\mathbf{r})\int{dt_{\scriptscriptstyle1}}d\mathbf{r}_{\scriptscriptstyle1}\hat{\Sigma}(\mathbf{r},t;\mathbf{r}_{\scriptscriptstyle1},t_{\scriptscriptstyle1})\gamma(\mathbf{r}_{\scriptscriptstyle1})\hat{G}(\mathbf{r}_{\scriptscriptstyle1},t_{\scriptscriptstyle1};\mathbf{r}',t').
\end{align}
The collision integral contains $\gamma(\mathbf{r}_{\scriptscriptstyle1})$ because each integration over time is accompanied by the gravitational field. At this stage, if to write the gravitational field as $\gamma(\mathbf{r})=\gamma_0+\mathbf{r\boldsymbol{\nabla}}\gamma$, it is possible to expand straightforwardly the kinetic equation up to the linear order in $\boldsymbol{\nabla}\gamma$. However, a radical simplification may be achieved by applying the following transformation to the kinetic equation:
\begin{align}\label{eq:QKET-GaugeTrans}
\hat{Y}(\mathbf{r},t;\mathbf{r}',t')=\gamma^{-1/2}(\mathbf{r})\underline{\underline{\hat{Y}}}(\mathbf{r},t;\mathbf{r}',t')\gamma^{-1/2}(\mathbf{r}'),
\end{align}
where $\hat{Y}$ can be the Green's function or the self-energy. When terms of the order $(\boldsymbol{\nabla}\gamma)^2$ and $\boldsymbol{\nabla}^2\gamma$ are neglected, the equation for the Green's function $\underline{\underline{\hat{G}}}(\mathbf{r},t;\mathbf{r}',t')$ acquires the form:
\begin{align}\label{eq:QKET-DEG2}\nonumber
&\gamma^{1/2}(\mathbf{r})\left[i\frac{\partial}{\gamma(\mathbf{r})\partial{t}}+\frac{\boldsymbol{\nabla}^2}{2m}-V_{imp}(\mathbf{r})+\mu\right]\gamma^{-1/2}(\mathbf{r}')\\&\times\underline{\underline{\hat{G}}}(\mathbf{r},t;\mathbf{r}',t')
=\delta(\mathbf{r}-\mathbf{r}')\delta(t-t')\\\nonumber
&+\gamma^{1/2}(\mathbf{r})\int{dt_{\scriptscriptstyle1}d\mathbf{r}_{\scriptscriptstyle1}}\underline{\underline{\hat{\Sigma}}}(\mathbf{r},t;\mathbf{r}_{\scriptscriptstyle1},t_{\scriptscriptstyle1})\underline{\underline{\hat{G}}}(\mathbf{r}_{\scriptscriptstyle1},t_{\scriptscriptstyle1};\mathbf{r}',t')\gamma^{-1/2}(\mathbf{r}').
\end{align}
Multiplying the above equation by $\gamma^{-1/2}(\mathbf{r})$ and $\gamma^{1/2}(\mathbf{r}')$, we eliminate the gravitational field from all terms beside the time derivative.

After Fourier transforming the relative time in Eq.~\ref{eq:QKET-DEG2}, the kinetic equation in the regime of linear response becomes:
\begin{align}\label{eq:QKET-DEGWig}\nonumber
&\left\{\left(1-\frac{\mathbf{R}\boldsymbol{\nabla}\gamma}{\gamma_0}-\frac{\boldsymbol{\rho\boldsymbol{\nabla}}\gamma}{\gamma_0}\right)\frac{\epsilon}{\gamma_0}
+\frac{\boldsymbol{\nabla}^2}{2m}\right.\\
&\left.-V_{imp}(\mathbf{R}+\boldsymbol{\rho}/2)+\mu
\phantom{\frac{1}{1}}\hspace{-2mm}\right\}\hat{\underline{\underline{G}}}(\mathbf{R};\boldsymbol{\rho},\epsilon)
=\delta(\boldsymbol{\rho})\\\nonumber
&+\int{d\mathbf{r}_{\scriptscriptstyle1}}\hat{\underline{\underline{\Sigma}}}\left(\mathbf{R}+\frac{\mathbf{r}_{\scriptscriptstyle1}}{2};\boldsymbol{\rho}-\mathbf{r}_{\scriptscriptstyle1},\epsilon\right)
\hat{\underline{\underline{G}}}\left(\mathbf{R}-\frac{\boldsymbol{\rho}-\mathbf{r}_{\scriptscriptstyle1}}{2};\mathbf{r}_{\scriptscriptstyle1},\epsilon\right).
\end{align}
Here we separated the dependence on the center of mass and the relative coordinates. Once again, we used  $\boldsymbol{\nabla}=\frac{1}{2}\boldsymbol{\nabla}_{R}+\boldsymbol{\nabla}_{\rho}$. Since the gravitational field is independent of time, and owing to the fact that we are interested in the stationary state, we may omit the center of mass time. It is worth paying attention to the difference in the way the electric and gravitational fields enter the kinetic equation. The electric field in the gauge invariant kinetic equation appears only with the relative coordinate $\boldsymbol{\rho}$. As a result, the gauge invariant Green's function depends on the center of mass coordinate only due to the scattering by the impurities. [Recall that we postpone the averaging over the disorder till the end of the calculation.] The essential feature of the quantum kinetic equation in the presence of a gravitational field is that $\boldsymbol{\nabla}\gamma$ multiplies not only $\boldsymbol{\rho}$ but also $\mathbf{R}$. Consequently, the gravitational field induces an additional dependence of the Green's function on the center of mass coordinate.

We shall write the Green's function $\hat{\underline{\underline{G}}}$ as a sum of the equilibrium Green's function $\hat{g}_{eq}$ and the $\boldsymbol{\nabla}\gamma$-dependent Green's function. The kinetic equation describing $\hat{g}_{eq}$ can be obtained by setting $\boldsymbol{\nabla}\gamma=0$:
\begin{align}\label{eq:QKET-G_eq}
&\left[\frac{\epsilon}{\gamma_0}+\frac{\boldsymbol{\nabla}^2}{2m}-V_{imp}(\mathbf{R}+\boldsymbol{\rho}/2)+\mu
\right]\hat{g}_{eq}(\boldsymbol{\rho},\epsilon;imp)\\\nonumber
&\hspace{5mm}=\delta(\boldsymbol{\rho})
+\int{d\mathbf{r}_{\scriptscriptstyle1}}\hat{\sigma}_{eq}\left(\boldsymbol{\rho}-\mathbf{r}_{\scriptscriptstyle1},\epsilon;imp\right)
\hat{g}_{eq}\left(\mathbf{r}_{\scriptscriptstyle1},\epsilon;imp\right).
\end{align}
The above equation practicably coincides with Eq.~\ref{eq:EC-Geq}; the only difference is that here the frequency appears as $\epsilon/\gamma_0$. As we have already done in the preceding section, we separate the dependence on $\mathbf{R}$ due to the driving force from the one caused by the impurity potential. In our notation, the latter has been incorporated into $imp$. Before averaging, the dependence of the Green's function on the disorder is described by the set of equations which was already given in Eq.~\ref{eq:EC-OpenDisorderG}.
After averaging over the disorder, the equilibrium Green's function becomes:
\begin{align}\label{eq:QKET-G_eqSol}
&\left\langle{g_{eq}^{R,A}(\boldsymbol{\rho},\epsilon)}\right\rangle_{imp}\\\nonumber
&\hspace{15mm}=\left[\frac{\epsilon}{\gamma_0}+\frac{\boldsymbol{\nabla}_{\rho}^2}{2m}+\mu\pm\frac{i}{2\tau}-\sigma_{eq}^{R,A}\left(\boldsymbol{\rho},{\epsilon}/{\gamma_0}\right)\right]^{-1}\hspace{-3mm};\\\nonumber
&\left\langle{g_{eq}^{K}(\boldsymbol{\rho},\epsilon)}\right\rangle_{imp}=\left(1-2n_F\left({\epsilon}/{\gamma_0}\right)\right)
\left[g_{eq}^{R}(\boldsymbol{\rho},\epsilon)-g_{eq}^{A}(\boldsymbol{\rho},\epsilon)\right],
\end{align}
where $\tau$ is the mean free time. These Green's functions can be interpreted as describing the quasiparticles in the  equilibrium state with temperature $T$, and effective frequency $\epsilon/\gamma_0$.

As we already discussed, the explicit dependence on $\mathbf{R}$ in  $\hat{\underline{\underline{G}}}(\mathbf{R};\boldsymbol{\rho},\epsilon;,imp)$ is induced by the gradient of the gravitational field.  Therefore, in the process of linearizing the kinetic equation with respect to $\boldsymbol{\nabla}\gamma$, we expand $\hat{\underline{\underline{G}}}$ and $\hat{\underline{\underline{\Sigma}}}$ in the collision integral with respect to this explicit dependence on $\mathbf{R}$. In other words, we may rewrite the last term in Eq.~\ref{eq:QKET-DEGWig} as:
\begin{widetext}
\begin{align}\label{eq:QKE-Expand}\nonumber
&\int{dr_{\scriptscriptstyle1}}\hat{\underline{\underline{\Sigma}}}\left(\mathbf{R}+\frac{\mathbf{r}_{\scriptscriptstyle1}}{2};\boldsymbol{\rho}-\mathbf{r}_{\scriptscriptstyle1},\epsilon;imp\right)
\hat{\underline{\underline{G}}}\left(\mathbf{R}-\frac{\boldsymbol{\rho}-\mathbf{r}_{\scriptscriptstyle1}}{2};\mathbf{r}_{\scriptscriptstyle1},\epsilon;imp\right)
\approx \int\hspace{-1mm}{dr_{\scriptscriptstyle1}}\hat{\underline{\underline{\Sigma}}}(\mathbf{R};\boldsymbol{\rho}-\mathbf{r}_{\scriptscriptstyle1},\epsilon;imp)\hat{\underline{\underline{G}}}(\mathbf{R};\mathbf{r}_{\scriptscriptstyle1},\epsilon;imp)\\
&\hspace{-1mm}+\hspace{-1mm}\int\hspace{-1mm}{dr_{\scriptscriptstyle1}}\frac{\mathbf{r}_{\scriptscriptstyle1}}{2}\frac{\partial\hat{\underline{\underline{\Sigma}}}(\mathbf{R};\boldsymbol{\rho}-\mathbf{r}_{\scriptscriptstyle1},\epsilon;imp)}{\partial\mathbf{R}}\hat{\underline{\underline{G}}}(\mathbf{R};\mathbf{r}_{\scriptscriptstyle1},\epsilon;imp)
-\hspace{-1mm}\int\hspace{-1mm}{dr_{\scriptscriptstyle1}}\hat{\underline{\underline{\Sigma}}}(\mathbf{R};\boldsymbol{\rho}-\mathbf{r}_{\scriptscriptstyle1},\epsilon;imp)\frac{\boldsymbol{\rho}-\mathbf{r}_{\scriptscriptstyle1}}{2}\frac{\partial\hat{\underline{\underline{G}}}(\mathbf{R};\mathbf{r}_{\scriptscriptstyle1},\epsilon;imp)}{\partial\mathbf{R}}.
\end{align}
\end{widetext}
We will see that the last two terms are indeed proportional to $\boldsymbol{\nabla}\gamma$.

We separate the part of $\hat{\underline{\underline{G}}}$  depending on $\boldsymbol{\nabla}\gamma$ into two pieces. The equation for the first one, $\hat{G}_{loc-eq}$, is:
\begin{align}\label{eq:QKET-G_LocEq}\nonumber
&\left[\frac{\epsilon}{\gamma_0}+\frac{\boldsymbol{\nabla}^2}{2m}-V_{imp}+\mu\right]\hat{G}_{loc-eq}(\mathbf{R};\boldsymbol{\rho},\epsilon;imp)\\
&-\frac{\mathbf{R}\boldsymbol{\nabla}\gamma}{\gamma_0^2}\epsilon\hat{g}_{eq}(\boldsymbol{\rho},\epsilon;imp)\\\nonumber&=
\int{d\mathbf{r}_{\scriptscriptstyle1}}\hat{\sigma}_{eq}\left(\boldsymbol{\rho}-\mathbf{r}_{\scriptscriptstyle1},\epsilon;imp\right)\hat{G}_{loc-eq}\left(\mathbf{R};\mathbf{r}_{\scriptscriptstyle1},\epsilon;imp\right)\\\nonumber
&+
\int{d\mathbf{r}_{\scriptscriptstyle1}}\hat{\Sigma}_{loc-eq}\left(\mathbf{R};\boldsymbol{\rho}-\mathbf{r}_{\scriptscriptstyle1},\epsilon;imp\right)\hat{g}_{eq}\left(\mathbf{r}_{\scriptscriptstyle1},\epsilon;imp\right).
\end{align}
Following the steps presented in Appendix~\ref{App:ElecricCond} for the derivation of the $\mathbf{E}$-dependent Green's function, we may rewrite the  expression for $\hat{G}_{loc-eq}\left(\mathbf{R};\boldsymbol{\rho},\epsilon;imp\right)$ as:
\begin{align}\label{eq:QKET-G_LocEq2}
\hat{G}_{loc-eq}(\epsilon)
&=(\mathbf{R}\boldsymbol{\nabla}\gamma)\hat{g}_{eq}(\epsilon)\frac{\epsilon}{\gamma_0^2}\hat{g}_{eq}(\epsilon)\\\nonumber
&+(\mathbf{R}\boldsymbol{\nabla}\gamma)\hat{g}_{eq}(\epsilon)\hat{\Sigma}_{loc-eq}(\epsilon)\hat{g}_{eq}(\epsilon).
\end{align}
Once again, one should understand the product as a convolution of the coordinates. The solution of this equation is
\begin{subequations}\label{eq:QKET-G_LocEq3}
\begin{align}\label{eq:QKET-G_LocEq3A}
&\hat{G}_{loc-eq}(\mathbf{R};\boldsymbol{\rho},\epsilon;imp)=\left(\mathbf{R}\boldsymbol{\nabla}\gamma\right)\frac{\partial\hat{g}_{eq}(\boldsymbol{\rho},\epsilon;imp)}{\partial\gamma_0};
\end{align}
\begin{align}\label{eq:QKET-G_LocEq3B}
&\hat{\Sigma}_{loc-eq}(\mathbf{R};\boldsymbol{\rho},\epsilon;imp)=\left(\mathbf{R}\boldsymbol{\nabla}\gamma\right)\frac{\partial\hat{\sigma}_{eq}(\boldsymbol{\rho},\epsilon;imp)}{\partial\gamma_0}.
\end{align}
\end{subequations}

We see that the local equilibrium Green's function, $\hat{G}_{loc-eq}$, is a straightforward extension of the equilibrium Green's function for a non-uniform gravitational field/temperature.  This part of the Green's function  describes the readjustment of quasiparticles to the non-uniform gravitational field/temperature when the system is trying to maintain a local equilibrium. This response of the electrons to the non-uniform perturbation  tempts to induce a spatial modulation of the density. Since for charged particles it  is impossible to have a large scale charge modulation, the gradient of the gravitational field transfers into a gradient of the electro-chemical potential.  Therefore,  $\mathbf{j}_{e}=\hat\sigma(\mathbf{E}+\boldsymbol{\nabla}\mu/e)=\hat{\sigma}\mathbf{E}^{*}$ where the effective field $\mathbf{E}^{*}$ is the one measured in experiments. In other words, although this contribution to the current is initiated by the temperature gradient, it reveals itself  through  the electric conductivity. We wish to remark that for a constant electric field an  equivalent for the local equilibrium Green's function does not appear. [The role of the local-equilibrium Green's function is most peculiar when the response to $\boldsymbol{\nabla}\gamma$ is considered in the presence of a magnetic field. Under these conditions, $\hat{G}_{loc-eq}$ is responsible for the non-vanishing contribution to $\mathbf{j}_{e}$ from the magnetization current.~\cite{KM2008}]

The equation for the second term of the $\boldsymbol{\nabla}\gamma$-dependent part of the Green's function, denoted by  $\hat{G}_{\boldsymbol{\nabla}\gamma}(\mathbf{R};\boldsymbol{\rho},\epsilon;imp)$, includes all the terms in Eq.~\ref{eq:QKET-DEGWig} that did not enter the equations for $\hat{g}_{eq}$ and $\hat{G}_{loc-eq}$:
\begin{widetext}
\begin{align}\label{eq:QKET-G_TransInv}
&\int{d\mathbf{r}_{\scriptscriptstyle1}}\hat{g}_{eq}^{-1}\left(\boldsymbol{\rho}-\mathbf{r}_{\scriptscriptstyle1},\epsilon;imp\right)
\hat{G}_{\boldsymbol{\nabla}\gamma}\left(\mathbf{R};\mathbf{r}_{\scriptscriptstyle1},\epsilon;imp\right)-\frac{\boldsymbol{\rho\boldsymbol{\nabla}}\gamma}{2\gamma_0^2}\epsilon\hat{g}_{eq}(\boldsymbol{\rho},\epsilon;imp)
+\frac{1}{2m}\frac{\partial^2\hat{G}_{loc-eq}(\mathbf{R};\boldsymbol{\rho},\epsilon;imp)}{\partial\mathbf{R}\partial\boldsymbol{\rho}}\\\nonumber
&=\int{d\mathbf{r}_{\scriptscriptstyle1}}\hat{\Sigma}_{\boldsymbol{\nabla}\gamma}\left(\mathbf{R};\boldsymbol{\rho}-\mathbf{r}_{\scriptscriptstyle1},\epsilon;imp\right)
\hat{g}_{eq}\left(\mathbf{r}_{\scriptscriptstyle1},\epsilon;imp\right)
+\int{d\mathbf{r}_{\scriptscriptstyle1}}\frac{\mathbf{r}_{\scriptscriptstyle1}}{2}\frac{\partial\hat{\Sigma}_{loc-eq}\left(\mathbf{R};\boldsymbol{\rho}-\mathbf{r}_{\scriptscriptstyle1},\epsilon;imp\right)}{\partial\mathbf{R}}
\hat{g}_{eq}\left(\mathbf{r}_{\scriptscriptstyle1},\epsilon;imp\right)\\\nonumber&-
\int{d\mathbf{r}_{\scriptscriptstyle1}}\hat{\sigma}_{eq}\left(\boldsymbol{\rho}-\mathbf{r}_{\scriptscriptstyle1},\epsilon;imp\right)
\frac{\boldsymbol{\rho}-\mathbf{r}_{\scriptscriptstyle1}}{2}\frac{\partial\hat{G}_{loc-eq}\left(\mathbf{R};\mathbf{r}_{\scriptscriptstyle1},\epsilon;imp\right)}{\partial\mathbf{R}}.
\end{align}
\end{widetext}
In the above equation, the derivatives with respect to the center of mass coordinate act only on the explicit dependence of $\hat{G}_{loc-eq}(\mathbf{R},\boldsymbol{\rho},\epsilon;imp)$ and $\hat{\Sigma}_{loc-eq}(\mathbf{R},\boldsymbol{\rho},\epsilon;imp)$ on $\mathbf{R}$ (i.e., through the spatial dependent gravitational field). Note that the derivatives with respect to the center of mass coordinate which act on $V_{imp}$ in the local equilibrium Green's function was already included in Eq.~\ref{eq:QKET-G_LocEq}.

Once the explicit expressions for $\hat{G}_{loc-eq}$ and $\hat{\Sigma}_{loc-eq}$  are inserted into Eq.~\ref{eq:QKET-G_TransInv},  the kinetic equation for $\hat{G}_{\boldsymbol{\nabla}\gamma}$ becomes similar to the one for $\hat{G}_{\mathbf{E}}$ analyzed in Appendix.~\ref{App:ElecricCond} (see Eq.~\ref{eq:App-EC-QKERealSpace1}). After similar manipulations, the kinetic equation acquires a simple form resembling Eq.~\ref{eq:EC-GE}:
\begin{align}\label{eq:QKET-G_TransInv2}
&\hat{G}_{\boldsymbol{\nabla}\gamma}(\boldsymbol{\rho},\epsilon;,imp)=\hat{g}_{eq}\left(\epsilon\right)
\hat{\Sigma}_{\boldsymbol{\nabla}\gamma}\left(\epsilon\right)\hat{g}_{eq}\left(\epsilon\right)\\\nonumber
&+\frac{i\boldsymbol{\nabla}\gamma}{2}\left[\frac{\partial\hat{g}_{eq}\left(\epsilon\right)}{\partial\gamma_0}\hat{\mathbf{v}}_{eq}(\epsilon)\hat{g}_{eq}\left(\epsilon\right)
-\hat{g}_{eq}\left(\epsilon\right)\hat{\mathbf{v}}_{eq}(\epsilon)\frac{\partial\hat{g}_{eq}\left(\epsilon\right)}{\partial\gamma_0}
\right].
\end{align}
The expression for the renormalized velocity was already defined in Eq.~\ref{eq:EC-velocityQP}. Let us emphasize that despite of the similarity in the structure, the equations for $\hat{G}_{\mathbf{E}}$ and $\hat{G}_{\boldsymbol{\nabla}\gamma}$ are not identical. The derivative with respect to the frequency in Eq.~\ref{eq:EC-GE} is replaced by a derivative with respect to $\gamma_0$ in Eq.~\ref{eq:QKET-G_TransInv2}.  Since the derivative with respect to $\gamma_0$ acts on the equilibrium Green's functions for which $\gamma_0$ accompanies the frequency as $\epsilon/\gamma_0$, this derivative can be replaced by $-(\epsilon/\gamma_0^2)\partial/\partial\epsilon$. We see that according to the quantum kinetic equation for $\hat{G}_{\boldsymbol{\nabla}\gamma}$, the derivative with respect to the frequency is multiplied by the same frequency. As we have pointed out in Sec.~\ref{sec:ElectricCond}, the derivative $\partial/\partial\epsilon$ corresponds to the expansion with respect to the external frequency in the Kubo formula. In the simplified version of the Kubo formula given by Eq.~\ref{eq:CurrentNonIntFrequency}, the frequency  multiplies the unrenormalized velocity vertex, which is not connected to the expansion with respect to the external frequency. As a result, in the presence of interactions, the frequency in the derivative and the one multiplying the velocity vertex are not necessarily the same. The difference between the expressions obtained by using the two methods may seem minor. However, it will become clear that this subtle point actually leads to different expressions for the thermal conductivity even in the Fermi-liquid theory.

To complete the derivation of the response of the system to $\boldsymbol{\nabla}\gamma$, we have to find the dependence of the propagator of the interaction $\hat{V}$ on the gravitational field. Let us return to the action presented in Eq.~\ref{eq:QKET-S}. In general, for a non-local in space interaction, one should be concerned about which of the coordinates, $\mathbf{r}$ or $\mathbf{r}'$, should be prescribed to $\gamma$. This question is relevant for the last term in the action which describes the bare interaction via the field $\phi$. In this paper we consider electrons interacting only through the Coulomb interaction. If one recalls that for the Coulomb interaction, $\int{d\mathbf{r}'}\phi(\mathbf{r},t)U^{-1}(\mathbf{r}-\mathbf{r}')\phi(\mathbf{r}',t)=e^2(\boldsymbol{\nabla}\phi(\mathbf{r},t))^2/8\pi$, it is clear that the problem of attributing the coordinate to $\gamma$ does not exist.

The Dyson equation for the propagator of the interaction in the presence of a gravitational field is:
\begin{align}\label{eq:QKET-DEV}
&-\frac{e^2}{8\pi}\boldsymbol{\nabla}\left(\gamma(\mathbf{r})\boldsymbol{\nabla}\hat{V}(\mathbf{r},t;\mathbf{r}',t')\right)=\delta(\mathbf{r}-\mathbf{r}')\delta(t-t')\\\nonumber
&-\gamma(\mathbf{r})\int{d\mathbf{r}_{\scriptscriptstyle1}dt_{\scriptscriptstyle1}}\hat{\Pi}(\mathbf{r},t;\mathbf{r}_{\scriptscriptstyle1},t_{\scriptscriptstyle1})\gamma(\mathbf{r}_{\scriptscriptstyle1})\hat{V}(\mathbf{r}_{\scriptscriptstyle1},t_{\scriptscriptstyle1};\mathbf{r}',t').
\end{align}
In the derivation of the above equation we used the specific expression for the Coulomb interaction. An important feature of the kinetic equation for the propagator of the instantaneous interaction is that it does not include any time derivatives. As a consequence of this fact, the explicit dependence on the gravitational field may be eliminated from the equation in the linear response by transforming to the propagator $\underline{\underline{\hat{V}}}$:
\begin{align}\label{eq:QKET-DEVTransform}\nonumber
-\frac{e^2}{8\pi}{\boldsymbol{\nabla}}^2\underline{\underline{\hat{V}}}&(\mathbf{r},t;\mathbf{r}',t')=\delta(\mathbf{r}-\mathbf{r}')\\
&-\int{d\mathbf{r}_{\scriptscriptstyle1}dt_{\scriptscriptstyle1}}\hat{\underline{\underline{\Pi}}}(\mathbf{r},t;\mathbf{r}_{\scriptscriptstyle1},t_{\scriptscriptstyle1})\hat{\underline{\underline{V}}}(\mathbf{r}_{\scriptscriptstyle1},t_{\scriptscriptstyle1};\mathbf{r}',t'),
\end{align}
Here we employ the same transformation as in Eq.~\ref{eq:QKET-GaugeTrans}. As one can see, the entire dependence of the propagator $\hat{\underline{\underline{V}}}$ on the gravitational field is through the quasiparticle Green's functions that enter the self-energy $\hat{\underline{\underline{\Pi}}}$. Note the similarity between the kinetic equation for the propagator of the interaction in the density channel in the presence of a gravitational field and Eq.~\ref{eq:EC-DEV}.

Let us separate the solution of Eq.~\ref{eq:QKET-DEVTransform} into the equilibrium and $\boldsymbol{\nabla}\gamma$-dependent propagators, $\hat{\underline{\underline{V}}}=\hat{V}_{eq}+\hat{V}_{loc-eq}+\hat{V}_{\boldsymbol{\nabla}\gamma}$. The propagator at equilibrium satisfies the equation:
\begin{align}\label{eq:QKET-V_eq}
\hat{V}_{eq}(\mathbf{R};\boldsymbol{\rho},\omega)=\left[U^{-1}(\boldsymbol{\rho})+\hat{\Pi}_{eq}(\mathbf{R};\boldsymbol{\rho},\omega)\right]^{-1}.
\end{align}
The entire dependence of $\hat{V}_{eq}(\mathbf{R};\boldsymbol{\rho},\omega)$ on the frequency is due to the quasiparticle Green's functions in $\hat{\Pi}$.  Hence, the frequency  enters only in the combination $\omega/\gamma_0$, because the frequency in $\hat{g}_{eq}$ appears as $\epsilon/\gamma_0$ (see Eq.~\ref{eq:QKET-DEGWig}).  The equations for the $\boldsymbol{\nabla}\gamma$-dependent propagators are
\begin{subequations}\label{eq:QKET-V_GradT}
\begin{align}
\hat{V}_{loc-eq}(\mathbf{R};\boldsymbol{\rho},\omega)=-\hat{V}_{eq}(\omega)\hat{\Pi}_{loc-eq}(\omega)\hat{V}_{eq}(\omega);
\end{align}
\begin{align}
\hat{V}_{\boldsymbol{\nabla}\gamma}(\mathbf{R};\boldsymbol{\rho},\omega)=-\hat{V}_{eq}(\omega)\hat{\Pi}_{\boldsymbol{\nabla}\gamma}(\omega)\hat{V}_{eq}(\omega).
\end{align}
\end{subequations}

We will give a detailed discussion regarding the contribution of the interaction field to the heat transport in the following section. In addition, in Appendix~\ref{sec:ElectricCond-SCF} we present the kinetic equation for an interaction field describing the fluctuations in the Cooper channel.

\section{The electric and heat currents in the presence of a temperature gradient}\label{sec:HeatCurrent}

In this section we present the derivation of the heat and electric currents in terms of the Green's functions as a response to a temperature gradient. We also find the heat current induced by an electric field which will be used in the next section for the verification of the Onsager relations.

We first consider the electric current. Just like in the derivation of the electric conductivity, we start from the charge density given by Eq.~\ref{eq:EC-ChargeDensity}, and extract the expression for the electric current from the continuity equation (see Eq.~\ref{eq:EC-ContinuityEq}). Further calculations using the continuity equation requires the equations of motion for the field variables which can be obtained from the action presented in Eq.~\ref{eq:QKET-S}. At this point, the derivation of the electric current as a response to a gravitational field departs from the one for the electric field, because the dynamics of the field variables is different. According to the action given in Eq.~\ref{eq:QKET-S}, the divergence of the electric current is
\begin{align}\label{eq:HC-ChargeContinuity}
&\boldsymbol{\nabla}\mathbf{j}_e=2e\lim_{\begin{array}{c}\scriptstyle{ \mathbf{r}'\rightarrow\mathbf{r}} \\ \scriptstyle{t'\rightarrow{t}^{+}} \\
\end{array}}\left(\frac{\partial}{\partial{t}}+\frac{\partial}{\partial{t'}}\right)\left\langle{\psi^{\dag}(\mathbf{r}',t')\psi(\mathbf{r},t)}\right\rangle\\\nonumber
&=e\lim_{\mathbf{r}'\rightarrow\mathbf{r}}\left[\frac{\boldsymbol{\nabla}\gamma(\mathbf{r})\boldsymbol{\nabla}-\boldsymbol{\nabla}'\gamma(\mathbf{r}')\boldsymbol{\nabla}'}{2m}\hat{G}(\mathbf{r},t;\mathbf{r}',t)\right.\\\nonumber
&\left.-\gamma(\mathbf{r})\int{d\mathbf{r}_{\scriptscriptstyle1}}dt_{\scriptscriptstyle1}\hat{\Sigma}(\mathbf{r},t;\mathbf{r}_{\scriptscriptstyle1},t_{\scriptscriptstyle1})\gamma(\mathbf{r}_{\scriptscriptstyle1})\hat{G}(\mathbf{r}_{\scriptscriptstyle1},t_{\scriptscriptstyle1};\mathbf{r}',t)\right.\\\nonumber
&\left.+\int{d\mathbf{r}_{\scriptscriptstyle1}}dt_{\scriptscriptstyle1}\hat{G}(\mathbf{r},t;\mathbf{r}_{\scriptscriptstyle1},t_{\scriptscriptstyle1})\gamma(\mathbf{r}_{\scriptscriptstyle1})\hat{\Sigma}(\mathbf{r}_{\scriptscriptstyle1},t_{\scriptscriptstyle1};\mathbf{r}',t)\gamma(\mathbf{r}')
\right]^{<}.
\end{align}
As before, the factor of $2$ is due to the summation over the spin index. Beside the dependence on $\gamma$ through the Green's functions and self-energies, the RHS of the above equation contains the gravitational field explicitly. We may eliminate this explicit dependence on $\gamma$ by transforming to $\hat{\underline{\underline{G}}}$ and $\hat{\underline{\underline{\Sigma}}}$ (using the transformation described in Eq.~\ref{eq:QKET-GaugeTrans}):
\begin{align}\label{eq:HC-ChargeContinuity2}\nonumber
\boldsymbol{\nabla}\mathbf{j}_e&=-e\lim_{\scriptstyle{ \mathbf{r}'\rightarrow\mathbf{r}}} \left[-\frac{\boldsymbol{\nabla}^2-\boldsymbol{\nabla}'^2}{2m}\hat{\underline{\underline{G}}}(\mathbf{r}_1,t_1;\mathbf{r}',t)\right.\\\nonumber
&\left.-\int{d\mathbf{r}_{\scriptscriptstyle1}dt_{\scriptscriptstyle1}}\hat{\underline{\underline{\Sigma}}}(\mathbf{r},t;\mathbf{r}_{\scriptscriptstyle1},t_{\scriptscriptstyle1})\hat{\underline{\underline{G}}}(\mathbf{r}_{\scriptscriptstyle1},t_{\scriptscriptstyle1};\mathbf{r}',t)\right.\\
&\left.+\int{d\mathbf{r}_{\scriptscriptstyle1}dt_{\scriptscriptstyle1}}\hat{\underline{\underline{G}}}(\mathbf{r},t;\mathbf{r}_{\scriptscriptstyle1},t_{\scriptscriptstyle1})\hat{\underline{\underline{\Sigma}}}(\mathbf{r}_{\scriptscriptstyle1},t_{\scriptscriptstyle1};\mathbf{r}',t)
\right]^{<}.
\end{align}
In the above equation we used the fact that the product $\gamma^{1/2}(\mathbf{r})\gamma^{-1/2}(\mathbf{r}')$ in the limit $\mathbf{r}'\rightarrow\mathbf{r}$ becomes $1$.

We arrived at an equation which is identical to the one obtained for the electric current as a response to an electric field in Sec.~\ref{sec:ElectricCond}. Therefore, we get the following expression for the current:
\begin{align}\label{eq:HC-J_e}
\mathbf{j}_e=ie\int{d\mathbf{r}'dt'}\left[\hat{\underline{\underline{\mathbf{v}}}}(\mathbf{r},t;\mathbf{r}',t')\hat{\underline{\underline{G}}}(\mathbf{r}',t';\mathbf{r},t)\right]^{<}+h.c.
\end{align}
The above expression for the current is valid in the regime of linear response. One may check that the second non-vanishing term in the expansion described in Eq.~\ref{eq:EC-Expanssion} yields a contribution to the current, $\mathbf{j}_{e}(\mathbf{R})\propto\int{d\mathbf{r}_1}d\epsilon\mathbf{r}_1(\mathbf{r}_1\boldsymbol{\nabla}_{\mathbf{R}})^2[\hat{\underline{\underline{G}}}(\mathbf{R};-\mathbf{r}_1,\epsilon)\hat{\underline{\underline{\Sigma}}}(\mathbf{R};\mathbf{r}_1,\epsilon)]^{<}$,
which is already beyond the linear response.

Next, we shall concentrate on the expression for the heat current, $\mathbf{j}_{h}$. Following the prescription used for  the electric current, we find $\mathbf{j}_{h}$ from the continuity equation for the heat density, $\dot{Q}+\boldsymbol{\nabla}\mathbf{j}_h=0$. We start from the heat density, which in  in the presence of the gravitational field is:
\begin{align}\label{eq:HC-HeatDensity}
Q(\mathbf{r},t)&=\gamma(\mathbf{r})\left[h(\mathbf{r},t)-\mu{n(\mathbf{r},t)}\right],
\end{align}
where the energy and particle densities, $h(\mathbf{r},t)$ and $n(\mathbf{r},t)$,  are defined in the absence of the gravitational field. Extracting the Hamiltonian density from the action, we may write the heat density as:
\begin{align}\label{eq:HC-HeatDensityT}
&Q(\mathbf{r},t)=\frac{1}{2}\hspace{-2mm}\lim_{\begin{array}{c}\scriptstyle{ \mathbf{r}'\rightarrow\mathbf{r}} \\ \scriptstyle{t'\rightarrow{t}^{+}} \\
\end{array}}\hspace{-2mm}\left\langle
\left[-\frac{\boldsymbol{\nabla}\gamma(\mathbf{r})\boldsymbol{\nabla}}{2m}-\frac{\boldsymbol{\nabla}'\gamma(\mathbf{r}')\boldsymbol{\nabla}'}{2m}\right.\right.\\\nonumber
&\left.\left.+\gamma(\mathbf{r})\left(V_{imp}(\mathbf{r})+\phi(\mathbf{r},t)\right)+\gamma(\mathbf{r}')\left(V_{imp}(\mathbf{r}')+\phi(\mathbf{r}',t')\right)\right.\right.\\\nonumber
&\left.\left.-(\gamma(\mathbf{r})+\gamma(\mathbf{r'}))\mu
\phantom{\frac{1}{1}}\hspace{-2mm}
\right]\sum_{s}\psi_s^{\dag}(\mathbf{r}',t')\psi_s(\mathbf{r},t)\right.\\\nonumber
&\left.-\frac{1}{2}(\gamma(\mathbf{r})+\gamma(\mathbf{r}'))\int{d\mathbf{r}_{\scriptscriptstyle1}}\phi(\mathbf{r},t)U^{-1}(\mathbf{r}'-\mathbf{r}_{\scriptscriptstyle1})\phi(\mathbf{r}_{\scriptscriptstyle1},t')
\right\rangle.\\\nonumber
\end{align}
Due to the fact that the average heat density is evaluated under the path-integral, Eq.~\ref{eq:HC-HeatDensityT} can be rewritten in the following way:
\begin{align}\label{eq:HC-HeatDensityT1}\nonumber
Q(\mathbf{r},t)&=i\lim_{\begin{array}{c}\scriptstyle{ \mathbf{r}'\rightarrow\mathbf{r}} \\ \scriptstyle{t'\rightarrow{t}^{+}} \\
\end{array}}\left[\left(\frac{\partial}{\partial{t}}-\frac{\partial}{\partial{t}'}\right)\left\langle
\psi^{\dag}(\mathbf{r}',t')\psi(\mathbf{r},t)\right\rangle\right.\\
&\left.-2\delta(\mathbf{r}-\mathbf{r}')\delta(t-t')\phantom{\frac{\partial}{\partial{t}}}\hspace{-4mm}\right].
\end{align}
When inserted into the continuity equation, the term in the heat density proportional to $\delta(t-t')$ vanishes.  The resulting continuity equation for the heat current acquires a rather simple form:
\begin{align}\label{eq:HC-ContinuityHeatDensity}
\boldsymbol{\nabla}\mathbf{j}_{h}&=-i\hspace{-2mm}\lim_{\begin{array}{c}\scriptstyle{ \mathbf{r}'\rightarrow\mathbf{r}} \\ \scriptstyle{t'\rightarrow{t}^{+}} \\
\end{array}}\hspace{-2mm}\left(\frac{\partial}{\partial{t}}+\frac{\partial}{\partial{t}'}\right)\left(\frac{\partial}{\partial{t}}-\frac{\partial}{\partial{t}'}\right)\left\langle
\psi^{\dag}(\mathbf{r}',t')\psi(\mathbf{r},t)\right\rangle.
\end{align}
We have already met the derivative with respect to the center of mass time acting on $\langle{\psi^{\dag}(\mathbf{r}',t')\psi(\mathbf{r},t)}\rangle$ in the calculation of the electric current as a response to the gravitational field (see Eq.~\ref{eq:HC-ChargeContinuity}). Following the same route as in the transition from Eq.~\ref{eq:HC-ChargeContinuity} to Eq.~\ref{eq:HC-J_e}, we may express the heat current in terms of $\underline{\underline{G}}$:
\begin{align}\label{eq:HC-HeatCurrentT}
\mathbf{j}_h&=\frac{1}{2}\lim_{t'\rightarrow{t}^{+}}\left(\frac{\partial}{\partial{t}}-\frac{\partial}{\partial{t}'}\right)
\int{d\mathbf{r}_{\scriptscriptstyle1}dt_{\scriptscriptstyle1}}\hat{\underline{\underline{\mathbf{v}}}}(\mathbf{r},t;\mathbf{r}_{\scriptscriptstyle1},t_{\scriptscriptstyle1})\\\nonumber&\times\hat{\underline{\underline{G}}}(\mathbf{r}_{\scriptscriptstyle1},t_{\scriptscriptstyle1};\mathbf{r},t')+h.c.\\\nonumber
&=
-i\int\frac{d\epsilon}{2\pi}{d\mathbf{r}'}\left[\epsilon\hat{\underline{\underline{\mathbf{v}}}}(\mathbf{r},\mathbf{r}',\epsilon)\hat{\underline{\underline{G}}}(\mathbf{r}',\mathbf{r},\epsilon)\right]^{<}+h.c.
\end{align}
In the last equality we performed the Fourier transform with respect to the relative time coordinate.

It is worth pointing out that although the interaction renormalizes both the velocity and the Green's function in Eq.~\ref{eq:HC-HeatCurrentT}, for an instantaneous interaction there is no direct contribution to the heat current from the interaction propagator. In the scheme developed here, the heat density and current were expressed in terms of the quasiparticle Green's function alone.  This observation is consistent with the kinetic equation for $\underline{\underline{\hat{V}}}$ that reveals that the propagator of the interaction depends on the gravitational field only through the quasiparticle Green's functions in the self-energy $\hat{\underline{\underline{\Pi}}}$. The physical picture behind this two results is connected to the fact that it is the quasiparticles that are actually coupled to the heat bath and have a well defined temperature, while the interaction field does not have a temperature of its own. Our expression for the heat current is different from the one given in Ref.~\onlinecite{Aleiner2005} where the heat current has been taken as a sum of two contributions, one from the quasiparticles and the other from the collective modes. The authors of Ref.~\onlinecite{Aleiner2005} introduced the two terms because their purpose was to construct the kinetic equation in terms of the local distribution functions of the quasiparticles and collective modes. Since we keep the kinetic equation for the propagators to be non-local, the quasiparticle Green's function entering the current carries the information about the collective modes as well. Another difference is related to the kinetic equation.  In Ref.~\onlinecite{Aleiner2005}, the temperature gradient enters through the derivative with respect to the spatial coordinate, rather than the time. Despite all the differences, the calculation of the thermal conductivity presented in Sec.~\ref{sec:WF} produces the same result as in Ref.~\onlinecite{Aleiner2005}.

Extracting the electric and heat currents from the continuity equations is similar to deriving the Ward-identities. Therefore, it is not surprising that the expressions for the currents contain the renormalized velocity. A similar approach for finding the vertex correction to the heat current has been applied by Langer~\cite{Langer1962}.  Our expression for the heat current coincides with the first term given in Eq.~3.30 of Ref.\onlinecite{Langer1962}. Langer's  heat current includes also a non-trivial term with a derivative of the interaction amplitude with respect to the momentum. In the present scheme we succeeded to bypass this complication.  We believe that in our scheme this term is hidden in the $\boldsymbol{\nabla}\gamma$-dependent self-energy that contains also contributions in which the gradient of the gravitational field enters through the propagator of the interaction.

For completeness, we derive the expression for the heat current as a response to an electric field. In general, the procedure is similar to that of finding the heat current generated by a gravitational field, but there is one important modification. Here, the continuity equation acquires an additional term due to the work performed by the electric field on the electrons:
\begin{align}\label{eq:HC-ContEqHeatE}
\dot{Q}(\mathbf{r},t)+\boldsymbol{\nabla}\mathbf{j}_h(\mathbf{r},t)=\mathbf{j}_e(\mathbf{r},t)\mathbf{E}(\mathbf{r}).
\end{align}
Unlike the heat density in Eq.~\ref{eq:HC-HeatDensity}, that is a function of the chemical potential $\mu$, in the presence of an electric field the heat density is a function of the electro-chemical potential. Therefore, the continuity equation for the heat density should be written as:
\begin{align}\label{eq:HC-ContEqHeatE1}
&\dot{h}(\mathbf{r},t)-(\mu-e\varphi(\mathbf{r}))\dot{n}(\mathbf{r},t)+\boldsymbol{\nabla}\mathbf{j}_h(\mathbf{r},t)=-\mathbf{j}_e(\mathbf{r},t)\boldsymbol{\nabla}\varphi(\mathbf{r})\\\nonumber
\end{align}
Using the continuity equation for the $n(\mathbf{r},t)$, one gets:
\begin{align}\label{eq:HC-ContEqHeatE2}
\dot{h}(\mathbf{r},t)-\mu\dot{n}(\mathbf{r},t)+\boldsymbol{\nabla}(\varphi(\mathbf{r})\mathbf{j}_e(\mathbf{r},t))+\boldsymbol{\nabla}\mathbf{j}_h(\mathbf{r},t)=0,
\end{align}
where the charge current is given in Eq.~\ref{eq:EC-Current2}. The additional term in the heat continuity equation makes the expression for the heat current generated by an electric field to be  gauge invariant. Following the same steps as in the derivation of the response to a gravitational field, we obtain:
\begin{align}\label{eq:HC-HeatCurrentE}
\mathbf{j}_h\hspace{-1mm}=&\frac{1}{2}\lim_{\begin{array}{c}\scriptstyle{ \mathbf{r}'\rightarrow\mathbf{r}} \\ \scriptstyle{t'\rightarrow{t}^{+}} \\
\end{array}}\left(\frac{\partial}{\partial{t}}+ie\varphi(\mathbf{r})-\frac{\partial}{\partial{t}'}+ie\varphi(\mathbf{r}')\right)\\\nonumber
&\int{d\mathbf{r}_{\scriptscriptstyle1}dt_{\scriptscriptstyle1}}\hat{\mathbf{v}}(\mathbf{r},t;\mathbf{r}_{\scriptscriptstyle1},t_{\scriptscriptstyle1})\hat{G}(\mathbf{r}_{\scriptscriptstyle1},t_{\scriptscriptstyle1};\mathbf{r}',t')+h.c.
\end{align}
Applying the gauge invariant Fourier transform as defined in  Eq.~\ref{eq:EC-Wigner}, the expression for the heat current in terms of the gauge invariant Green's function becomes:
\begin{align}\label{eq:HC-HeatCurrentE2}
\mathbf{j}_h=-i\int\frac{d\epsilon}{2\pi}{d\mathbf{r}'}&\left[\epsilon\hat{\underline{\mathbf{v}}}(\mathbf{r},\mathbf{r}',\epsilon)\hat{\underline{G}}(\mathbf{r}',\mathbf{r},\epsilon)\right]^{<}+h.c.
\end{align}
Here, the frequency carried by the flow multiplies the renormalized velocity.

\begin{figure}[pt]
\begin{flushright}\begin{minipage}{0.5\textwidth}  \centering
        \includegraphics[width=0.7\textwidth]{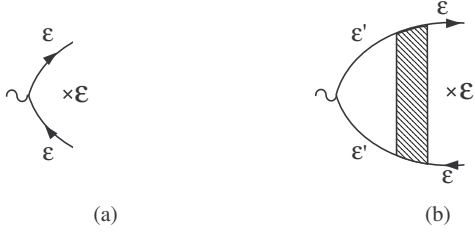}
                 \caption[0.4\textwidth]{\small (a) The heat current vertex for non-interacting electrons. (b) The heat current vertex in the presence of interactions is a product of the renormalized velocity multiplied by the frequency of the incoming and outgoing legs.}
                 \label{fig:velocity}
\end{minipage}\end{flushright}
\end{figure}

We wish to emphasize the most essential feature of the expressions for the heat current obtained in our scheme, Eqs.~\ref{eq:HC-HeatCurrentT} and~\ref{eq:HC-HeatCurrentE2}. That is, that the frequency factor corresponds to the legs of the renormalized vertex, but not to the frequency of the two Green's functions connected to the bare velocity inside the vertex, for illustration see  Fig.~\ref{fig:velocity}. In the diagrammatic technique, one usually starts with the bare vertex and then adds the interaction. In Appendix~\ref{App:FermiLiquidKubo}, we show that for the simplified Kubo formula, where the bare heat current vertex is $\epsilon\mathbf{v}_0$, dressing this vertex with interactions is not enough to reproduce the correct answer. We believe that this is the reason why the simplified Kubo formula produces  wrong results in the presence of interactions.

We like to point out the similarity in the structure of the four currents obtained in our scheme (Eqs.~\ref{eq:EC-AverageCurrent},~\ref{eq:HC-J_e},~\ref{eq:HC-HeatCurrentT} and~\ref{eq:HC-HeatCurrentE2}). One may notice  that after using the proper transformations described by Eqs.~\ref{eq:EC-Wigner} and~\ref{eq:QKET-GaugeTrans} (for the response to an electric field and a gravitational field, respectively), all the currents acquire the universal and amazingly simple form of Eq.~\ref{eq:current}. In particular, the entire dependence of the currents on the external fields is only through the renormalized velocity and Green's function.

From now on we will work only with the temperature gradient. This is possible because we already found the kinetic equations and the currents and therefore, the gravitational field fulfilled its role. We now present the expression for the heat current generated by a temperature gradient. First we have to adjust the $\boldsymbol{\nabla}\gamma$-dependent parts of the Green's function to describe the response to a non-uniform temperature. For that we replace $\boldsymbol{\nabla}\gamma$ by $\boldsymbol{\nabla}T/T$ and set $\gamma_0=1$ in Eqs.~\ref{eq:QKET-G_LocEq3} and~\ref{eq:QKET-G_TransInv2}.  Next, we insert the expressions for $\hat{G}_{loc-eq}$, $\hat{G}_{\boldsymbol{\nabla}T}$ as well as the $\boldsymbol{\nabla}T$-dependent velocities into Eq.~\ref{eq:HC-HeatCurrentT}. In the regime of linear response, the local equilibrium Green's function and the corresponding velocity do not contribute to the longitudinal current, because the dependence on the center of mass coordinate makes such contribution  vanish after averaging over the volume. Thus, the heat current is entirely determined by $\hat{G}_{\boldsymbol{\nabla}T}$ and $\hat{\Sigma}_{\boldsymbol{\nabla}T}$:
\begin{widetext}
\begin{align}\label{eq:FL-Jh}\nonumber
j_h^{i}&=\frac{\boldsymbol{\nabla}_jT}{2T}\int\frac{d\epsilon}{2\pi}\epsilon^2\frac{\partial{n_F(\epsilon)}}{\partial\epsilon}\left[v_{i}^R(\epsilon)g_{eq}^{R}(\epsilon)v_{j}^A(\epsilon)g_{eq}^A(\epsilon)
+v_{i}^R(\epsilon)g_{eq}^{R}(\epsilon)v_{j}^R(\epsilon)g_{eq}^A(\epsilon)-v_{i}^R(\epsilon)g_{eq}^{R}(\epsilon)v_{j}^R(\epsilon)g_{eq}^R(\epsilon)\right.\\\nonumber
&\left.
-
g_{eq}^{R}(\epsilon)v_{j}^R(\epsilon)g_{eq}^R(\epsilon)v_{i}^A(\epsilon)
\right]+
\frac{\boldsymbol{\nabla}_jT}{T}\int\frac{d\epsilon}{2\pi}\epsilon^2{n}_F(\epsilon)\left[v_{i}^R(\epsilon)\frac{\partial{g}_{eq}^R(\epsilon)}{\partial\epsilon}v_{j}^R(\epsilon)g_{eq}^R(\epsilon)-
v_{i}^R(\epsilon){g}_{eq}^R(\epsilon)v_{j}^R(\epsilon)\frac{\partial{g}_{eq}^R(\epsilon)}{\partial\epsilon}\right]\\
&+
i\int\frac{d\epsilon}{2\pi}\epsilon{v}_{i}^R(\epsilon)g_{eq}^R(\epsilon)\left[\Sigma_{\boldsymbol{\nabla}T}^{<}(\epsilon)(1-n_F(\epsilon))+\Sigma_{\boldsymbol{\nabla}T}^{>}(\epsilon)n_F(\epsilon)\right](g_{eq}^{R}(\epsilon)-g_{eq}^A(\epsilon))
+c.c.
\end{align}
\end{widetext}
We will use this expression to analyze the thermal conductivity in the following sections.

\section{electric and Thermal conductivity in the Fermi-liquid theory and the Wiedemann-Franz law}\label{sec:FermiLiquidQKE}

In this section we apply the microscopic scheme developed in this paper in order to calculate the electric and thermal conductivities in the framework of the Fermi-liquid theory. We demonstrate that the two conductivities are related through the Wiedemann-Franz law as it should be according to phenomenology. We will consider a sufficiently clean  system, so that $\varepsilon_F\tau\gg1$. On the other hand, we will not deal with the hydrodynamic aspect of transport in clean systems and we will assume $T\tau<1$.

We start with the calculation of the longitudinal electric conductivity.  For this purpose, we use the expression for the electric current presented in Eq.~\ref{eq:EC-JeFinal}. We have to perform the averaging over disorder. In the Fermi-liquid theory, for a short range disorder, each Green's function is averaged separately. Since on average the disordered medium is uniform, the equilibrium Green's functions of the quasiparticles become diagonal in momentum space after averaging:
\begin{align}\label{eq:EC-FL-g_eq}
&g_{eq}^{R,A}(\mathbf{k},\epsilon)=\frac{z}{\epsilon-\frac{m}{m^{*}}\xi_{\mathbf{k}}\pm\frac{i}{2\tau}};\\\nonumber
&z=\left[1-\frac{\partial\Re\sigma_{eq}(\mathbf{k}_F,\epsilon)}{\partial\epsilon}\Big|_{\epsilon=0}\right]^{-1}.
\end{align}
From now one, $\tau$ denotes the renormalized scattering time of the quasiparticles. The diagonal form of the Green's functions allows us to rewrite the electric current in momentum space. As mentioned above, the second term in Eq.~\ref{eq:EC-JeFinal} vanishes for the longitudinal current. The last term is proportional to the imaginary part of the interaction propagator. This term yields a correction to the Drude conductivity that is smaller~\cite{Langer1962b,Eliashberg1962} by a factor of  $(T\tau)^2$. (In fact, this correction is canceled out by other terms. However, as we have already mentioned, in this paper we will not study contributions that vanish in the limit $T\rightarrow0$.) Therefore, for the calculation within the framework of the Fermi-liquid theory, only the first part of Eq.~\ref{eq:EC-JeFinal} is needed. Let us concentrate on this term:
\begin{align}\label{eq:EC-FL-Je}\nonumber
&j_e^{i}=-\frac{e^2E_j}{2}\int\frac{d\mathbf{k}d\epsilon}{(2\pi)^{d+1}}\frac{\partial{n_F(\epsilon)}}{\partial\epsilon}
\left[(v_{i}^R(\mathbf{k},\epsilon)+v_{i}^A(\mathbf{k},\epsilon))\right.\\
&\left.\times{g}_{eq}^{R}(\mathbf{k},\epsilon)(v_{j}^R(\mathbf{k},\epsilon)+v_{j}^A(\mathbf{k},\epsilon))g_{eq}^A(\mathbf{k},\epsilon)\right.\\\nonumber
&\left.-(v_{i}^R(\mathbf{k},\epsilon)+v_{i}^A(\mathbf{k},\epsilon))g_{eq}^{R}(\mathbf{k},\epsilon)v_{j}^R(\mathbf{k},\epsilon)g_{eq}^R(\mathbf{k},\epsilon)\right.\\\nonumber
&\left.-g_{eq}^{A}(\mathbf{k},\epsilon)(v_{j}^R(\mathbf{k},\epsilon)+v_{j}^A(\mathbf{k},\epsilon))g_{eq}^A(\mathbf{k},\epsilon)v_{i}^A(\mathbf{k},\epsilon)
\right].
\end{align}

\begin{figure}[pt]
\begin{flushright}\begin{minipage}{0.5\textwidth}  \centering
        \includegraphics[width=0.7\textwidth]{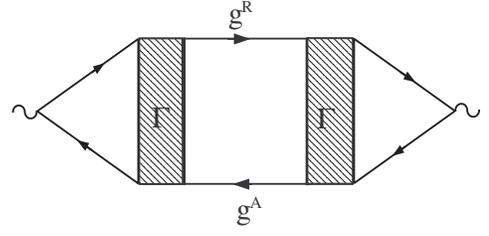}
                 \caption[0.4\textwidth]{\small The diagrammatic representation of the electric conductivity in the Fermi-liquid theory.} \label{fig:Fermi-Liquid}
\end{minipage}\end{flushright}
\end{figure}

A non zero contribution in the Fermi-liquid theory is generated by products in which one of the two Green's functions is retarded, while the other one is advanced (see Fig.~\ref{fig:Fermi-Liquid}). In the above expression, the last two terms which contain $g_{eq}^Rg_{eq}^R$ or $g_{eq}^Ag_{eq}^A$ vanish after the momentum integration. Next, we use the fact that the constant part of the renormalized velocity can be written as $\mathbf{v}^{R,A}=\mathbf{v_F}(1+\partial\Re\sigma_{eq}^{R,A}/\partial\xi_{\mathbf{k}}\big|_{\xi_{\mathbf{k}}=0})$, where $v_F=k_F/m$ is the unrenormalized Fermi velocity. Then, the longitudinal electric current acquires the form:
\begin{align}\label{eq:EC-FL-Je2}
j_e=-2e^2E&\left(1+\frac{\partial\Re\sigma_{eq}^{R}}{\partial\xi_{\mathbf{k}}}\Big|_{\xi_{\mathbf{k}}=0}\right)^2\frac{v_F^{2}}{d}\\\nonumber
&\times
\int\frac{d\mathbf{k}d\epsilon}{(2\pi)^{d+1}}\frac{\partial{n_F(\epsilon)}}{\partial\epsilon}
{g}_{eq}^{R}(\mathbf{k},\epsilon)g_{eq}^A(\mathbf{k},\epsilon),
\end{align}
where $d$ denotes the dimensionality. At this stage, one may neglect the dependence of the Green's functions on the frequency in the integration over $\epsilon$. After integration over the frequency  and  momentum, one obtains for  the electric conductivity:
\begin{align}\label{eq:EC-FL-Je3}
&\sigma&=\left(\frac{
1+\partial{\Re}\sigma_{eq}(\mathbf{k},0)/\partial\xi_{\mathbf{k}}\big|_{\xi_{\mathbf{k}}=0}}{1-\partial{\Re\sigma_{eq}(\mathbf{k}_F,\epsilon)}/\partial\epsilon\big|_{\epsilon=0}}\right)^{2}
\frac{m^{*}}{m}\frac{e^2n\tau}{m},
\end{align}
where $m^{*}$ is the renormalized mass $m^{*}/m=[1-\partial{\Re\sigma_{eq}(\mathbf{k=k}_F,\epsilon)}/\partial\epsilon\big|_{\epsilon=0}]/
[1+\partial{\Re}\sigma_{eq}(\mathbf{k},\epsilon=0)/\partial\xi_{\mathbf{k}}\big|_{\xi_{\mathbf{k}}=0}]$.
Thus, we have reproduced the known expression for the Drude electric conductivity in the Fermi-liquid theory, $\sigma=e^2n\tau/m^{*}$.

We turn now to the calculation of the thermal conductivity. Within the framework of the Fermi-liquid theory the calculation of the electric and thermal conductivities are practically parallel. Exactly as  in the calculation of the electric conductivity, the second term in the expression for the heat current given in Eq.~\ref{eq:FL-Jh} vanishes. The last term is proportional to the imaginary part of the interaction and it yields a contribution to $\kappa/T$ of the order $(T\tau)^2$ that can be neglected in the Fermi-liquid theory. [We will come back to this term in the next section in the discussion following Eq.~\ref{eq:WF-j_ha1}.] Therefore, the main contribution to the heat current arises from the first term in Eq.~\ref{eq:FL-Jh}. Written in  momentum space, this term acquires the form:
\begin{align}\label{eq:FL-Jh2}\nonumber
j_h^{i}&=\frac{\boldsymbol{\nabla}_jT}{2T}\int\frac{d\mathbf{k}d\epsilon}{(2\pi)^{d+1}}\epsilon^2\frac{\partial{n_F(\epsilon)}}{\partial\epsilon}
\left[(v_{i}^R(\mathbf{k},\epsilon)+v_{i}^A(\mathbf{k},\epsilon))\right.\\
&\left.\times{g}_{eq}^{R}(\mathbf{k},\epsilon)(v_{j}^R(\mathbf{k},\epsilon)+v_{j}^A(\mathbf{k},\epsilon))g_{eq}^A(\mathbf{k},\epsilon)\right.\\\nonumber
&\left.-(v_{i}^R(\mathbf{k},\epsilon)+v_{i}^A(\mathbf{k},\epsilon))g_{eq}^{R}(\mathbf{k},\epsilon)v_{j}^R(\mathbf{k},\epsilon)g_{eq}^R(\mathbf{k},\epsilon)\right.\\\nonumber
&\left.-g_{eq}^{A}(\mathbf{k},\epsilon)(v_{j}^R(\mathbf{k},\epsilon)+v_{j}^A(\mathbf{k},\epsilon))g_{eq}^A(\mathbf{k},\epsilon)v_{i}^A(\mathbf{k},\epsilon)
\right].
\end{align}
Since the last two lines do not contain a pair of retarded and advanced Green's functions, they do not contribute to the current. Using the fact that the renormalized velocity can be taken outside of the integrals, the longitudinal heat current can be written as
\begin{align}\label{eq:FL-Jh4}
j_h=2\frac{\boldsymbol{\nabla}{T}}{T}&\left(1+\frac{\partial\Re\sigma_{eq}^{R}}{\partial\xi_{\mathbf{k}}}\Big|_{\xi_{\mathbf{k}}=0}\right)^2\frac{v_F^{2}}{d}\\\nonumber
&\times\int\frac{d\mathbf{k}d\epsilon}{(2\pi)^{d+1}}\epsilon^2\frac{\partial{n_F(\epsilon)}}{\partial\epsilon}
g_{eq}^{R}(\mathbf{k},\epsilon)g_{eq}^A(\mathbf{k},\epsilon).
\end{align}
The only difference between Eq.~\ref{eq:EC-FL-Je2} and the above equation is that the heat current contains a factor of $\epsilon^2$ while the electric current includes the coefficient $e^2$. The obtained expression is exactly what one expects to get for the heat current transported by quasiparticles. It can be interpreted as if each quasiparticle contributes to the heat current its energy $\epsilon$. This energy is flowing with the velocity of the quasiparticle carrying it.

As a result of integrating over $\epsilon$, the thermal conductivity $\kappa$ becomes:
\begin{align}\label{eq:FL-Jh5}
\kappa=\frac{\pi^2T}{3}&\left(1+\frac{\partial\Re\sigma_{eq}^{R}}{\partial\xi_{\mathbf{k}}}\Big|_{\xi_{\mathbf{k}}=0}\right)^2\\\nonumber
&\times\frac{v_F^{2}}{d}\int\frac{d\mathbf{k}}{(2\pi)^{d+1}}
g_{eq}^{R}(\mathbf{k},0)g_{eq}^A(\mathbf{k},0).
\end{align}
The remaining integral over the momentum is identical to the one encountered while calculating the electric conductivity. Eventually, we get that the ratio of the thermal and electric conductivities in the framework of the Fermi-liquid, is proportional to  the Lorentz number:
\begin{align}\label{eq:FL-WF}
\frac{\kappa}{\sigma{T}}=\frac{\pi^2}{3e^2}.
\end{align}
This natural result was obtained in our scheme almost automatically, mostly due to the fact that the currents have been expressed in terms of the renormalized velocities from the very beginning. On the contrary, since the Kubo formula starts from the bare vertices, the derivation of the thermal conductivity in the Kubo formalism is not a trivial task.~\cite{Langer1962}  The situation with the simplified Kubo formula is even worse.  As we show in Appendix~\ref{App:FermiLiquidKubo}, the simplified Kubo formula generates terms that violate the Wiedemann-Franz law in the framework of the Fermi-liquid.

\section{Diffusion corrections to the thermal conductivity}\label{sec:WF}

In the previous section we have shown that for a Fermi-liquid system the ratio between the thermal and the electric conductivities is determined by the Lorentz number. One may wonder whether the Wiedemann-Franz law still holds for a disordered system when one goes beyond the framework of the Fermi-liquid theory in the diffusive regime, $T\tau\ll1$. This question was the subject of a long lasting debate. While in Ref.~\onlinecite{Castellani1987} it was concluded that the Wiedemann-Franz law remains valid, later studies show that, in fact, it is violated.~\cite{Livanov1991,Raimondi2004,Smith2005,Aleiner2005} To find the answer to this question using the scheme presented here, we study the leading order corrections with respect to $(\varepsilon_F\tau)^{-1}$ to the thermal conductivity. In the electric conductivity these are the Altshuler-Aronov corrections.~\cite{Altshuler1985} In two dimensions, the Altshuler-Aronov corrections logarithmically diverge as the temperature goes to zero, $\delta\sigma\propto{e}^2\ln1/T\tau$. Correspondingly, we wish to examine the singular corrections to the thermal conductivity for which $\kappa/T$ logarithmically diverges  in the limit $T\rightarrow0$. We will show that some of the corrections, originating from the long range  Coulomb interaction, violate the Wiedemann-Franz law. These corrections emerge from the third term in Eq.~\ref{eq:FL-Jh}, which is related to the imaginary part of the polarization operator. While in the Fermi-liquid calculation they are proportional to $(T\tau)^2$, when dressed by the diffusion propagators (diffusons), they are dramatically increased and do not contain this smallness anymore. Our result agrees with the one given in Refs.~\onlinecite{Livanov1991,Raimondi2004,Smith2005,Aleiner2005}.

\begin{figure}[pt]
\begin{flushright}\begin{minipage}{0.5\textwidth}  \centering
        \includegraphics[width=0.4\textwidth]{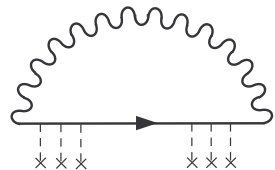}
                 \caption[0.4\textwidth]{\small The self-energy before averaging over the disorder. After averaging, it will be decorated with the diffusons.}
                 \label{fig:SelfEnergy1st}
\end{minipage}\end{flushright}
\end{figure}

To study the corrections to the transport coefficients in the diffusive limit, $T\tau\ll1$, one has to perform properly the averaging over the disorder. For that one has to construct the diffusion propagators in the following way:
\begin{align}\label{eq:WF-Diffuson}
&\langle{g_{eq}^{R,A}(\mathbf{r},\mathbf{r}_1,\epsilon)g_{eq}^{A,R}(\mathbf{r}_{\scriptscriptstyle2},\mathbf{r},\epsilon-\omega)}\rangle_{imp}\\\nonumber
&=\int{d}\mathbf{r}'
D^{R,A}(\mathbf{r},\mathbf{r}',\omega)
g_{eq}^{R,A}(\mathbf{r}',\mathbf{r}_1,\epsilon)g_{eq}^{A,R}(\mathbf{r}_{\scriptscriptstyle2},\mathbf{r}',\epsilon-\omega),
\end{align}
where the diffusion propagator is:
\begin{align}\label{eq:WF-Diffuson2}
D^{R,A}(\boldsymbol{\rho},\omega)=\left[\mp{i}\omega\tau-D\tau\boldsymbol{\nabla}_{\rho}^{2}\right]^{-1}.
\end{align}
Here, $D$ is the diffusion coefficient. As we shall see, we need to consider contributions which may include up to four diffusons.  To obtain the lowest order corrections in $(\varepsilon_F\tau)^{-1}$, all the diffusons must carry the same momentum. This is because each integration over the momentum of the diffusion propagators produces a small factor $(\varepsilon_F\tau)^{-1}$. This argument implies that all the diffusons must be affiliated with the same self-energy.  Therefore, we generate the singular corrections by choosing one of the self-energies to be decorated by the diffusons. When the chosen self-energy is the equilibrium one, we denote it by $\hat{\sigma}^{diff}$, while for the $\boldsymbol{\nabla}T$-dependent self-energy we use the notation $\hat{\Sigma}_{\boldsymbol{\nabla}T}^{diff}$.

\begin{figure}[pb]
\begin{flushright}\begin{minipage}{0.5\textwidth}  \centering
        \includegraphics[width=0.9\textwidth]{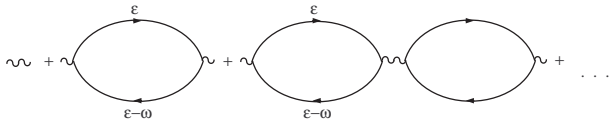}
                 \caption[0.4\textwidth]{\small The series describing the interaction propagator.}
                 \label{fig:PolarizationOperator}
\end{minipage}\end{flushright}
\end{figure}

Next, we expand the expression for the heat current given in Eq.~\ref{eq:FL-Jh} with respect to the choosen  self-energies. In the first term of the heat current, we expand both the equilibrium Green's functions and velocities with respect to $\hat{\sigma}^{diff}$. For example, the expansion of the Green's function yields $\hat{g}_{eq}=\hat{g}+\hat{g}\hat{\sigma}^{diff}\hat{g}$, where $\hat{g}$ is the equilibrium Green's function which already incorporates the Fermi-liquid renormalizations. As has been mentioned before, the second term of the heat current does not contribute to the longitudinal conductivity. Unlike the calculation in the previous section, the third term in Eq.~\ref{eq:FL-Jh} is no longer negligible. In order to compensate the smallness initially associated with the third term, it is $\hat{\Sigma}_{\boldsymbol{\nabla}T}$ that must be decorated by the diffusons. Since the number of integration over the momentum of the diffusons is restricted to one, the self-energy chosen to contain the diffusons can have only one \textit{effective} interaction amplitude as illustrated in Fig.~\ref{fig:SelfEnergy1st}. [Note, that the expansion is with respect to $(\varepsilon_F\tau)^{-1}$, and not over the interaction.] For the Coulomb interaction, the effective interaction amplitude can be approximated by the propagator $\hat{V}$. Hence,
\begin{align}\label{eq:OR-Sigma}
&\Sigma_{diff}^{<,>}(\mathbf{R};\boldsymbol{\rho},\tau)=iG^{<,>}(\mathbf{R};\boldsymbol{\rho},\tau)V^{<,>}(\mathbf{R};\boldsymbol{\rho},\tau);\\\nonumber
&\Sigma_{diff}^{R,A}(\mathbf{R};\boldsymbol{\rho},\tau)=iG^{R,A}(\mathbf{R};\boldsymbol{\rho},\tau)V^{<}(\mathbf{R};\boldsymbol{\rho},\tau)\\\nonumber
&\hspace{25mm}+
iG^{>}(\mathbf{R};\boldsymbol{\rho},\tau)V^{R,A}(\mathbf{R};\boldsymbol{\rho},\tau).
\end{align}
The interaction propagator is described by the infinite geometrical series presented in Fig.~\ref{fig:PolarizationOperator}. Each term in this series is a convolution in space. Symbolically, the result can be written as:
\begin{align}\label{eq:OR-V}
\hat{V}(\omega)=\left[U^{-1}+\hat{\Pi}(\omega)\right]^{-1}.
\end{align}
The polarization operator $\hat{\Pi}$ has the following analytic structure:
\begin{align}\label{eq:OR-PolarizationOperator}
&\Pi^{<,>}(\mathbf{R};\boldsymbol{\rho},\tau)=iG^{<,>}(\mathbf{R};\boldsymbol{\rho},\tau)G^{>,<}(\mathbf{R};-\boldsymbol{\rho},-\tau);\\\nonumber
&\Pi^{R,A}(\mathbf{R};\boldsymbol{\rho},\tau)=iG^{R,A}(\mathbf{R};\boldsymbol{\rho},\tau)G^{<}(\mathbf{R};-\boldsymbol{\rho},-\tau)\\\nonumber
&\hspace{25mm}+iG^{<}(\mathbf{R};\boldsymbol{\rho},\tau)G^{A,R}(\mathbf{R};-\boldsymbol{\rho},-\tau).
\end{align}
In Eq.~\ref{eq:OR-Sigma}, $\Sigma$ denotes either the equilibrium or the $\boldsymbol{\nabla}T$-dependent self-energy. For $\hat{\Sigma}_{\boldsymbol{\nabla}T}$ we should consider all possibilities to linearize the expression in Eq.~\ref{eq:OR-Sigma} with respect to $\boldsymbol{\nabla}T$ including the Green's functions inside the polarization operator.

\begin{figure}[pt]
\begin{flushright}\begin{minipage}{0.5\textwidth}  \centering
        \includegraphics[width=0.7\textwidth]{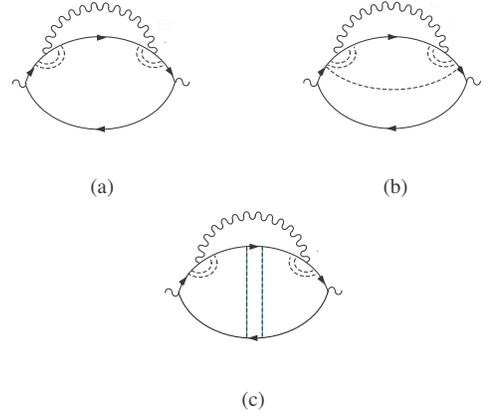}
                 \caption[0.4\textwidth]{\small The terms with a horizontal interaction line after averaging over the disorder.}
                 \label{fig:DOSHorizontal}
\end{minipage}\end{flushright}
\end{figure}

We are now fully equipped to study the leading order corrections to the thermal conductivity induced by disorder.
We separate the corrections to the current into three groups. The first group includes all the contributions which arise from the expansion of the equilibrium Green's function in the first term in Eq.~\ref{eq:FL-Jh}:
\begin{align}\label{eq:WF-j_hb1}\nonumber
&j_{h\hspace{1.5mm}hor}^{i}=2\frac{\boldsymbol{\nabla}_iT}{dT}\int\frac{d\epsilon}{2\pi}d\mathbf{r}_{\scriptscriptstyle1}...d\mathbf{r}_{\scriptscriptstyle4}
\epsilon^2\frac{\partial{n_{F}(\epsilon)}}{\partial\epsilon}{v}^{j}(\mathbf{r}_{\scriptscriptstyle6},\mathbf{r}_{\scriptscriptstyle1})\\\nonumber
&\times{g}^{R}(\mathbf{r}_{\scriptscriptstyle1},\mathbf{r}_{\scriptscriptstyle2},\epsilon)
\sigma_{diff}^{R}(\mathbf{r}_{\scriptscriptstyle2},\mathbf{r}_{\scriptscriptstyle3},\epsilon)g^{R}(\mathbf{r}_{\scriptscriptstyle3},\mathbf{r}_{\scriptscriptstyle4},\epsilon){v}^{j}(\mathbf{r}_{\scriptscriptstyle4},\mathbf{r}_{\scriptscriptstyle5})\\
&\times\left[g^{A}(\mathbf{r}_{\scriptscriptstyle5},\mathbf{r}_{\scriptscriptstyle6},\epsilon)-g^{R}(\mathbf{r}_{\scriptscriptstyle5},\mathbf{r}_{\scriptscriptstyle6},\epsilon)\right]+c.c.
\end{align}
Here, we used the fact that for the longitudinal currents ($i=j$) the contribution from the term  $v^{i}g^R\sigma_{diff}^Rg^Rv^{i}g_R$ is the same as the one from $v^{i}g^Rv^{i}g^R\sigma_{diff}^Rg^R$. After performing the average over the disorder in $\mathbf{j}_{h\hspace{1.5mm}hor}$, we get:
\begin{align}\label{eq:WF-j_hb2}\nonumber
&j_{h\hspace{1.5mm}hor}^{i}=2i\frac{\boldsymbol{\nabla}_iT}{T}\hspace{-1mm}\int\hspace{-1mm}\frac{d\mathbf{k}d\epsilon}{(2\pi)^{d+1}}\frac{d\mathbf{q}d\omega}{(2\pi)^{d+1}}
\epsilon^2V_{eq}^R(\mathbf{q},\omega)({D}^R(\mathbf{q},\omega))^2\\\nonumber
&\left[(n_F(\epsilon)-n_F(\epsilon-\omega))\frac{\partial{n_P(-\omega)}}{\partial\omega}+\frac{\partial{n}_F(\epsilon)}{\partial\epsilon}n_P(-\omega)\right]\\\nonumber
&\left\{\phantom{\frac{1}{1}}\hspace{-3mm}\frac{v_F^2}{d}(g^R(\mathbf{k},\epsilon))^2g^A(\mathbf{k-q},\epsilon-\omega)\left[g^R(\mathbf{k},\epsilon)-g^A(\mathbf{k},\epsilon)\right]
\right.\\\nonumber
&\left.-
\frac{v_F^2}{2\pi\nu\tau{d}}\int\frac{d\mathbf{k}'}{(2\pi)^d}(g^R(\mathbf{k},\epsilon))^2g^A(\mathbf{k},\epsilon)(g^R(\mathbf{k}',\epsilon))^2\right.\\\nonumber
&\left.\times{g}^A(\mathbf{k}'-\mathbf{q},\epsilon-\omega)
+\frac{1}{2\pi\nu\tau{d}}\int\frac{d\mathbf{k}'}{(2\pi)^d}\frac{k_jk_j'}{m^2}(g^R(\mathbf{k},\epsilon))^2\right.\\\nonumber
&\left.\times{g}^A(\mathbf{k-q},\epsilon-\omega)(g^R(\mathbf{k}',\epsilon))^2
{g}^A(\mathbf{k}'-\mathbf{q},\epsilon-\omega)D_R(\mathbf{q},\omega)
\phantom{\frac{1}{2}}\hspace{-3.5mm}\right\}\\
&+c.c.
\end{align}

The contributions to the current $\mathbf{j}_{h\hspace{1.5mm}hor}$ can be interpreted in terms of the diagrams with a horizontal interaction as shown in Fig.~\ref{fig:DOSHorizontal}. The only difference between the expression written above and the corresponding contributions to the electric conductivity is that $\epsilon^2$ should be replaced by $-e^2$. This is the proper place to remind that our scheme does not require a diagrammatic calculation, rather all the contributions are generated using the quantum kinetic equation. In this method the analytic structure of each of the terms and their numerical coefficients are determined by the kinetic equation. We give a diagrammatic interpretation of the different term for the purpose of illustration only.

\begin{figure}[pt]
\begin{flushright}\begin{minipage}{0.5\textwidth}  \centering
        \includegraphics[width=0.7\textwidth]{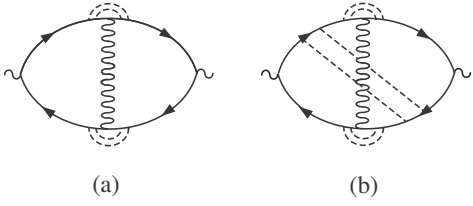}
                 \caption[0.4\textwidth]{\small The diagrams with a vertical interaction line. The obvious counterpart of diagram (b) is not shown.}
                 \label{fig:DOSVertical}
\end{minipage}\end{flushright}
\end{figure}

In the second group we collect terms related to the renormalization of the velocity.  These contributions originate from the first and the third terms in Eq.~\ref{eq:FL-Jh}. In the first term we expand one of the velocities with respect to $\sigma^{diff}$. In the last term, we consider the contributions in which the the temperature gradient enters $\Sigma_{\boldsymbol{\nabla}T}^{diff}$ through the quasiparticle Green's function. [We relate the other contributions to the Coulomb drag that will be discussed afterward.] Performing these operations, we get:
\begin{align}\label{eq:WF-j_hc1}\nonumber
&j_{h\hspace{1.5mm}ver}^{i}=-\int\frac{d\epsilon{d}\omega}{(2\pi)^2}d\mathbf{r}_{\scriptscriptstyle1}...d\mathbf{r}_{\scriptscriptstyle4}\epsilon{v}^{i}(\mathbf{r}_{\scriptscriptstyle4},\mathbf{r}_{\scriptscriptstyle1})g^{R}(\mathbf{r}_{\scriptscriptstyle1},\mathbf{r}_{\scriptscriptstyle2},\epsilon)\\\nonumber
&\times\left[(1-n_F(\epsilon))V_{eq}^{<}(\mathbf{r}_{\scriptscriptstyle2},\mathbf{r}_{\scriptscriptstyle3},\omega)G_{\boldsymbol{\nabla}T}^{<}(\mathbf{r}_{\scriptscriptstyle2},\mathbf{r}_{\scriptscriptstyle3},\epsilon-\omega)\right.\\\nonumber
&\left.+n_F(\epsilon)V_{eq}^{>}(\mathbf{r}_{\scriptscriptstyle2},\mathbf{r}_{\scriptscriptstyle3},\omega)G_{\boldsymbol{\nabla}T}^{>}(\mathbf{r}_{\scriptscriptstyle2},\mathbf{r}_{\scriptscriptstyle3},\epsilon-\omega)\right]\\\nonumber
&\times\left[g^{R}(\mathbf{r}_{\scriptscriptstyle3},\mathbf{r}_{\scriptscriptstyle4},\epsilon)-g^{A}(\mathbf{r}_{\scriptscriptstyle3},\mathbf{r}_{\scriptscriptstyle4},\epsilon)\right]
+\frac{\boldsymbol{\nabla}_iT}{2Td}\int\frac{d\epsilon}{2\pi}d\mathbf{r}_{\scriptscriptstyle1}...d\mathbf{r}_{\scriptscriptstyle4}\epsilon^2\\\nonumber
&\times\left[4\delta{v}_{j}^{R}(\mathbf{r}_{\scriptscriptstyle4},\mathbf{r}_{\scriptscriptstyle1},\epsilon)g^{R}(\mathbf{r}_{\scriptscriptstyle1},\mathbf{r}_{\scriptscriptstyle2},\epsilon)
v^{j}(\mathbf{r}_{\scriptscriptstyle2},\mathbf{r}_{\scriptscriptstyle3})g^{A}(\mathbf{r}_{\scriptscriptstyle3},\mathbf{r}_{\scriptscriptstyle4},\epsilon)\right.\\\nonumber
&\left.-3\delta{v}_{j}^{R}(\mathbf{r}_{\scriptscriptstyle4},\mathbf{r}_{\scriptscriptstyle1},\epsilon)g^{R}(\mathbf{r}_{\scriptscriptstyle1},\mathbf{r}_{\scriptscriptstyle2},\epsilon)
v^{j}(\mathbf{r}_{\scriptscriptstyle2},\mathbf{r}_{\scriptscriptstyle3})g^{R}(\mathbf{r}_{\scriptscriptstyle3},\mathbf{r}_{\scriptscriptstyle4},\epsilon)\right.\\\nonumber
&\left.-g^{R}(\mathbf{r}_{\scriptscriptstyle1},\mathbf{r}_{\scriptscriptstyle2},\epsilon)
v^{j}(\mathbf{r}_{\scriptscriptstyle2},\mathbf{r}_{\scriptscriptstyle3})g^{R}(\mathbf{r}_{\scriptscriptstyle3},\mathbf{r}_{\scriptscriptstyle4},\epsilon)\delta{v}_{j}^{A}(\mathbf{r}_{\scriptscriptstyle4},\mathbf{r}_{\scriptscriptstyle1},\epsilon)\right]\frac{\partial{n_{F}(\epsilon)}}{\partial\epsilon}\\
&+c.c.
\end{align}
Here, $\delta{v}_i^{R,A}(\mathbf{r},\mathbf{r}',\epsilon)=-i(r_i-r_i')\sigma_{diff}^{R,A}(\mathbf{r},\mathbf{r}',\epsilon)$. Since in this calculation $G_{\boldsymbol{\nabla}T}$ includes only the Fermi-liquid renormalizations, it is described by the last term in Eq.~\ref{eq:QKET-G_TransInv2}, $\hat{G}_{\boldsymbol{\nabla}T}(\epsilon)=-i\epsilon\boldsymbol{\nabla}{T}/T[\partial{\hat{g}}(\epsilon)/\partial\epsilon\hat{\mathbf{v}}(\epsilon)\hat{g}(\epsilon)-
\hat{g}(\epsilon)\hat{\mathbf{v}}(\epsilon)\partial{\hat{g}}(\epsilon)/\partial\epsilon]$. As a result of averaging over the disorder we get:
\begin{align}\label{eq:WF-j_hc2}
&j_{h\hspace{1.5mm}ver}^{i}=i\frac{\boldsymbol{\nabla}_iT}{2T}\int\frac{d\mathbf{k}d\epsilon}{(2\pi)^{d+1}}\frac{d\mathbf{q}{d}\omega}{(2\pi)^{d+1}}(D^R(\mathbf{q},\omega))^2\\\nonumber
&\left\{4\epsilon^2V_{eq}^R(\mathbf{q},\omega)
\left[(n_F(\epsilon)-n_F(\epsilon-\omega))\frac{\partial{n_P(-\omega)}}{\partial\omega}\right.\right.\\\nonumber
&\left.\left.+\frac{\partial{n}_F(\epsilon)}{\partial\epsilon}n_P(-\omega)\right]
+\epsilon\omega\left[n_F(\epsilon)-n_F(\epsilon-\omega)\right]\frac{\partial{n_P(\omega)}}{\partial\omega}\right.\\\nonumber
&\left.\times\left[V_{eq}^R(\mathbf{q},\omega)-V_{eq}^A(\mathbf{q},\omega)\right]\phantom{\frac{2}{2}}\hspace{-3mm}\right\}
\left[\frac{k_j(k_j-q_j)}{m^2}(g^R(\mathbf{k},\epsilon))^2\right.\\\nonumber&\left.\times(g^A(\mathbf{k-q},\epsilon-\omega))^2
+\frac{2}{2\pi\nu\tau}\int\frac{d\mathbf{k}'}{(2\pi)^{d}}\frac{k_jk_j'}{m^2}(g^R(\mathbf{k},\epsilon))^2
\right.\\\nonumber&\left.\times{g}^A(\mathbf{k-q},\epsilon-\omega)
(g^A(\mathbf{k}',\epsilon-\omega))^2g^R(\mathbf{k'}+\mathbf{q},\epsilon)D_R(\mathbf{q},\omega)
\phantom{\frac{1}{1}}\hspace{-3mm}\right]
\\\nonumber&+c.c.
\end{align}
In the transition between the last two equations as well as between Eq.~\ref{eq:WF-j_hc1} and Eq.~\ref{eq:WF-j_hc2} we  used the standard identities for the product of Fermi distribution functions:
\begin{subequations}\label{eq:OR-identities}
\begin{align}\label{eq:OR-identities1}\nonumber
&\frac{\partial{n_F(\epsilon)}}{\partial\epsilon}n_F(\epsilon-\omega)=-\frac{\partial{n_P(-\omega)}}{\partial\omega}[n_F(\epsilon)-n_F(\epsilon-\omega)]\\
&\hspace{30mm}-n_P(-\omega)\frac{\partial{n_F(\epsilon)}}{\partial\epsilon};
\end{align}
\begin{align}\label{eq:OR-identities2}
&n_F(\epsilon)(1-n_F(\epsilon-\omega))=n_P(\omega)[n_F(\epsilon-\omega)-n_F(\epsilon)].
\end{align}
\end{subequations}

The  diagrams corresponding to the second group are presented in Fig.~\ref{fig:DOSVertical}. One can observed that the second group contain the contributions with a vertical interaction line. The terms containing $\epsilon^2$ in Eq.~\ref{eq:WF-j_hc2} have their counterparts in the corrections to the electric conductivity. Together with the first group they give a correction to the thermal conductivity which satisfies the Wiedemann-Franz law:
\begin{align}\label{eq:WF-JWF}\nonumber
&\delta{\kappa}_{WF}=4i\nu{D}\int\frac{d\mathbf{q}{d}\omega}{(2\pi)^{d+1}}d\epsilon
\epsilon^2\frac{\partial{n_F(\epsilon)}}{\partial\epsilon}\frac{\partial(\omega{n_P(-\omega)})}{\partial\omega}\\
&Dq^2V_{eq}^R(\mathbf{q},\omega)D_R^3(\mathbf{q},\omega)\tau^3
+c.c.=\frac{\pi^2T}{3e^2}\delta\sigma_{AA},
\end{align}
where $\delta\sigma_{AA}$ is the Altshuler-Aronov corrections to the electric conductivity. The additional terms that are proportional to $\epsilon\omega$ in Eq.~\ref{eq:WF-j_hc2} are responsible for the deviation from  the Wiedemann-Franz law. Only the diagram with two diffusons gives a singular contribution  which in two dimensions is accumulated in the region of small momenta: $\omega/D\kappa_{screen}<q<\sqrt{\omega/D}$, where $1/\kappa_{screen}=1/(2\pi{e}^2\nu)$ is the inverse screening length in $d=2$. Eventually, the non Wiedemann-Franz law correction to the thermal conductivity is
\begin{align}\label{eq:WF-JCorr}\nonumber
&\delta\kappa_{non-WF}=-i\nu{D}\tau^2\int\frac{d\mathbf{q}{d}\omega}{(2\pi)^3}d\epsilon
\epsilon\omega\left[n_F(\epsilon)-n_F(\epsilon-\omega)\right]\\\nonumber&\frac{\partial{n}_P(\omega)}{\partial\omega}
\left[V_{eq}^R(\mathbf{q},\omega)-V_{eq}^{A}(\mathbf{q},\omega)\right]\left[D_R^2(\mathbf{q},\omega)+D_A^2(\mathbf{q},\omega)\right]\\
&=\frac{T}{12}\ln\left(\frac{D\kappa_{screen}^2}{T}\right).
\end{align}
Although both $\delta\kappa_{WF}$ and $\delta\kappa_{non-WF}$ logarithmically diverge in $d=2$, the origin of the singularities are different. The logarithmic correction that do not violate the Wiedemann-Franz law accumulate over a wide region of momenta $\sqrt{T/D}<q<1/v_F\tau$.

\begin{figure}[pt]
\begin{flushright}\begin{minipage}{0.5\textwidth}  \centering
        \includegraphics[width=0.55\textwidth]{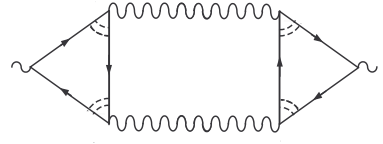}
                 \caption[0.4\textwidth]{\small The drag diagrams  after the averaging over the disorder.}
                 \label{fig:Drag}
\end{minipage}\end{flushright}
\end{figure}

The third group corresponds to the Coulomb drag, see Fig.~\ref{fig:Drag}. These terms are generated when the temperature gradient enters $\Sigma_{\boldsymbol{\nabla}T}^{diff}$ through the interaction propagator. To exploit the symmetry related to the Coulomb drag, we shall use the fact that the interaction field $\phi$ is real. Therefore, the corresponding propagator satisfies the relation: $\hat{V}(\mathbf{r},t;\mathbf{r}',t')=-i\langle{T_c\{\phi(\mathbf{r},t)\phi(\mathbf{r}',t')\}}\rangle=\hat{V}^{T}(\mathbf{r}',t';\mathbf{r},t)$,
and we may write the lesser and greater components of the interaction propagator as:
\begin{align}\label{eq:WF-V}
V^{<,>}(\boldsymbol{\rho},\tau)=\frac{1}{2}\left[V^{<,>}(\boldsymbol{\rho},\tau)+V^{>,<}(-\boldsymbol{\rho},-\tau)\right].
\end{align}
Consequently,  the self-energy given in Eq.~\ref{eq:OR-Sigma} can be split into two parts:
\begin{align}\label{eq:OR-Sigma2}\nonumber
\Sigma^{<,>}&(\mathbf{R},\boldsymbol{\rho},\epsilon)=\frac{i}{2}\int\frac{d\omega}{2\pi}
\left[G^{<,>}(\mathbf{R},\boldsymbol{\rho},\epsilon-\omega)V^{<,>}(\mathbf{R};\boldsymbol{\rho},\omega)\right.\\
&\left.+
G^{<,>}(\mathbf{R};\boldsymbol{\rho},\epsilon+\omega)V^{>,<}(\mathbf{R};-\boldsymbol{\rho},\omega)\right].
\end{align}
This way of writing the propagator $V$ and the self-energy turns out to be highly useful in the derivation of the Coulomb drag terms as well as for the proof of the Onsager relation presented in the next section.

A full analysis of the Coulomb drag contributions to the thermal conductivity, starting from Eq.~\ref{eq:OR-Sigma2}, is presented in Appendix.~\ref{App:Drag}. We show there that in the thermal conductivity, unlike the electric conductivity~\cite{Oreg1995}, the Coulomb drag can be decorated by four diffusons (see Eq.~\ref{eq:Drag-Survivng}). After averaging over the impurities, the correction to the heat current from the Coulomb drag is
\begin{align}\label{eq:WF-j_ha1}
&j_{h\hspace{1.5mm}drag}^{i}=-\frac{\boldsymbol{\nabla}_i{T}}{4T}\int\frac{d\mathbf{q}{d}\omega}{(2\pi)^{d+1}}|V_{eq}(\mathbf{q},\omega)|^2\omega^2\frac{\partial{n_P(\omega)}}{\partial\omega}\\\nonumber
&\left\{\frac{\mathbf{\partial}}{\partial{q}_j}\int\frac{d\mathbf{k}d\epsilon}{(2\pi)^{d+1}}[n_F(\epsilon-\omega)-n_F(\epsilon)]
\left[g^{R}(\mathbf{k}-\mathbf{q},\epsilon-\omega)\right.\right.\\\nonumber&\left.\left.\times{g}^{A}(\mathbf{k},\epsilon)D^A(\mathbf{q},\omega)-
g^{A}(\mathbf{k}-\mathbf{q},\epsilon-\omega)g^{R}(\mathbf{k},\epsilon)D^R(\mathbf{q},\omega)\right]\phantom{\frac{1}{1}}\hspace{-4mm}\right\}^{2}\hspace{-2mm}.
\end{align}
Owing to the structure of this term, there are no divergencies related to the region of small momenta indicated above. Moreover, the integration over the frequency $\omega$ is restricted to $|\omega|\lesssim{T}$. Therefore, this contribution to $\kappa/T$ is regular. [Note, that the structure of the Coulomb drag term presented here differs from the one in Ref.~\onlinecite{Smith2005} obtained from the simplified Kubo formula.]  In the previous section we argued that in the framework of the Fermi-liquid theory the third term in Eq.~\ref{eq:FL-Jh} generates contributions that are proportional to $(T\tau)^2$.  Using the expressions in Eqs.~\ref{eq:WF-JCorr} and~\ref{eq:WF-j_ha1}, one may check that in the absence of the diffusons the contributions related to the third term indeed acquire this small factor.

To conclude the section, we return to the singular correction to the Wiedemann-Franz law presented in Eq.~\ref{eq:WF-JCorr}.  The infrared divergency of this kind does not occur in the electric conductivity due to gauge invariance.~\cite{Raimondi2004,Finkelstein1988,Finkelstein1990,Finkelstein1994} To understand this argument, one has to go back to Eq.~\ref{eq:EC-S-HS} describing the action of interacting electrons. Writing the action in momentum space and considering the limit $\mathbf{q}\rightarrow0$, one may neglect the dependence of the quasiparticle Green's functions on the momentum transferred by the interaction. Then, after the integration over the transferred momentum, the field $\phi$ that describes the effective electron-electron interaction becomes only a function of time. An interaction field of the form $\phi(t)$ can be gauged out by redefining the quasiparticle field $\psi(\mathbf{r},t)\rightarrow{e}^{-i\int^{t}dt'\phi(t')}\psi(\mathbf{r},t)$. In the presence of a temperature gradient the situation is more complicated. The correction to thermal conductivity violating the Wiedemann-Franz law are proportional to the imaginary part of the interaction propagator $\hat{V}$. In other words, these terms are sensitive to the decay of the interaction into particle-hole pairs. Therefore, once the inner structure of the interaction becomes important, gauging out the interaction field on the level of the action is no longer justified.

\section{Onsager relations}\label{sec:Onsager}

In Sec.~\ref{sec:HeatCurrent}, we derived the expressions for the electric current as a response to $\boldsymbol{\nabla}T$ and the heat current generated by $\mathbf{E}$. The two expressions were obtained independently from each other. On the other hand, these two currents must be related through the Onsager relations. In this section, as an additional test for our scheme, we verify that the expressions given in Eqs.~\ref{eq:HC-J_e} and~\ref{eq:HC-HeatCurrentE2} indeed satisfy the Onsager relation.

The Onsager relations~\cite{Onsager1931} connect between the off-diagonal elements of the conductivity tensor: $\sigma_{ij}(\mathbf{B})=\sigma_{ji}(-\mathbf{B})$, $\kappa_{ij}(\mathbf{B})=\kappa_{ji}(-\mathbf{B})$, and $\tilde{\alpha}_{ij}(\mathbf{B})=T{\alpha}_{ji}(-\mathbf{B})$. In the absence of magnetic field, the Onager relations reduce to $\tilde{\alpha}_{xx}=T{\alpha}_{xx}$. Here, we restrict our demonstration of this relation to  a system of electrons interacting through the Coulomb interaction as has been discussed in the previous sections. [In this paper, our treatment of the electric and heat currents does not include the modifications needed to account for the effect of a magnetic field. The generalization of the present scheme for calculating the thermoelectric transport coefficients in the presence of a magnetic field is straightforward, and was applied by us in the analysis of the Nernst effect in disordered films above the superconducting transition temperature~\cite{KM2008}.]

To find the two thermoelectric currents, we have (i) to insert the $\mathbf{E}$-dependent velocity and Green's function into the expression for the heat current given in Eq.~\ref{eq:HC-HeatCurrentE2}, and similarly (ii) to insert the $\boldsymbol{\nabla}T$-dependent velocity and Green's function into the electric current described by Eq.~\ref{eq:HC-J_e}. [As has been already argued in the derivation of the thermal conductivity (see the discussion above Eq.~\ref{eq:FL-Jh}), the local equilibrium Green's function does not contribute to the longitudinal currents.] Due to the similarity between $\hat{G}_{\mathbf{E}}$ and $\hat{G}_{\boldsymbol{\nabla}T}$ and the common structure of the four currents (Eqs.~\ref{eq:EC-AverageCurrent},~\ref{eq:HC-J_e},~\ref{eq:HC-HeatCurrentT} and~\ref{eq:HC-HeatCurrentE2}), all the responses determining the \textit{longitudinal} components of the conductivity tensor can be written as:
\begin{widetext}
\begin{align}\label{eq:OR-Currents}\nonumber
&j_{e,h}^{i}(\mathbf{F})=\frac{1}{2d}\int\frac{d\epsilon}{2\pi}\chi_{e,h}(\epsilon)F_{i}(\epsilon)\frac{\partial{n_F(\epsilon)}}{\partial\epsilon}\left[v_{j}^R(\epsilon)g_{eq}^{R}(\epsilon)v_{j}^A(\epsilon)g_{eq}^A(\epsilon)
+v_{j}^R(\epsilon)g_{eq}^{R}(\epsilon)v_{j}^R(\epsilon)g_{eq}^A(\epsilon)-
v_{j}^R(\epsilon)g_{eq}^{R}(\epsilon)v_{j}^R(\epsilon)g_{eq}^R(\epsilon)\right.\\
&\left.-
g_{eq}^{R}(\epsilon)v_{j}^R(\epsilon)g_{eq}^R(\epsilon)v_{j}^A(\epsilon)
\right]+i\int\frac{d\epsilon}{2\pi}\chi_{e,h}(\epsilon)v_{i}^R(\epsilon)g_{eq}^R(\epsilon)\left[\Sigma_{\mathbf{F}}^{<}(\epsilon)(1-n_F(\epsilon))+\Sigma_{\mathbf{F}}^{>}(\epsilon)n_F(\epsilon)\right](g_{eq}^{R}(\epsilon)-g_{eq}^A(\epsilon))
+c.c.
\end{align}
\end{widetext}
Here, $\mathbf{F}(\epsilon)$ is equal to $e\mathbf{E}$ for the response to an electric field and to $\epsilon\boldsymbol{\nabla}T/T$ for the response to a temperature gradient. For the electric current, $\chi_{e}(\epsilon)=-e$, while for the heat current, $\chi_{h}(\epsilon)=\epsilon$. Comparing the expressions for $\mathbf{j}_{e}(\boldsymbol{\nabla}T)$ and $\mathbf{j}_{h}(\mathbf{E})$, one may immediately see that the contributions described by the first term in the above expressions satisfy the Onsager relation, ${j_{h}^{i}}/{E_i}=-T{j_{e}^{i}}/{\boldsymbol{\nabla}_iT}$. Therefore, it remains to show that the same holds for the last term. The dependence of the last term in Eq.~\ref{eq:OR-Currents} on the external perturbation $\mathbf{F}$ enters through the self-energy. As has been mentioned in the previous section, we need to consider all the possibilities to take one of the propagators ($\hat{G}$ or $\hat{V}$) in $\Sigma_{\mathbf{F}}$ to depend on $\mathbf{F}$.

In general, the fulfilment of the Onsager relation demands microscopic reversibility, which in our case implies that $\hat{G}(\mathbf{r},\mathbf{r}',\epsilon)=\hat{G}(\mathbf{r}',\mathbf{r},\epsilon)$ and $\hat{V}(\mathbf{r},\mathbf{r}',\epsilon)=\hat{V}(\mathbf{r}',\mathbf{r},\epsilon)$. The Onsager relation is satisfied if by reading the contributions to $\mathbf{j}_{e}(\boldsymbol{\nabla}T)$ described in Eq.~\ref{eq:OR-Currents} from  right to left instead of left to right, one gets $\mathbf{j}_{h}(\mathbf{E})$, and vice versa. Besides the microscopic reversibility, in order to get the desired relation the thermoelectric currents should have a proper symmetry related to the frequency. There should be a well defined symmetry (for each of the currents) under the exchange of the frequencies carried by the Green's functions attached to the two velocity vertices. The second velocity vertex appears in the last term of Eq.~\ref{eq:OR-Currents}, because, as can be seen from Eqs.~\ref{eq:EC-GE} and ~\ref{eq:QKET-G_TransInv2}, the external perturbations are accompanied by the velocity.

\begin{figure}[pb]
\begin{flushright}\begin{minipage}{0.5\textwidth}  \centering
        \includegraphics[width=1\textwidth]{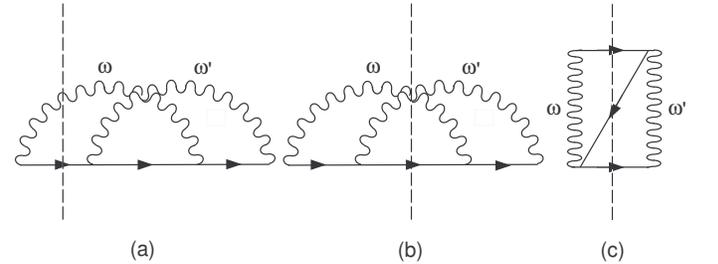}
                 \caption[0.4\textwidth]{\small The self-energy in the second order expansion with respect to  the interaction.}
                 \label{fig:SelfEnergy2st}
\end{minipage}\end{flushright}
\end{figure}

We will show now that this symmetry is embedded in the structure of the combination $\mathcal{Z}$ entering the last term in Eq.~\ref{eq:OR-Currents}:
\begin{align}\label{eq:OR-Struct}
\mathcal{Z}=\Sigma_{\mathbf{F}}^{<}(\epsilon)(1-n_F(\epsilon))+\Sigma_{\mathbf{F}}^{>}(\epsilon)n_F(\epsilon).
\end{align}
Recall that $\Sigma_{\mathbf{F}}$ contains one $\mathbf{F}$-dependent propagator, $\hat{G}_{\mathbf{F}}$ or $\hat{V}_{\mathbf{F}}$, that by itself can include a self-energy that depends on the external perturbation, $\hat{\Sigma}_{\mathbf{F}}$ or $\hat{\Pi}_{\mathbf{F}}$, i.e., $\Sigma_{\mathbf{F}}$ is determined iteratively. We start our analysis of $\mathcal{Z}$  at the point when the iterative process was already terminated. In other word, we will use the expression for $\Sigma_{\mathbf{F}}$ at a stage when  the Green's function depending on the external perturbation is equal to:
\begin{align}\label{eq:OR-G_F}
\hat{G}_{\mathbf{F}s}&(\epsilon+\Omega)\\\nonumber
&=-\frac{i}{2}\mathbf{F}(\epsilon+\Omega)\left[\frac{\partial{\hat{g}_{eq}}(\epsilon+\Omega)}{\partial{\epsilon}}\hat{\mathbf{v}}(\epsilon+\Omega)\hat{g}_{eq}(\epsilon+\Omega)\right.\\\nonumber
&\left.-\hat{g}_{eq}(\epsilon+\Omega)\hat{\mathbf{v}}(\epsilon+\Omega)\frac{\partial{\hat{g}_{eq}}(\epsilon+\Omega)}{\partial{\epsilon}}\right].
\end{align}
Here the argument of the Green's function reflects the fact that in the  proof of the Onsager relation we will apply the standard trick used in the derivation of the Ward identities. Namely, we will arrange the arguments of the Green's functions inside $\Sigma_{\mathbf{F}}(\epsilon)$ in such a way that they all include $\epsilon$ (even those inside the loops).

Any given contribution to $\Sigma^{<,>}$ is a sum of few terms with a different analytic structure of the propagators. Each of these terms can be cut into two pieces in such a way that the propagators along the cut are the lesser or greater components only. [For a self-energy with two crossed interaction lines all possible cuts are shown in Fig.~\ref{fig:SelfEnergy2st}.] Therefore, each term can be written as a product $A\cdot{B}$, where $B$ includes all the propagators along the cut, while all the rest of the propagators are collected in $A$. Then, each term in  $\Sigma^{<}=A\cdot{B}^{<}$ has a counterpart in $\Sigma^{>}=A\cdot{B}^{>}$ with the same $A$ but "opposite" $B$. "Opposite" means that if we substitute in $B^{<}$ the lesser propagators by the greater ones and simultaneously all the greater by lesser, we get $B^{>}$.

To demonstrate how this prescription works, let us look at the expression corresponding to the diagram presented in Fig.~\ref{fig:SelfEnergy2st}(b):
\begin{align}\label{eq:OR-Sigma2nd}\nonumber
&\Sigma^{<,>}(\mathbf{r},\mathbf{r}',\epsilon)=-\frac{1}{2}\int\frac{d\omega{d\omega'}}{(2\pi)^2}d\mathbf{r}_{\scriptscriptstyle1}d\mathbf{r}_{\scriptscriptstyle2}
G^{R}(\mathbf{r},\mathbf{r}_{\scriptscriptstyle1},\epsilon-\omega)\\\nonumber
&\times{G}^{<,>}(\mathbf{r}_{\scriptscriptstyle1},\mathbf{r}_{\scriptscriptstyle2},\epsilon-\omega-\omega')
V^{<,>}(\mathbf{r},\mathbf{r}_{\scriptscriptstyle2},\omega)V^{<,>}(\mathbf{r}_{\scriptscriptstyle1},\mathbf{r}',\omega')\\\nonumber
&\times{G}^{A}(\mathbf{r}_{\scriptscriptstyle2},\mathbf{r}',\epsilon-\omega')-\frac{1}{2}
\int\frac{d\omega{d\omega'}}{(2\pi)^2}d\mathbf{r}_{\scriptscriptstyle1}d\mathbf{r}_{\scriptscriptstyle2}
G^{R}(\mathbf{r},\mathbf{r}_{\scriptscriptstyle1},\epsilon+\omega)\\\nonumber
&\times{G}^{<,>}(\mathbf{r}_{\scriptscriptstyle1},\mathbf{r}_{\scriptscriptstyle2},\epsilon+\omega+\omega')
V^{>,<}(\mathbf{r}_{\scriptscriptstyle2},\mathbf{r},\omega)V^{>,<}(\mathbf{r}',\mathbf{r}_{\scriptscriptstyle1},\omega')\\
&\times{G}^{A}(\mathbf{r}_{\scriptscriptstyle2},\mathbf{r}',\epsilon+\omega').
\end{align}
In this example $A=G^RG^A$, $B^{<}=G^{<}V^{<}V^{<}$, and $B^{>}=G^{>}V^{>}V^{>}$. The above expression contains two contributions in which the frequencies $\omega$ and $\omega'$ appear with opposite signs. To get the two contributions we split the lesser and greater components of the interaction propagator into two pieces according to Eq.~\ref{eq:WF-V}.  We will use this representation of $\hat{V}$ throughout the proof of the Onsager relation in order to write the self-energy as a sum of two terms in which the $\mathbf{F}$-dependent Green's function appears either as $G_{\mathbf{F}}(\epsilon-\Omega)$ or $G_{\mathbf{F}}(\epsilon+\Omega)$.

Let us return to the  discussion of the general properties of $\mathcal{Z}$. After a pair, $AB^{<}$ and $AB^{>}$, is inserted into Eq.~\ref{eq:OR-Struct}, one has to linearize it with respect to the external perturbation. Recall that in the end of the iterative process, $\mathbf{F}$ enters the self-energy through the quasiparticle Green's functions $\hat{G}_{\mathbf{F}s}$. When this Green's function belongs to $A$, the corresponding contribution to Eq.~\ref{eq:OR-Struct}, $A_{\mathbf{F}}[(1-n_F(\epsilon))B_{eq}^{<}(\epsilon)+n_F(\epsilon)B_{eq}^{>}(\epsilon)]$, vanishes. This is because the total frequency transferred along the cut is equal to $\epsilon$, and at equilibrium $\hat{B}(\epsilon)$ has the same properties as any other fermionic propagator in the Keldysh technique. Therefore, the only non-vanishing contribution is of the form  $\mathcal{Z}=A_{eq}[(1-n_F(\epsilon))B_{\mathbf{F}}^{<}+n_F(\epsilon)B_{\mathbf{F}}^{>}]$. Before we start analyzing $B_{\mathbf{F}}$, we wish to note that we may extend $A_{eq}$ to include all the terms with the same cut. The Green's function through which the external perturbation enters $B_{\mathbf{F}}$, $G_{\mathbf{F}s}(\epsilon\pm\Omega)$, is described by Eq.~\ref{eq:OR-G_F}. The rest of the propagators in $B$ which are at equilibrium carry altogether the bosonic frequency $\Omega$ and will be denote as $P_{eq}(\Omega)$. Then, we can write $B$ as $B^{<}(\epsilon)=G_{\mathbf{F}s}^{<}(\epsilon-\Omega)P_{eq}(\Omega)n_P(\Omega)+G_{\mathbf{F}s}^{<}(\epsilon+\Omega)P_{eq}(\Omega)(1+n_P(\Omega))$, and correspondingly, $B^{>}(\epsilon)=G_{\mathbf{F}s}^{>}(\epsilon-\Omega)P_{eq}(\Omega)(1+n_P(\Omega))+G_{\mathbf{F}s}^{>}(\epsilon+\Omega)P_{eq}(\Omega)n_P(\Omega)$. In the last two identities we assumed that the $\mathbf{F}$-dependent Green's function in $B^{<}$ is $G_{\mathbf{F}s}^{<}$, while in $B^>$ it is $G_{\mathbf{F}s}^{>}$. [The possibility that the external perturbation enters $B^{<}$ $(B^{>})$ through $G_{\mathbf{F}s}^{>}$ ($G_{\mathbf{F}s}^{<}$)  will be discussed below.] Inserting the above expression for $B^{<,>}$ into $\mathcal{Z}$, we get:
\begin{align}\label{eq:OR-Struct2}
\mathcal{Z}&=[(1-n_F(\epsilon))n_P(\Omega)G_{\mathbf{F}s}^{<}(\epsilon-\Omega)\\\nonumber
&+n_F(\epsilon)(1+n_P(\Omega))G_{\mathbf{F}s}^{>}(\epsilon-\Omega)]{A}_{eq}P_{eq}(\Omega)\\\nonumber
&+[(1-n_F(\epsilon))(1+n_P(\Omega))G_{\mathbf{F}s}^{<}(\epsilon+\Omega)\\\nonumber
&+n_F(\epsilon)n_P(\Omega)G_{\mathbf{F}s}^{>}(\epsilon+\Omega)]{A}_{eq}P_{eq}(\Omega).
\end{align}
Applying the Keldysh rules on the product of matrices in Eq.~\ref{eq:OR-G_F}, we obtain that $G_{\mathbf{F}s}^{<}(\epsilon-\Omega)=n_F(\epsilon-\Omega)\mathcal{X}+\partial{n_F(\epsilon-\Omega)}/\partial\epsilon\mathcal{Y}$, and $G_{\mathbf{F}s}^{>}(\epsilon-\Omega)=(n_F(\epsilon-\Omega)-1)\mathcal{X}+\partial{n_F(\epsilon-\Omega)}/\partial\epsilon\mathcal{Y}$.  From the identity given in Eq.~\ref{eq:OR-identities2}, it follows that the terms proportional to $\mathcal{X}$ in Eq.~\ref{eq:OR-Struct2} vanish. Then, using the identity for the product of distribution functions given in Eq.~\ref{eq:OR-identities1},  we get that the discussed contributions to the thermoelectric currents are of the form:
\begin{align}\label{eq:generalJ}
\mathbf{j}_{e,h}&=\int\frac{d\epsilon{d}\omega}{(2\pi)^{2}}\frac{\partial{n_{P}(\Omega)}}{\partial{\Omega}}[n_F(\epsilon)-n_F(\epsilon-\Omega)]\\\nonumber
&\times\left[{f}_{e,h}(\epsilon)I_1\mathbf{F}(\epsilon-\Omega)+{f}_{e,h}(\epsilon-\Omega)I_{2}\mathbf{F}(\epsilon)\right].
\end{align}
In the second term we shifted the frequency $\epsilon$ by $\Omega$. This shift of the frequency affects $I_{2}$ in such a way that under the condition of microscopic reversibility, the functions $I_1$ and $I_{2}$ transform one into another when read in the opposite directions. Here we rely on the fact that all the contributions with the same cut have been included in $I_1$ and $I_2$. Then, it follows from Eq.~\ref{eq:generalJ} that all the contributions to the thermoelectric current under discussion satisfy the Onsager relation.

It remains to examine the case when the external perturbation enters $B^{<}$ through $G_{\mathbf{F}s}^{>}$ and $B^{>}$ through $G_{\mathbf{F}s}^{<}$. Then, $B$ can be written as $B^{<}=G_{\mathbf{F}s}^{>}(\epsilon-\Omega)P_{eq}(\Omega)(1+n_P(\Omega))+G_{\mathbf{F}s}^{>}(\epsilon+\Omega)P_{eq}(\Omega)n_P(\Omega)$, and correspondingly, $B^{>}=G_{\mathbf{F}s}^{<}(\epsilon-\Omega)P_{eq}(\Omega)n_P(\Omega)+G_{\mathbf{F}s}^{>}(\epsilon+\Omega)P_{eq}(\Omega)(1+n_P(\Omega))$. One can check, in the same way as before,  that the contributions to the current from such terms are also described by Eq.~\ref{eq:generalJ}. Thus, we have shown that the last term in Eq.~\ref{eq:OR-Currents} is compatible with the Onsager relation. In Appendix~\ref{App:OR}, we give representative examples illustrating the general arguments presented here.

To summarize, we proved that the expressions for the longitudinal thermoelectric currents obtained using the quantum kinetic approach (presented in Eq.~\ref{eq:OR-Currents}) satisfy the Onsager relations. We use the Onsager relations as a check for the correctness of our scheme. In the presence of a magnetic field the transverse thermoelectric currents contain additional contributions and, hence, the proof of the Onsager relations should be modified accordingly. A demonstration of the validity of the Onsager relations for the thermoelectric transport coefficients in the presence of a magnetic field for electrons interacting in the Cooper channel will be given in a separated manuscript.

\section{Summary}

We developed a new scheme for studying thermal and thermoelectric transport in interacting electron systems in the presence of disorder. The kinetic equation in the presence of a temperature gradient was derived directly from the action. One of the novel aspects of the scheme regards the expressions for the four currents corresponding to the different components of the conductivity tensor, see Eqs.~\ref{eq:EC-AverageCurrent},~\ref{eq:HC-J_e},~\ref{eq:HC-HeatCurrentT} and~\ref{eq:HC-HeatCurrentE2}. These currents, derived from the continuity equations for the charge and the energy, share a uniform and compact structure summarized in Eq.~\ref{eq:current}. This equation reveals that the expressions for both the electric and heat currents include the renormalized velocity. The frequency factor in the expression for heat current corresponds to the legs of the renormalized velocity as illustrated in Fig.~\ref{fig:velocity} (but not to the frequency of the two Green's functions connected to the bare velocity inside the vertex). This observation, which is the main advantage of the scheme presented here, implies that in the heat current the flow of energy occurs with the renormalized velocity. As we  demonstrated in Sec.~\ref{sec:FermiLiquidQKE}, this structure of the heat current guarantees that the Wiedemann-Franz law is satisfied for Fermi-liquid systems.

In this paper we considered the Coulomb interaction, which is instantaneous in time. Applying our scheme for electron-electron interaction mediated by phonons that by themselves carry energy is straitforward.  An example for electron-electron interaction of a different kind has been briefly described in Appendix~\ref{sec:ElectricCond-SCF}. There we consider an interaction mediated by superconducting fluctuations. In this case, the electric current acquires an additional contribution because the fluctuations in the Cooper channel carry charge. Nevertheless, this additional contribution does not ruin the general structure of the current.

The main strength of our scheme is in its generality and compactness. This is the reason why we believe that it is an adequate alternative to the Kubo formula, which for the thermal transport is rather cumbersome.

\begin{acknowledgments}
We thank Carlo~DiCastro, Georg~Schwiete, and the condensed matter theory group in TAMU for their interest in this work and for the extended discussions. The research was supported by the US-Israel BSF.
\end{acknowledgments}

\appendix
\section{The electric field dependent Green's function}\label{App:ElecricCond}

In this appendix we wish to present additional details of the derivation of the quantum kinetic equation in the presence of an electric field. In particular,  we show how to obtain the electric field dependent Green's function which solves the kinetic equation, see Eq.~\ref{eq:EC-GE}.

We start from the Dyson equation for the Green's function of the quasiparticles given in Eq.~\ref{eq:EC-DEG}.
After performing the gauge invariant Fourier transform introduced in Eq.~\ref{eq:EC-Wigner}, we obtain the quantum kinetic equation for the gauge invariant Green's function:
\begin{widetext}
\begin{align}\label{eq:App-EC-QKE}\nonumber
&\left\{\frac{i}{2}\frac{\partial}{\partial{\mathcal{T}}}+\left(\epsilon+\frac{1}{2m}\frac{\partial^2}{\partial\boldsymbol{\rho}^2}\right)+\frac{1}{8m}\frac{\partial^2}{\partial\mathbf{R}^2}+\frac{e^2E^2}{8m}\frac{\partial^2}{\partial{\epsilon}^2}-
\frac{e\mathbf{E}}{4m}\frac{\partial^2}{\partial\epsilon\partial\mathbf{R}}+\frac{1}{2m}\frac{\partial^2}{\partial\boldsymbol{\rho}\partial\mathbf{R}}-\frac{e\mathbf{E}}{2m}\frac{\partial^2}{\partial\boldsymbol{\rho}\partial\epsilon}-V_{imp}(\mathbf{R}+\boldsymbol{\rho}/2)
-\frac{e\boldsymbol{\rho}\mathbf{E}}{2}\right\}\\\nonumber
&\times\hat{\underline{G}}(\mathbf{R},\mathcal{T};\boldsymbol{\rho},\epsilon)=\delta(\boldsymbol{\rho})
+\int{dt_{\scriptscriptstyle1}d\mathbf{r}_{\scriptscriptstyle1}}d\tau{d}\epsilon_{\scriptscriptstyle1}d\epsilon_{\scriptscriptstyle2}{e}^{i\epsilon\tau-i\epsilon_{\scriptscriptstyle1}(\tau-t_{\scriptscriptstyle1})-i\epsilon_{\scriptscriptstyle2}\tau}
\hat{\underline{\Sigma}}\left(\mathbf{R}+{\mathbf{r}_{\scriptscriptstyle1}}/{2},\mathcal{T}+t_{\scriptscriptstyle1}/2;\boldsymbol{\rho}-\mathbf{r}_{\scriptscriptstyle1},\epsilon_{\scriptscriptstyle1}\right)\\
&\times{e}^{-ie\mathbf{E}/{2}[\mathbf{r}_{\scriptscriptstyle1}(\tau-t_{\scriptscriptstyle1})-(\boldsymbol{\rho}-\mathbf{r}_{\scriptscriptstyle1})t_{\scriptscriptstyle1}]}
\hat{\underline{G}}\left(\mathbf{R}-(\boldsymbol{\rho}-\mathbf{r}_{\scriptscriptstyle1})/{2},\mathcal{T}-(\tau-t_{\scriptscriptstyle1})/2;\mathbf{r}_{\scriptscriptstyle1},\epsilon_{\scriptscriptstyle2}\right).
\end{align}
The proof that the Green's function transformed in this way is gauge invariant as well as other useful technical details can be found in Ch.~$7$ of Ref.~\onlinecite{Haug}. The quantum kinetic equation is considerably simplified if to restrict the calculation to the steady state response (i.e., a DC electric field). Then, the dependence of the Green's functions and self-energies on time is only through the relative time-coordinate $\tau$. After expanding the phases in the RHS with respect to the electric field, the kinetic equation acquires the form presented in Eq.~\ref{eq:EC-QKE-E}.

Next, we write the Green's function as $\hat{\underline{G}}=\hat{g}_{eq}+\hat{G}_{\mathbf{E}}$ and, similarly, we replace the self-energy by $\hat{\underline{\Sigma}}=\hat{\sigma}_{eq}+\hat{\Sigma}_{\mathbf{E}}$. As the properties of the equilibrium Green's function have been already described in Eq.~\ref{eq:GeqMatrix}, here we shall examine only the electric field dependent part of the Green's function. For this purpose, we collect all the terms  in Eq.~\ref{eq:EC-QKE-E}, which are linear in the electric field:
\begin{align}\label{eq:App-EC-QKERealSpace1}
&\left(\epsilon+\frac{\boldsymbol{\nabla}^2}{2m}+\mu-V_{imp}\right)\hat{G}_{\mathbf{E}}(\boldsymbol{\rho},\epsilon;imp)
-\int{d\mathbf{r}_{\scriptscriptstyle1}}\hat{\sigma}_{eq}(\boldsymbol{\rho}-\mathbf{r}_{\scriptscriptstyle1},\epsilon;imp)
\hat{G}_{\mathbf{E}}(\mathbf{r}_{\scriptscriptstyle1},\epsilon;imp)\\\nonumber
&=
\frac{e\mathbf{E}}{2}\int{d\mathbf{r}_{\scriptscriptstyle1}}\left\{
\left[\delta(\boldsymbol{\rho}-\mathbf{r}_{\scriptscriptstyle1})-\frac{\partial\hat{\sigma}_{eq}\left(\boldsymbol{\rho}-\mathbf{r}_{\scriptscriptstyle1},\epsilon;imp\right)}
{\partial\epsilon}\right]\mathbf{r}_{\scriptscriptstyle1}
+\left[\frac{\delta(\boldsymbol{\rho}-\mathbf{r}_{\scriptscriptstyle1})}{m}\frac{\partial}{\partial\mathbf{r}_{\scriptscriptstyle1}}+(\boldsymbol{\rho}-\mathbf{r}_{\scriptscriptstyle1})\hat{\sigma}_{eq}\left(\boldsymbol{\rho}-\mathbf{r}_{\scriptscriptstyle1},\epsilon;imp\right)\right]
\frac{\partial}{\partial\epsilon}\right\}\\\nonumber&\times\hat{g}_{eq}\left(\mathbf{r}_{\scriptscriptstyle1},\epsilon;imp\right)
+\int{d\mathbf{r}_{\scriptscriptstyle1}}\hat{\Sigma}_{\mathbf{E}}\left(\boldsymbol{\rho}-\mathbf{r}_{\scriptscriptstyle1},\epsilon;imp\right)\hat{g}_{eq}\left(\mathbf{r}_{\scriptscriptstyle1},\epsilon;imp\right)
.
\end{align}
As we have already mentioned, the entire dependence of the gauge invariant Green's functions and self-energies on the center of mass coordinates is due to the scattering by the impurity potential. This dependence has been incorporated in our notation into $imp$. We will use the fact that the equilibrium Green's function is
$\hat{g}_{eq}^{-1}(\boldsymbol{\rho},\epsilon;imp)=\epsilon+\frac{\boldsymbol{\nabla}^2}{2m}+\mu-V_{imp}-\sigma_{eq}(\boldsymbol{\rho},\epsilon;imp)$ (see Eq.~\ref{eq:EC-Geq}). Then, we may rewrite Eq.~\ref{eq:App-EC-QKERealSpace1} as:
\begin{align}\label{eq:App-EC-QKERealSpace2}
&\int{d\mathbf{r}_{\scriptscriptstyle1}}\hat{g}_{eq}^{-1}(\boldsymbol{\rho}-\mathbf{r}_{\scriptscriptstyle1},\epsilon;imp)\hat{G}_{\mathbf{E}}(\mathbf{r}_{\scriptscriptstyle1},\epsilon;imp)
=\int{d\mathbf{r}_{\scriptscriptstyle1}d}
\hat{\Sigma}_{\mathbf{E}}\left(\boldsymbol{\rho}-\mathbf{r}_{\scriptscriptstyle1},\epsilon;imp\right)\hat{g}_{eq}\left(\mathbf{r}_{\scriptscriptstyle1},\epsilon;imp\right)\\\nonumber
&+\frac{e\mathbf{E}}{2}\hspace{-1mm}\int\hspace{-1mm}{d\mathbf{r}_{\scriptscriptstyle1}}\left\{
\left[\delta(\boldsymbol{\rho}-\mathbf{r}_{\scriptscriptstyle1})-\frac{\partial\hat{\sigma}_{eq}\left(\boldsymbol{\rho}-\mathbf{r}_{\scriptscriptstyle1},\epsilon;imp\right)}
{\partial\epsilon}\right]\mathbf{r}_{\scriptscriptstyle1}\hspace{-1mm}+\hspace{-1mm}\left[\frac{\delta(\boldsymbol{\rho}-\mathbf{r}_{\scriptscriptstyle1})}{m}\frac{\partial}{\partial\mathbf{r}_{\scriptscriptstyle1}}+(\boldsymbol{\rho}-\mathbf{r}_{\scriptscriptstyle1})\hat{\sigma}_{eq}\left(\boldsymbol{\rho}-\mathbf{r}_{\scriptscriptstyle1},\epsilon;imp\right)\right]
\frac{\partial}{\partial\epsilon}\right\}\hat{g}_{eq}\left(\mathbf{r}_{\scriptscriptstyle1},\epsilon;imp\right).
\end{align}
One may check that the following identities for the equilibrium Green's function hold:
\begin{align}\label{eq:App-EC-derivatives}
\frac{\partial\hat{g}_{eq}(\boldsymbol{\rho},\epsilon;imp)}{\partial\epsilon}&=
-\int{d\mathbf{r}_{\scriptscriptstyle1}d\mathbf{r}_{\scriptscriptstyle2}}\hat{g}_{eq}\left(\mathbf{r}_{\scriptscriptstyle1},\epsilon;imp\right)
\left[\delta(\boldsymbol{\rho}-\mathbf{r}_{\scriptscriptstyle1}-\mathbf{r}_{\scriptscriptstyle2})-\frac{\partial\hat{\sigma}_{eq}\left(\boldsymbol{\rho}-\mathbf{r}_{\scriptscriptstyle1}-\mathbf{r}_{\scriptscriptstyle2},\epsilon;imp\right)}
{\partial\epsilon}\right]\hat{g}_{eq}\left(\mathbf{r}_{\scriptscriptstyle2},\epsilon;imp\right);
\\\nonumber
-i\boldsymbol{\rho}\hat{g}_{eq}(\boldsymbol{\rho},\epsilon;imp)&=-i
\int{d\mathbf{r}_{\scriptscriptstyle1}d\mathbf{r}_{\scriptscriptstyle2}}\hat{g}_{eq}\left(\mathbf{r}_{\scriptscriptstyle1},\epsilon;imp\right)
\left[\frac{\delta(\boldsymbol{\rho}-\mathbf{r}_{\scriptscriptstyle1}-\mathbf{r}_{\scriptscriptstyle2})}{m}\frac{\partial}{\partial\mathbf{r}_{\scriptscriptstyle2}}+(\boldsymbol{\rho}-\mathbf{r}_{\scriptscriptstyle1}-\mathbf{r}_{\scriptscriptstyle2})\hat{\sigma}_{eq}\left(\boldsymbol{\rho}-\mathbf{r}_{\scriptscriptstyle1}-\mathbf{r}_{\scriptscriptstyle2},\epsilon;imp\right)\right]
\hat{g}_{eq}\left(\mathbf{r}_{\scriptscriptstyle2},\epsilon;imp\right)\\\nonumber
&=\hat{g}(\epsilon)\hat{\mathbf{v}}(\epsilon)\hat{g}(\epsilon).
\end{align}
\end{widetext}
Using the above identities, one may convert Eq.~\ref{eq:App-EC-QKERealSpace2} into  Eq.~\ref{eq:EC-GE} describing $\hat{G}_{\mathbf{E}}$.

The expression for $\hat{G}_{\mathbf{E}}$ is used in order to transform from Eq.~\ref{eq:EC-AverageCurrent} to the final formula for the electric current presented in Eq.~\ref{eq:EC-JeFinal}. Before inserting Eq.~\ref{eq:EC-GE} into Eq.~\ref{eq:EC-AverageCurrent}, one should recall that the renormalized velocity which appears in the expression for the current also depends on the electric field through the self-energy term. Therefore, the current linearized with respect to $\mathbf{E}$ contains two terms:
\begin{align}\label{eq:App-EC-Current}\nonumber
\mathbf{j}_e&=-\frac{ie}{2}\int{d\mathbf{r}'dt'}\left[\hat{\mathbf{v}}_{eq}(\mathbf{r},t;\mathbf{r}',t')\hat{G}_{\mathbf{E}}(\mathbf{r}',t';\mathbf{r},t)\right.\\
&\left.+\hat{\mathbf{v}}_{\mathbf{E}}(\mathbf{r},t;\mathbf{r}',t')\hat{g}_{eq}(\mathbf{r}',t';\mathbf{r},t)
\right]^{<}+h.c,
\end{align}
where $\hat{\mathbf{v}}_{\mathbf{E}}(\mathbf{r},t;\mathbf{r}',t')=-i(\mathbf{r-r}')\Sigma_{\mathbf{E}}(\mathbf{r},t;\mathbf{r}',t')$. One may notice that in the limit $\mathbf{r}'\rightarrow\mathbf{r}$ and $t'\rightarrow{t}$ (i.e., under the trace) the term with the electric field dependent velocity can be arranged as follows:
\begin{align}\label{eq:App-EC-VEterm}
&-i(\mathbf{r-r}')\Sigma_{\mathbf{E}}(\mathbf{r},t;\mathbf{r}',t')\hat{g}_{eq}(\mathbf{r}',t';\mathbf{r},t)\\\nonumber
&=\Sigma_{\mathbf{E}}(\mathbf{r},t;\mathbf{r}',t')i(\mathbf{r}'-\mathbf{r})\hat{g}_{eq}(\mathbf{r}',t';\mathbf{r},t)
=-\hat{\Sigma}_{\mathbf{E}}g_{eq}\hat{\mathbf{v}}_{eq}\hat{g}_{eq}.
\end{align}
Finally, one should replace $\hat{G}_{\mathbf{E}}$ in Eq.~\ref{eq:App-EC-Current} with its explicit expression given in Eq.~\ref{eq:EC-GE}. To get Eq.~\ref{eq:EC-JeFinal}, it remains to extract the lesser component. For this purpose, we use the fact that the current is diagonal in the energy (convolution in time), and thus in the basis of the retarded, advanced and Keldysh components the product of matrices obeys the following rules:
\begin{align}\label{eq:App-EC-Product}
&(\hat{A}\cdot\hat{B})^{R}=A^{R}B^{R};\\\nonumber
&(\hat{A}\cdot\hat{B})^{A}=A^{A}B^{A};\\\nonumber
&(\hat{A}\cdot\hat{B})^{K}=A^{R}B^{K}+A^{K}B^{A}.
\end{align}
In addition, we use the relation $A^{<}=(A^{K}-A^{R}+A^{A})/2$.

\section{Calculation of the thermal conductivity using the simplified Kubo formula}\label{App:FermiLiquidKubo}

\begin{figure}[pb]
\begin{flushright}\begin{minipage}{0.5\textwidth}  \centering
        \includegraphics[width=0.9\textwidth]{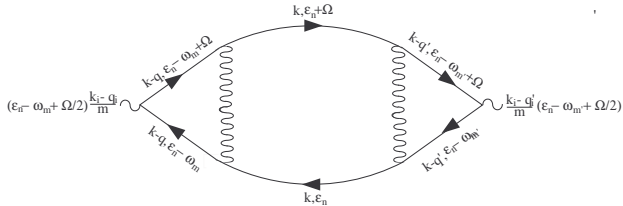}
                 \caption[0.4\textwidth]{\small A contribution to the thermal conductivity calculated using the simplified Kubo formula given in Eq.~\ref{eq:CurrentNonIntFrequency}.}
                 \label{fig:FermiLiquidKubo}
\end{minipage}\end{flushright}
\end{figure}

In this appendix we wish to examine the thermal conductivity calculated using the simplified Kubo formula presented in Eq.~\ref{eq:CurrentNonIntFrequency}. Let us consider one representative term in the perturbation expansion of the Kubo formula with respect to the electron-electron interaction. For this purpose, we choose the second order term drawn in Fig.~\ref{fig:FermiLiquidKubo}. In the Fermi-liquid theory, the main contribution to the transport coefficients arises from diagrams which contain one pair of Green's's functions with the same arguments but opposite analytic structure, i.e., one is retarded while the other is advanced. Then, the quasiparticles corresponding to these two Green's's functions are both on the mass shell. This pair yields a contribution to the transport coefficients that is proportional to $\tau$.

According to Eq.~\ref{eq:CurrentNonIntFrequency}, the expression corresponding to the diagram presented in Fig.~\ref{fig:FermiLiquidKubo} is
\begin{align}\label{eq:FLK-TheramlCond}\nonumber
\kappa&=-\frac{1}{\Omega{d}}T^{3}\sum_{\epsilon_{n},\omega_{m},\omega_{m'}}\sum_{\mathbf{k},\mathbf{q},\mathbf{q}'}
\frac{k_i-q_i}{m}i\left(\epsilon_n+\omega_m+\frac{\Omega}{2}\right)\\\nonumber
&\times\frac{k_i-q_i'}{m}i\left(\epsilon_n+\omega_{m'}+\frac{\Omega}{2}\right)
g(\mathbf{k-q},\epsilon_{n}-\omega_{m}+\Omega)\\\nonumber
&\times{g}(\mathbf{k-q},\epsilon_{n}-\omega_{m}){V}(\mathbf{q},\omega_{m})
g(\mathbf{k},\epsilon_{n}+\Omega)g(\mathbf{k},\epsilon_{n})\\
&\times{V}(\mathbf{q}',\omega_{m}')g(\mathbf{k-q}',\epsilon_{n}-\omega_{m'}+\Omega)g(\mathbf{k-q}',\epsilon_{n}-\omega_{m'}).
\end{align}
Notice, that in the above expression we use the heat current operators of free electrons. Here, $\epsilon_n=2\pi{T}(n+1/2)$ and $\omega_{m}=2\pi{T}m$ are the Matsubara frequencies, and the Green's function of the quasiparticles is $g(\mathbf{k},\epsilon_n)=[i\epsilon_n-\xi_{\mathbf{k}}+\frac{i}{2\tau}sign(\epsilon_n)]^{-1}$. In search for the DC conductivity, we should expand the expression to a linear order in the frequency of the external field, $\Omega$. Keeping only the terms with one $g^{R}g^{A}$ pair (\textit{R}-\textit{A} section), we get:
\begin{widetext}
\begin{align}\label{eq:FLK-TheramlCond2}
\kappa&=\frac{1}{\Omega{d}}T^{3}\sum_{\mathbf{k},\mathbf{q},\mathbf{q}'}
\sum_{-\Omega<\epsilon<0}\left[\sum_{\omega_{m}=-\infty}^{-\Omega}\sum_{\omega_{m'}=-\infty}^{-\Omega}
+\sum_{\omega_{m}=-\infty}^{-\Omega}\sum_{\omega_{m'}=\Omega}^{\infty}
+\sum_{\omega_{m}=\Omega}^{\infty}\sum_{\omega_{m'}=-\infty}^{-\Omega}
+\sum_{\omega_{m}=\Omega}^{\infty}\sum_{\omega_{m'}=\Omega}^{\infty}\right]\\\nonumber
&\times\left[\frac{k_i-q_i}{m}\left(\epsilon_n-\omega_{m}+\frac{\Omega}{2}\right)
\frac{k_i-q_i'}{m}\left(\epsilon_n-\omega_{m'}+\frac{\Omega}{2}\right)
V(\mathbf{q},\omega_m)V(\mathbf{q}',\omega_{m'})\right.\\\nonumber&\left.
+2
\frac{k_i}{m}\left(\epsilon_n+\frac{\Omega}{2}\right)
\frac{k_i-q_i'}{m}\left(\epsilon_n-\omega_{m'}+\frac{\Omega}{2}\right)
V(-\mathbf{q},-\omega_m)V(\mathbf{q}+\mathbf{q}',\omega_m+\omega_{m'})\right]g(\mathbf{k},\epsilon_{n}+\Omega)g(\mathbf{k},\epsilon_{n})\\\nonumber
&\times{g}(\mathbf{k-q},\epsilon_{n}-\omega_{m}+\Omega)g(\mathbf{k-q},\epsilon_{n}-\omega_{m})
{g}(\mathbf{k-q}',\epsilon_{n}-\omega_{m'}+\Omega)g(\mathbf{k-q}',\epsilon_{n}-\omega_{m'}).
\end{align}
\end{widetext}
In the above expression, the first line describes all the cases where the \textit{R}-\textit{A} section is in the middle of the diagram, while in the second line the \textit{R}-\textit{A}  section is located in one of the sides. The arguments in Eq.~\ref{eq:FLK-TheramlCond2}  are arranged in such a way that each time the frequency of the effective \textit{R}-\textit{A}  section is denoted by $\epsilon_n$. Then, the sum over the frequency $\epsilon_n$ is restricted to a narrow window of the width $\Omega$. Therefore, in the limit $\Omega\rightarrow0$, the external frequency in the Green's functions, the heat current vertices and the sum over the frequencies $\omega_m$ and $\omega_{m'}$  can be set to zero.

\begin{figure}[pb]
\begin{flushright}\begin{minipage}{0.5\textwidth}  \centering
        \includegraphics[width=0.75\textwidth]{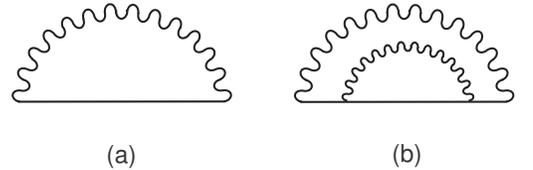}
                 \caption[0.4\textwidth]{\small The contributions to the self-energy of the quasiparticles relevant for the discussion of the diagram presented in Fig.~\ref{fig:FermiLiquidKubo}.}
                 \label{fig:SelfEnergy}
\end{minipage}\end{flushright}
\end{figure}

In the next step we perform the analytic continuation from the Matsubara frequencies to the real axis, replacing the sums over the frequencies with integrals. In the present calculation, the only difference between the thermal and the electric conductivities is that instead of the frequencies in the heat current vertices the electric current vertices give a factor of $-e^2$. Therefore, for the Wiedemann-Franz law to be valid, the integral over the frequency $\epsilon$ should be of the form $\int{d\epsilon}(\partial\tanh(\epsilon/2T)/\partial\epsilon)\epsilon^{2}$. We are going to check whether all the contributions to the thermal conductivity given in Eq.~\ref{eq:FLK-TheramlCond2} are indeed of this form. With this in mind, we separate the contributions to the thermal conductivity into two group. The first contains all the terms in which the \textit{R}-\textit{A} section is connected to a bare vertex. Let us concentrate on the $\epsilon$ integration, which is the same for all the terms in this group. In the limit $T\tau\ll1$, we may neglect the dependence of all the Green's functions on $\epsilon$. Then, the integral over $\epsilon$ reduces to:
\begin{align}\label{eq:FLK-integral-P1}
\int{d\epsilon}\frac{\partial\tanh(\epsilon/2T)}{\partial\epsilon}\epsilon\left(
\epsilon-\omega'\right).
\end{align}
Only the $\epsilon^2$ part in the above integral results in a finite contribution, while the rest  being an odd function of the frequency vanishes. Naturally, the contribution of this group acquires the following form:
\begin{align}\label{eq:FLK-TheramlCond-P1-Simple}
\kappa&=\frac{2}{d}\sum_{\mathbf{k}}
\int\frac{d\epsilon}{4\pi}\frac{\partial\tanh(\epsilon/2T)}{\partial\epsilon}
\frac{k_i}{m}\epsilon^2{g}_R(\mathbf{k},\epsilon)g_A(\mathbf{k},\epsilon)\\\nonumber
&\times\frac{\partial\Re\sigma_{\scriptscriptstyle2}^R(\mathbf{k},\epsilon)}{\partial{k}_i},
\end{align}
where $\sigma_{2}$ is one of the contributions to the self-energy with two interaction lines (see Fig.~\ref{fig:SelfEnergy}(b)). Clearly, such a contribution is consistent with the Fermi-liquid expression for the thermal conductivity presented in Eq.~\ref{eq:FL-Jh4}, and as such satisfies the Wiedemann-Franz law.

We turn to the second group that, as we will see, appear to be inconsistent with the Wiedemann-Franz law.  Now the \textit{R}-\textit{A} section is located between the two interaction lines. Expanding with respect to the external frequency and performing the analytic continuation, one gets:
\begin{widetext}
\begin{align}\label{eq:FLK-TheramlCond-P2}\nonumber
\kappa&=\frac{1}{d}\sum_{\mathbf{k},\mathbf{q},\mathbf{q}'}
\int\frac{d\omega'}{4\pi{i}}\coth\left(\frac{\omega'}{2T}\right)
\int\frac{d\omega}{4\pi{i}}\coth\left(\frac{\omega}{2T}\right)
\frac{k_i-q_i}{m}\frac{k_i-q_i'}{m}
\int\frac{d\epsilon}{4\pi}\frac{\partial\tanh(\epsilon/2T)}{\partial\epsilon}(\epsilon-\omega)(\epsilon-\omega')
g_R(\mathbf{k},\epsilon)g_A(\mathbf{k},\epsilon)\\\nonumber
&\times\left[V_{A}(\mathbf{q},\omega)V_{A}(\mathbf{q}',\omega')g_R^2(\mathbf{k-q},\epsilon-\omega)g_R^2(\mathbf{k-q}',\epsilon-\omega')
-V_{A}(\mathbf{q},\omega)V_{R}(\mathbf{q}',\omega')g_R^2(\mathbf{k-q},\epsilon-\omega)g_A^2(\mathbf{k-q}',\epsilon-\omega')\right.\\
&\left.-V_{R}(\mathbf{q},\omega)V_{A}(\mathbf{q}',\omega')g_A^2(\mathbf{k-q},\epsilon-\omega)g_R^2(\mathbf{k-q}',\epsilon-\omega')
+V_{R}(\mathbf{q},\omega)V_{R}(\mathbf{q}',\omega')g_A^2(\mathbf{k-q},\epsilon-\omega)g_A^2(\mathbf{k-q}',\epsilon-\omega')\right].
\end{align}
\end{widetext}
The above expression can be arranged in a rather compact form:
\begin{align}\label{eq:FLK-TheramlCond-P2B}
&\kappa=\frac{1}{d}\sum_{\mathbf{k}}
\int\frac{d\epsilon}{4\pi}\frac{\partial\tanh(\epsilon/2T)}{\partial\epsilon}g_R(\mathbf{k},\epsilon)g_A(\mathbf{k},\epsilon)\\\nonumber
&\times\left[\frac{\partial}{\partial{k}_i}\sum_{\mathbf{q}}\int\frac{d\omega}{4\pi{i}}\coth\left(\frac{\omega}{2T}\right)
(\epsilon-\omega)\left(\phantom{\frac{}{}}\hspace{-1mm}V_{A}(\mathbf{q},\omega)\right.\right.\\\nonumber
&\left.\left.\times{g}_R(\mathbf{k-q},\epsilon-\omega)-V_{R}(\mathbf{q},\omega)g_A(\mathbf{k-q},\epsilon-\omega)\phantom{\frac{}{}}\hspace{-1mm}\right)\phantom{\frac{1}{1}}\hspace{-3mm}\right]^2.
\end{align}
Note that the expression in the squared brackets is real. Without the factor $\epsilon-\omega$, this expression is precisely the first order expansion with respect to the interaction of the renormalized velocity. For Eq.~\ref{eq:FLK-TheramlCond-P2B}  to be consistent with Eq.~\ref{eq:FL-Jh4}, the expression in the squared brackets must reduce into $\epsilon\partial\Re\sigma_1(\mathbf{k},\epsilon)/\partial{k}_i$ (where $\sigma_1(\mathbf{k},\epsilon)$ is the self-energy term with a single interaction line shown in Fig.~\ref{fig:SelfEnergy}(a)).

Below we shall examine the expression in the squared brackets following the calculation of the self-energy in the Fermi-liquid theory presented in Ref.~\onlinecite{AGD}. First, we rewrite this expression in the following way:
\begin{align}\label{eq:FLK-"SelfEnegy"}\nonumber
\Lambda&(\mathbf{k},\epsilon)=-\frac{\partial}{\partial{k}_i}\mathcal{R}\left[\int\frac{d\mathbf{q}d\epsilon'}{(2\pi)^{d+1}}\tanh\left(\frac{\epsilon'}{2T}\right)
\epsilon'V_R(\mathbf{q},\epsilon-\epsilon')\right.\\\nonumber
&\left.\times\mathcal{I}m{g}_{R}(\mathbf{k-q},\epsilon')+\int\frac{d\mathbf{q}d\omega}{(2\pi)^{d+1}}
\coth\left(\frac{\omega}{2T}\right)\right.\\
&\left.\times(\epsilon-\omega){g}_{R}(\mathbf{k-q},\epsilon-\omega){\mathcal{I}m}V_R(\mathbf{q},\omega)
\phantom{\int}\hspace{-2mm}\right].
\end{align}
Next, we use the Lehman representation for the Green's function and the interaction propagator:
\begin{align}\label{eq:FLK-"SelfEnegy"2}\nonumber
\Lambda&(\mathbf{k},\epsilon)=-\frac{\partial}{\partial{k}_i}\mathcal{R}\int\frac{d\mathbf{q}d\epsilon'}{(2\pi)^{d+1}}\int_{-\infty}^{\infty}\frac{d\omega}{\pi}
\left[\epsilon'\tanh\left(\frac{\epsilon'}{2T}\right)\right.\\
&\left.+
(\epsilon-\omega)\coth\left(\frac{\omega}{2T}\right)
\right]\frac{\mathcal{I}m{V}_R(\mathbf{q},\omega)\mathcal{I}m{g}_{R}(\mathbf{k-q},\epsilon')}{\omega+\epsilon'-\epsilon-i\delta}.
\end{align}
One may approximate the imaginary part of the quasiparticle Green's function with a $\delta$-function, $\mathcal{I}m{g}_R(\mathbf{k},\epsilon)\approx-\pi\delta(\epsilon-\xi_{\mathbf{k}})$. For the sake of simplicity, we consider a three dimensional system with a parabolic spectrum, $\xi_{\mathbf{k}}=k^2/2m-\mu$.  Then, the integration over the momentum $\mathbf{q}$ can be written as $\int{d\mathbf{q}}=\int_0^{\infty}2\pi{q}^2dq\int_{0}^{\pi}sin\theta{d\theta}$:
\begin{align}\label{eq:FLK-"SelfEnegy"3}\nonumber
\Lambda(\mathbf{k},\epsilon)&=\frac{\partial}{\partial{k}_i}\mathcal{R}\int\frac{d\epsilon'd\omega}{(2\pi)^{3}}\int_{0}^{\infty}q^2dq
\frac{\mathcal{I}m{V}_R(q,\omega)}{\omega+\epsilon'-\epsilon-i\delta}\\
&\times\left[\epsilon'\tanh\left(\frac{\epsilon'}{2T}\right)+
(\epsilon-\omega)\coth\left(\frac{\omega}{2T}\right)\right]\\\nonumber
&\times\int_{-1}^{1}dx\delta\left(\epsilon'-\frac{k^2-2kqx+q^2}{m}\right).
\end{align}
The integration over $x$ yields non-zero result when $(k-q)^2/2m<\epsilon'<(k+q)^2/2m$.

Next, we transform from the integral $\int_{-\infty}^{\infty}d\omega$ to $\int_{0}^{\infty}d\omega$:
\begin{align}\label{eq:FLK-"SelfEnegy"4}
&\Lambda(\mathbf{k},\epsilon)=\frac{1}{(2\pi)^3}\frac{\partial}{\partial{k}_i}\frac{m}{k}\int{d\epsilon'}\mathcal{P}\int_0^{\infty}{d\omega}\int_{0}^{\infty}qdq\\\nonumber
&\hspace{-1mm}\times\left\{\frac{\mathcal{I}m{V}_R(q,\omega)}{\omega+\epsilon'-\epsilon}
\times\left[\epsilon'\tanh\left(\frac{\epsilon'}{2T}\right)+
(\epsilon-\omega)\coth\left(\frac{\omega}{2T}\right)\right]\right.\\\nonumber
&\left.-\frac{\mathcal{I}m{V}_R(\mathbf{q},\omega)}{-\omega+\epsilon'-\epsilon}
\left[\epsilon'\tanh\left(\frac{\epsilon'}{2T}\right)-
(\epsilon+\omega)\coth\left(\frac{\omega}{2T}\right)\right]\right\},
\end{align}
where $\mathcal{P}\int$ denotes the principle value. Finally, simple manipulations of the integrals yield:
\begin{align}\label{eq:FLK-"SelfEnegy"5}\nonumber
&\Lambda(\mathbf{k},\epsilon)=\frac{1}{(2\pi)^3}\frac{\partial}{\partial{k}_i}\frac{m}{k}\int{d\epsilon'}\mathcal{P}\hspace{-1mm}
\int_0^{\infty}\hspace{-1mm}{d\omega}\hspace{-1mm}\int_{0}^{\infty}qdq
\frac{\mathcal{I}m{V}_R(q,\omega)}{\omega+\epsilon'}\\\nonumber
&\hspace{-1mm}\times\left\{\epsilon\left[\tanh\left(\frac{\epsilon+\epsilon'}{2T}\right)\hspace{-1mm}+
\tanh\left(\frac{\epsilon-\epsilon'}{2T}\right)\right]\hspace{-1mm}+\hspace{-0.5mm}
\epsilon'\hspace{-0.5mm}\left[\tanh\left(\frac{\epsilon+\epsilon'}{2T}\right)\right.\right.\\
&\left.\left.\hspace{4mm}
-\tanh\left(\frac{\epsilon-\epsilon'}{2T}\right)\right]
-2\omega\coth\left(\frac{\omega}{2T}\right)\right\}.
\end{align}
We may compare now between the expressions for the self-energy and $\Lambda$. In the self-energy the last two terms do not exist. The first term in the above equation for $\Lambda$ is equal to the self-energy multiplied by the frequency $\epsilon$, and hence it precisely coincides with the renormalized heat current vertex (see Eq.~\ref{eq:FL-Jh4}). Indeed, one can check that the integral over $\omega$ yields:
\begin{align}\label{eq:FLK-SE-FL2}
\Lambda_a&(\mathbf{k},\epsilon)=\frac{\epsilon}{(2\pi)^3}\frac{\partial}{\partial{k}_i}\frac{m}{k}\int{d\epsilon'}\int_{0}^{\infty}qdq
\mathcal{R}{V}_R(\mathbf{q},\epsilon')\\\nonumber
&\times\left[\tanh\left(\frac{\epsilon+\epsilon'}{2T}\right)+
\tanh\left(\frac{\epsilon-\epsilon'}{2T}\right)\right]=\epsilon\frac{\partial\mathcal{R}\sigma_1(\mathbf{k},\epsilon)}{\partial{k}_i}.
\end{align}
This expression is exactly what we anticipate to get in order to satisfy the Wiedemann-Franz law. Therefore, in the framework of the Fermi-liquid theory the contribution from the other two terms in $\Lambda$ must be zero. However, only the last term is zero, while the second one is not:
\begin{align}\label{eq:FLK-SE-NonFL1}
&\Lambda_b(\mathbf{k},\epsilon)=\frac{1}{(2\pi)^3}\frac{\partial}{\partial{k}_i}\frac{m}{k}\int{d\epsilon'}\mathcal{P}\int_0^{\infty}{d\omega}\int_{0}^{\infty}qdq
\\\nonumber
&\times\frac{\mathcal{I}m{V}_R(q,\omega)}{\omega+\epsilon'}
\left[\epsilon'\tanh\left(\frac{\epsilon+\epsilon'}{2T}\right)-\epsilon'
\tanh\left(\frac{\epsilon-\epsilon'}{2T}\right)\right]\\\nonumber
&=\frac{1}{(2\pi)^3}\frac{\partial}{\partial{k}_i}\frac{m}{k}\int{d\epsilon'}\int_{0}^{\infty}qdq
\mathcal{R}{V}_R(q,\epsilon')\\\nonumber
&\times\left[\epsilon'\tanh\left(\frac{\epsilon+\epsilon'}{2T}\right)-\epsilon'
\tanh\left(\frac{\epsilon-\epsilon'}{2T}\right)\right].
\end{align}
One can see that this integral is not zero (the integrand is an even function of $\epsilon'$). Moreover, the integration over the frequency $\epsilon'$ in Eqs.~\ref{eq:FLK-SE-FL2} and~\ref{eq:FLK-SE-NonFL1} accumulates over a range of frequencies that is restricted only by the typical scale of the interaction. We may conclude that the second term is comparable with the anticipated contribution of the first term (Eq.~\ref{eq:FLK-SE-FL2}) and, therefore, the difference between the renormalized heat current vertex and $\Lambda$ cannot be neglected. Other second order contributions can not save the situation because in all of them the \textit{R}-\textit{A} section is connected to one of the bare vertices. Such  contributions do not violate the Wiedemann-Franz law, see the discussion below Eq.~\ref{eq:FLK-integral-P1} where the first group was analyzed.

In this appendix we calculated a particular contribution to the thermal conductivity in the second order perturbation theory with respect to the electron-electron interaction. We showed that already on the level of the Fermi-liquid theory, there is a disagreement between the quantum kinetic approach and the simplified Kubo formula. While the quantum kinetic approach reproduces the known phenomenological behavior, the simplified Kubo formula fails this test. Therefore, the use of the simplified Kubo formula in the presence of electron-electron interactions is unjustified and may lead to erroneous results.

\section{The Coulomb drag}\label{App:Drag}

In this Appendix we concentrate on the contribution of the Coulomb drag to the electric and heat currents. The Coulomb drag term is generated in the quantum kinetic approach when the external perturbation (either the electric field or the temperature gradient) enters the self-energy given in Eq.~\ref{eq:OR-Sigma} through the propagator of the interaction.

\begin{figure}[pt]
\begin{flushright}\begin{minipage}{0.5\textwidth}  \centering
        \includegraphics[width=0.85\textwidth]{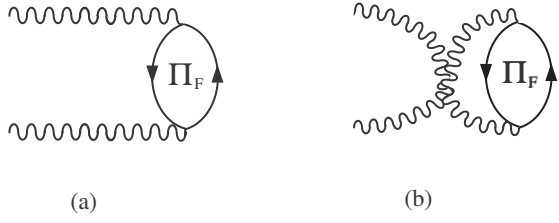}
                 \caption[0.4\textwidth]{\small The two contributions to the Coulomb drag corresponding to $\mathcal{U}=\left[n_P(-\omega)V_{\mathbf{F}}^{<}(\mathbf{q},\omega)+
n_P(\omega)V_{\mathbf{F}}^{>}(\mathbf{q},\omega)\right]$. (a) the diagrammatic interpretation for $V_{\mathbf{F}}^{<}(\mathbf{q},\omega)$ and (b) the diagrammatic interpretation for $V_{\mathbf{F}}^{>}(\mathbf{q},\omega)$ } \label{fig:VLessGreat}
\end{minipage}\end{flushright}
\end{figure}

For the sake of simplicity, we will write the expression for the Coulomb drag term when the averaging over the disorder was already performed and therefore we can use the momentum space representation. The explicit decoration of this term with diffusons can be easily done afterwards. Therefore, we may Fourier transform the spatial coordinates in the expression for the self-energy given in Eq.~\ref{eq:OR-Sigma2}:
\begin{align}\label{eq:Drag-Sigma2}\nonumber
&\Sigma^{<,>}(\mathbf{k},\epsilon)=\frac{i}{2}\hspace{-0.5mm}\int\hspace{-1mm}\frac{d\mathbf{q}d\omega}{(2\pi)^{d+1}}\left[G^{<,>}(\mathbf{k}-\mathbf{q},\epsilon-\omega)V^{<,>}(\mathbf{q},\omega)\right.\\
&\left.+G^{<,>}(\mathbf{k}+\mathbf{q},\epsilon+\omega)V^{>,<}(\mathbf{q},\omega)
\right].
\end{align}
Inserting the self-energy presented above into the third term in the expressions for the electric and heat currents, Eqs.~\ref{eq:EC-JeFinal} and~\ref{eq:FL-Jh}, one obtains:
\begin{widetext}
\begin{align}\label{eq:Drag-JE}
j_{e,h}^{i}=
&i\int\frac{d\mathbf{k}d\epsilon}{(2\pi)^{d+1}}v_{i}^R(\mathbf{k},\epsilon)g_{eq}^R(\mathbf{k},\epsilon)\chi_{e,h}(\epsilon)\left[\Sigma_{\mathbf{F}}^{<}(\mathbf{q},\epsilon)(1-n_F(\epsilon))+\Sigma_{\mathbf{F}}^{>}(\mathbf{q},\epsilon)n_F(\epsilon)\right](g_{eq}^{R}(\mathbf{k},\epsilon)-g_{eq}^A(\mathbf{k},\epsilon))
+c.c.\\\nonumber&=-\frac{1}{2}\int\frac{d\mathbf{k}d\epsilon}{(2\pi)^{d+1}}\frac{d\mathbf{q}{d}\omega}{(2\pi)^{d+1}}\chi_{e,h}(\epsilon)\left[v_{i}^R(\mathbf{k},\epsilon)g_{eq}^R(\mathbf{k},\epsilon)-
v_{i}^A(\mathbf{k},\epsilon)g_{eq}^A(\mathbf{k},\epsilon)\right]
\left[g_{eq}^A(\mathbf{k}-\mathbf{q},\epsilon-\omega)-g_{eq}^R(\mathbf{k}-\mathbf{q},\epsilon-\omega)\right]\\\nonumber
&\left[g_{eq}^R(\mathbf{k},\epsilon)-g_{eq}^A(\mathbf{k},\epsilon)\right]
\left[(1-n_F(\epsilon))n_F(\epsilon-\omega)V_{\mathbf{F}}^{<}(\mathbf{q},\omega)+
n_F(\epsilon)(n_F(\epsilon-\omega)-1)V_{\mathbf{F}}^{>}(\mathbf{q},\omega)\right]\\\nonumber
&-\frac{1}{2}\int\frac{d\mathbf{k}d\epsilon}{(2\pi)^{d+1}}\frac{d\mathbf{q}{d}\omega}{(2\pi)^{d+1}}\chi_{e,h}(\epsilon)\left[v_{i}^R(\mathbf{k},\epsilon)g_{eq}^R(\mathbf{k},\epsilon)
-v_{i}^A(\mathbf{k},\epsilon)g_{eq}^A(\mathbf{k},\epsilon)\right]
\left[g_{eq}^A(\mathbf{k}+\mathbf{q},\epsilon+\omega)-g_{eq}^R(\mathbf{k}+\mathbf{q},\epsilon+\omega)\right]\\\nonumber
&\left[g_{eq}^R(\mathbf{k},\epsilon)-g_{eq}^A(\mathbf{k},\epsilon)\right]
\left[(1-n_F(\epsilon))n_F(\epsilon+\omega)V_{\mathbf{F}}^{>}(\mathbf{q},\omega)+
n_F(\epsilon)(n_F(\epsilon+\omega)-1)V_{\mathbf{F}}^{<}(\mathbf{q},\omega)\right]
\end{align}
\end{widetext}
Recall that the difference between the electric and heat currents is absorbed into the function $\chi_{e,h}(\epsilon)$, where $\chi_{e}(\epsilon)=-e$ while $\chi_{h}(\epsilon)=\epsilon$. In addition, $\mathbf{F}$ indicates the dependence on the external perturbation. Note that Eq.~\ref{eq:Drag-JE} describes not only the electric and thermal conductivities, $\sigma$ and $\kappa$, but also the two off-diagonal transport coefficients, $\alpha$ and $\tilde{\alpha}$. One may connect the two contributions given above with the two possible locations of the velocity vertex inside the polarization operator. Using the identity for the products of the distribution functions given in Eq.~\ref{eq:OR-identities2}, the currents can be rewritten as:
\begin{widetext}
\begin{align}\label{eq:Drag-JE1}\nonumber
&j_{e,h}^{i}=-\frac{1}{2}\int\frac{d\mathbf{k}d\epsilon}{(2\pi)^{d+1}}\frac{d\mathbf{q}{d}\omega}{(2\pi)^{d+1}}
\left\{\chi_{e,h}(\epsilon)\left[v_{i}^R(\mathbf{k},\epsilon)g_{eq}^R(\mathbf{k},\epsilon)-
v_{i}^A(\mathbf{k},\epsilon)g_{eq}^A(\mathbf{k},\epsilon)\right]
\left[g_{eq}^A(\mathbf{k}-\mathbf{q},\epsilon-\omega)-g_{eq}^R(\mathbf{k}-\mathbf{q},\epsilon-\omega)\right]\right.\\\nonumber
&\left.\left[g_{eq}^R(\mathbf{k},\epsilon)-g_{eq}^A(\mathbf{k},\epsilon)\right]
-\chi_{e,h}(\epsilon-\omega)\left[v_{i}^R(\mathbf{k}-\mathbf{q},\epsilon-\omega)g_{eq}^R(\mathbf{k}-\mathbf{q},\epsilon-\omega)-
v_{i}^A(\mathbf{k}-\mathbf{q},\epsilon-\omega)g_{eq}^A(\mathbf{k}-\mathbf{q},\epsilon-\omega)\right]\right.\\
&\left.
\left[g_{eq}^A(\mathbf{k},\epsilon)-g_{eq}^R(\mathbf{k},\epsilon)\right]\left[g_{eq}^R(\mathbf{k}-\mathbf{q},\epsilon-\omega)-g_{eq}^A(\mathbf{k}-\mathbf{q},\epsilon-\omega)\right]\right\}
\left[n_F(\epsilon)-n_F(\epsilon-\omega)\right]
\mathcal{U}(\omega).
\end{align}
\end{widetext}
In the second term we have redefined the frequency $\epsilon+\omega\rightarrow\epsilon$. Here, $\mathcal{U}$ represents the  two interaction lines connected to the polarization operator through which the external perturbation enters (see Fig.~\ref{fig:Drag}):
\begin{align}\label{eq:Drag-V}
\mathcal{U}=[n_P(-\omega)V_{\mathbf{F}}^{<}+n_P(\omega)V_{\mathbf{F}}^{>}].
\end{align}
Let us analyze the analytic structure of $\mathcal{U}$. For  this purpose, we need the expression for the $\mathbf{F}$-dependent propagator of the interaction, $\hat{V}_{\mathbf{F}}(\mathbf{q},\omega)=-\hat{V}_{eq}(\mathbf{q},\omega)\hat{\Pi}_{\mathbf{F}}(\mathbf{q},\omega)\hat{V}_{eq}(\mathbf{q},\omega)$, see Eqs.~\ref{eq:EC-DEV} and~\ref{eq:QKET-V_GradT}. The lesser and greater components of $\hat{V}_\mathbf{F}$  are:
\begin{align}\label{eq:Drag-V2}
V_{\mathbf{F}}^{<,>}=-V_{eq}^{<,>}\Pi_{\mathbf{F}}^{A}V_{eq}^{A}-V_{eq}^{R}\Pi_{\mathbf{F}}^{<,>}V_{eq}^{A}
-V_{eq}^{R}\Pi_{\mathbf{F}}^{R}V_{eq}^{<,>}.
\end{align}
As a consequence of the standard relations between the different components of the propagator at equilibrium, $V_{eq}^{<}=n_P(\omega)[V_{eq}^{R}-V_{eq}^{A}]$ and  $V_{eq}^{>}=-n_P(-\omega)[V_{eq}^{R}-V_{eq}^{A}]$, the  non-vanishing contributions to  Eq.~\ref{eq:Drag-V} occur when the two interaction propagators have an opposite analytic structure, $V^{R}V^{A}$. [If one identifies $V_{\mathbf{F}}^{<}(\mathbf{q},\omega)$ with the diagram presented in Fig.~\ref{fig:VLessGreat}(a) and $V_{\mathbf{F}}^{>}(\mathbf{q},\omega)$ with the diagram in Fig.~\ref{fig:VLessGreat}(b), it becomes clear that the symmetry with respect to $\mathbf{q},\omega\longleftrightarrow-\mathbf{q}-\omega$  is already embedded in $\mathcal{U}$. As  shown in Ref.~\onlinecite{Oreg1995}, this symmetry is responsible for vanishing of the Coulomb drag terms with two retarded or two advanced propagators.]

Let us look at the lesser and greater components of the polarization operator, $\Pi_{\mathbf{F}}^{<,>}(\mathbf{q},\omega)=\int{d\mathbf{k}d\epsilon}/(2\pi)^{d+1}G^{<,>}(\mathbf{k},\epsilon)G^{>,<}(\mathbf{k}-\mathbf{q},\epsilon-\omega)$.
In the linear response, one has to exploit the two possibilities to replace one of the Green's functions in $\Pi_{\mathbf{F}}$ by the appropriate $G_{\mathbf{F}}$ presented in Eqs.~\ref{eq:EC-GE} and~\ref{eq:QKET-G_TransInv2}. Examining Eq.~\ref{eq:Drag-JE1}, one can recognize that the contribution for the electric and heat currents from the Coulomb drag can be presented in a very compact and symmetric way as a product of two polarization operators:
\begin{subequations}\label{eq:Drag-JE3}
\begin{align}
&\mathbf{j}_{e}\mathbf{E}=\int\frac{d\mathbf{q}{d}\omega}{(2\pi)^{d+1}}\left(\frac{\partial{n_P(\omega)}}{\partial\omega}\right)^{-1}
|V_{eq}^{R}(\mathbf{q},\omega)|^2\\\nonumber&\left[n_P(-\omega)\Pi_{\mathbf{E}}^{<}+n_P(\omega)\Pi_{\mathbf{E}}^{>}\right]
\left[n_P(-\omega)\Pi_{\mathbf{F}}^{<}+n_P(\omega)\Pi_{\mathbf{F}}^{>}\right];\\\nonumber
\end{align}
\begin{align}
&\mathbf{j}_{h}\boldsymbol{\nabla}T=-T\int\frac{d\mathbf{q}{d}\omega}{(2\pi)^{d+1}}\hspace{-1mm}\left(\frac{\partial{n_P(\omega)}}{\partial\omega}\right)^{-1}\hspace{-2mm}
|V_{eq}^{R}(\mathbf{q},\omega)|^2\\\nonumber&\left[n_P(-\omega)\Pi_{\boldsymbol{\nabla}T}^{<}+n_P(\omega)\Pi_{\boldsymbol{\nabla}T}^{>}\right]
\left[n_P(-\omega)\Pi_{\mathbf{F}}^{<}+n_P(\omega)\Pi_{\mathbf{F}}^{>}\right].
\end{align}
\end{subequations}
Next, we shall focus on one of the triangles in the Coulomb drag,  $\left[n_P(-\omega)\Pi_{\mathbf{F}}^{<}(\mathbf{q},\omega)+n_P(\omega)\Pi_{\mathbf{F}}^{>}(\mathbf{q},\omega)\right]$. We separate the triangle into two groups of terms, $\vartriangleleft_1$ and $\vartriangleleft_{\scriptscriptstyle2}$:
\begin{widetext}
\begin{subequations}
\begin{align}\label{eq:Drag-JE3}\nonumber
&\vartriangleleft_1=
\frac{1}{2}\int\frac{d\mathbf{k}d\epsilon}{(2\pi)^{d+1}}\frac{\partial{n}_P(\omega)}{\partial\omega}
\left\{\mathbf{F}(\epsilon)\left[\mathbf{v}^R(\mathbf{k},\epsilon)(g_{eq}^R(\mathbf{k},\epsilon))^2g_{eq}^A(\mathbf{k}-\mathbf{q},\epsilon-\omega)-
\mathbf{v}^A(\mathbf{k},\epsilon)(g_{eq}^A(\mathbf{k},\epsilon))^2g_{eq}^R(\mathbf{k}-\mathbf{q},\epsilon-\omega)\right]\right.\\\nonumber
&\left.-
\mathbf{F}(\epsilon-\omega)\left[\mathbf{v}^R(\mathbf{k}-\mathbf{q},\epsilon-\omega)(g_{eq}^R(\mathbf{k}-\mathbf{q},\epsilon-\omega))^2g_{eq}^A(\mathbf{k},\epsilon)-
\mathbf{v}^A(\mathbf{k}-\mathbf{q},\epsilon-\omega)(g_{eq}^A(\mathbf{k}-\mathbf{q},\epsilon-\omega))^2g_{eq}^R(\mathbf{k},\epsilon)\right]
\right\}\\&\times[n_F(\epsilon-\omega)-n_F(\epsilon)];
\end{align}
\begin{align}\nonumber
&\vartriangleleft_{\scriptscriptstyle2}=-
\int\frac{d\mathbf{k}d\epsilon}{(2\pi)^{d+1}}\frac{\partial{n}_P(\omega)}{\partial\omega}
\mathbf{F}(\epsilon)\left\{\mathbf{v}^R(\mathbf{k},\epsilon)(g_{eq}^R(\mathbf{k},\epsilon))^2g_{eq}^R(\mathbf{k}-\mathbf{q},\epsilon-\omega)-
\mathbf{v}^A(\mathbf{k},\epsilon)(g_{eq}^A(\mathbf{k},\epsilon))^2g_{eq}^A(\mathbf{k}-\mathbf{q},\epsilon-\omega)\right.\\&\left.
-\left[\mathbf{v}^A(\mathbf{k},\epsilon)+\mathbf{v}^R(\mathbf{k},\epsilon)\right]g_{eq}^R(\mathbf{k},\epsilon)g_{eq}^A(\mathbf{k},\epsilon)
\left[g_{eq}^A(\mathbf{k}-\mathbf{q},\epsilon-\omega)-g_{eq}^R(\mathbf{k}-\mathbf{q},\epsilon-\omega)\right]
\right\}[n_F(\epsilon-\omega)-n_F(\epsilon)],
\end{align}
\end{subequations}
\end{widetext}
where $\mathbf{F}(\epsilon)$ is equal $e\mathbf{E}$ for the response to an electric current and $\epsilon\boldsymbol{\nabla}T/T$ for the response to a temperature gradient. The expression for the second group, i.e., $\vartriangleleft_{\scriptscriptstyle2}$, was simplified using the symmetry with respect to $\mathbf{q},\omega\longleftrightarrow-\mathbf{q},-\omega$. The analytic structure of this group is the same as obtained in the case of electric conductivity using the Matsubara technique.~\cite{Oreg1995} We will show that the $\vartriangleleft_1$ vanishes for the response to an electric field. Furthermore, we examine the fate of this term for the response to a temperature gradient. With this in mind, we reformulate the products $g_{eq}^{R,A}(\mathbf{k},\epsilon)\mathbf{v}_{eq}^{R,A}(\mathbf{k},\epsilon)g_{eq}^{R,A}(\mathbf{k},\epsilon)$ as a derivative of a single Green's function $\partial{g}_{eq}^{R,A}(\mathbf{k},\epsilon)/\partial\mathbf{k}$. After integrating by parts we obtain:
\begin{align}\label{eq:Drag-Vanishing}\nonumber
&\vartriangleleft_1=\frac{1}{2}\frac{\partial}{\partial{\mathbf{q}}}\int\frac{d\mathbf{k}d\epsilon}{(2\pi)^{d+1}}
\left\{\mathbf{F}(\epsilon)\left[g_{eq}^R(\mathbf{k},\epsilon)g_{eq}^A(\mathbf{k}-\mathbf{q},\epsilon-\omega)\right.\right.\\\nonumber&\left.\left.-
g_{eq}^A(\mathbf{k},\epsilon)g_{eq}^R(\mathbf{k}-\mathbf{q},\epsilon-\omega)\right]
+\mathbf{F}(\epsilon-\omega)\left[g_{eq}^A(\mathbf{k},\epsilon)\right.\right.\\\nonumber&\left.\left.\times
{g}_{eq}^R(\mathbf{k}-\mathbf{q},\epsilon-\omega)
-
g_{eq}^A(\mathbf{k}-\mathbf{q},\epsilon-\omega)g_{eq}^R(\mathbf{k},\epsilon)\right]
\right\}\\&\times[n_F(\epsilon-\omega)-n_F(\epsilon)].
\end{align}

\begin{figure}[pb]
\begin{flushright}\begin{minipage}{0.5\textwidth}  \centering
        \includegraphics[width=0.85\textwidth]{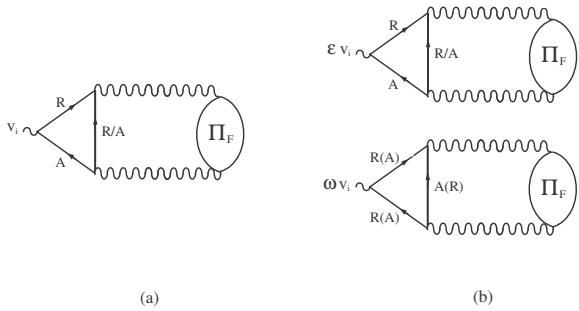}
                 \caption[0.4\textwidth]{\small The analytical structure of the different contributions to the drag diagram for (a) the electric current and (b) the heat current as a response to either an electric field or a temperature gradient. } \label{fig:dragAnalyt}
\end{minipage}\end{flushright}
\end{figure}

For the response to an electric field, when $\mathbf{F}(\epsilon)$ is independent of $\epsilon$, Eq.~\ref{eq:Drag-Vanishing} is identically zero. However, for the response to a temperature gradient, part of the RHS of Eq.~\ref{eq:Drag-Vanishing} survives, because $\mathbf{F}(\epsilon)$ and $\mathbf{F}(\epsilon-\omega)$ are not the same. The remaining term is proportional to $\omega$:
\begin{align}\label{eq:Drag-Survivng}
&\vartriangleleft_1=-\omega\frac{\boldsymbol{\nabla}T}{2T}\int\frac{d\mathbf{k}d\epsilon}{(2\pi)^{d+1}}[n_F(\epsilon-\omega)-n_F(\epsilon)]\\\nonumber
&\frac{\partial}{\partial\mathbf{q}}\left[{g}_{eq}^R(\mathbf{k}-\mathbf{q},\epsilon-\omega)g_{eq}^A(\mathbf{k},\epsilon)
-{g}_{eq}^A(\mathbf{k}-\mathbf{q},\epsilon-\omega)g_{eq}^R(\mathbf{k},\epsilon)\right]
\end{align}
In the presence of disorder, this triangle can be decorated by two diffusons and, hence, its contribution to the currents is significantly larger than the one from $\vartriangleleft_{\scriptscriptstyle2}$. The leading contribution from the Coulomb drag to the thermal conductivity has $\vartriangleleft_1$ on each side as illustrated in Fig.~\ref{fig:Drag}. This term is discussed at the end of Sec.~\ref{sec:WF}.

To conclude,  we showed that the analytic structure of the Coulomb drag terms is different for the responses to an electric field and a temperature gradient. In Fig.~\ref{fig:dragAnalyt} we present the corresponding drag diagrams. It is interesting to note that the combination $\vartriangleleft_1V^RV^A$ generated by the quantum kinetic approach cannot be obtained using the simplified Kubo formula described in Eq.~\ref{eq:CurrentNonIntFrequency}. We believe that the origin of this contribution is related to the part in Luttinger's heat current operator~\cite{Luttinger1964} that depends on the interaction.

\section{Examples illustrating the Onsager relation}\label{App:OR}

In this appendix we present several examples in which we demonstrate the fulfilment of Onsager relation. We prove the Onsager relation for the last term in Eq.~\ref{eq:OR-Currents} when the self-energy contains only one interaction line. This self-energy is described in Eq.~\ref{eq:OR-Sigma2} and illustrated in Fig.~\ref{fig:SelfEnergy1st}. Inserting the self-energy given in Eq.~\ref{eq:OR-Sigma2} into the last term in Eq.~\ref{eq:OR-Currents}, we get:
\begin{align}\label{eq:OR-Currents2}\nonumber
j_{e,h}^{i}&=-\frac{1}{2}\int\frac{d\epsilon{d}\omega}{(2\pi)^2}\chi_{e,h}(\epsilon){v}_{i}^R(\epsilon)g_{eq}^R(\epsilon)\left[G^{<}(\epsilon-\omega)\right.\\\nonumber
&\left.\times
V^{<}(\omega)(1-n_F(\epsilon))
+G^{<}(\epsilon+\omega)V^{>}(\omega)(1-n_F(\epsilon))\right.\\\nonumber
&\left.+
G^{>}(\epsilon-\omega)V^{>}(\omega)n_F(\epsilon)
+G^{>}(\epsilon+\omega)V^{<}(\omega)n_F(\epsilon)\right]\\&(g_{eq}^{R}(\epsilon)-g_{eq}^{A}(\epsilon))
+c.c.
\end{align}
Here either $G=G_{\mathbf{F}}$ is the field-dependent Green's function and $V=V_{eq}$ is the equilibrium propagator or the other way around.

\begin{figure}[pb]
\begin{flushright}\begin{minipage}{0.5\textwidth}  \centering
        \includegraphics[width=01\textwidth]{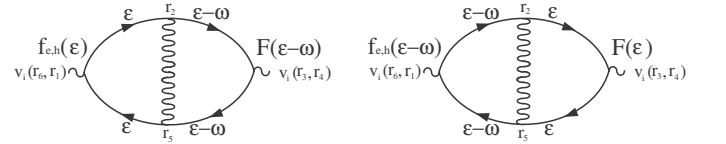}
                 \caption[0.4\textwidth]{\small The diagrammatic representation of the contributions to the currents described in Eq.~\ref{eq:OR-Currents3}. For simplicity we do not show here the dashed lines that indicate the scattering by the impurities.}
                 \label{fig:Je1}
\end{minipage}\end{flushright}
\end{figure}
\begin{figure}[pt]
\begin{flushright}\begin{minipage}{0.5\textwidth}  \centering
        \includegraphics[width=0.9\textwidth]{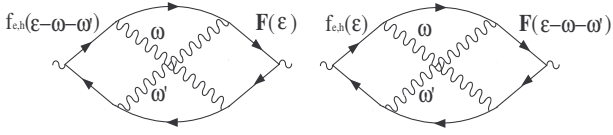}
                 \caption[0.4\textwidth]{\small The diagrammatic representation of the contributions to the currents written in Eqs.~\ref{eq:OR-JeMiddle} and~\ref{eq:OR-JhMiddle}.}
                 \label{fig:Je3}
\end{minipage}\end{flushright}
\end{figure}

We start with the case when the external perturbation enters through the Green's function of the quasiparticles. The $\mathbf{F}$-dependent quasiparticle Green's function is given by Eq.~\ref{eq:OR-G_F}. Using the identities for products of the distribution functions given in Eq.~\ref{eq:OR-identities}, we obtain that the contributions for the thermoelectric currents generated by Eq.~\ref{eq:OR-Currents2} are
\begin{widetext}
\begin{align}\label{eq:OR-Currents3}
&j_{e,h}^{i}=\frac{i}{4d}\int\frac{d\epsilon{d}\omega}{(2\pi)^2}d\mathbf{r}_{\scriptscriptstyle2}...d\mathbf{r}_{\scriptscriptstyle6}\frac{\partial{n_P(\omega)}}{\partial\omega}\chi_{e,h}(\epsilon)F_i(\epsilon-\omega)\left[n_F(\epsilon-\omega)-n_F(\epsilon)\right]
\left[V^{R}(\mathbf{r}_{\scriptscriptstyle2},\mathbf{r}_{\scriptscriptstyle5},\omega)-V^{A}(\mathbf{r}_{\scriptscriptstyle2},\mathbf{r}_{\scriptscriptstyle5},\omega)\right]\\\nonumber
&\left[{g}_{eq}^{A}(\mathbf{r}_{\scriptscriptstyle5},\mathbf{r}_{\scriptscriptstyle6},\epsilon)v_{j}(\mathbf{r}_{\scriptscriptstyle6},\mathbf{r}_{\scriptscriptstyle1})\left(g_{eq}^A(\mathbf{r}_{\scriptscriptstyle1},\mathbf{r}_{\scriptscriptstyle2},\epsilon)-g_{eq}^R(\mathbf{r}_{\scriptscriptstyle1},\mathbf{r}_{\scriptscriptstyle2},\epsilon)\right)
-
\left({g}_{eq}^{A}(\mathbf{r}_{\scriptscriptstyle5},\mathbf{r}_{\scriptscriptstyle6},\epsilon)-{g}_{eq}^{R}(\mathbf{r}_{\scriptscriptstyle5},\mathbf{r}_{\scriptscriptstyle6},\epsilon)\right)
v_{j}(\mathbf{r}_{\scriptscriptstyle6},\mathbf{r}_{\scriptscriptstyle1})g_{eq}^R(\mathbf{r}_{\scriptscriptstyle1},\mathbf{r}_{\scriptscriptstyle2},\epsilon)
\right]
\\\nonumber&
\left[\left(g_{eq}^A(\mathbf{r}_{\scriptscriptstyle2},\mathbf{r}_{\scriptscriptstyle3},\epsilon-\omega)-g_{eq}^R(\mathbf{r}_{\scriptscriptstyle2},\mathbf{r}_{\scriptscriptstyle3},\epsilon-\omega)\right)v_{j}(\mathbf{r}_{\scriptscriptstyle3},\mathbf{r}_{\scriptscriptstyle4},){g}_{eq}^{A}(\mathbf{r}_{\scriptscriptstyle4},\mathbf{r}_{\scriptscriptstyle5},\epsilon-\omega)
-
g_{eq}^R(\mathbf{r}_{\scriptscriptstyle2},\mathbf{r}_{\scriptscriptstyle3},\epsilon-\omega)v_{j}(\mathbf{r}_{\scriptscriptstyle3},\mathbf{r}_{\scriptscriptstyle4})\right.\\\nonumber&\left.
\left({g}_{eq}^{A}(\mathbf{r}_{\scriptscriptstyle4},\mathbf{r}_{\scriptscriptstyle5},\epsilon-\omega)-{g}_{eq}^{R}(\mathbf{r}_{\scriptscriptstyle4},\mathbf{r}_{\scriptscriptstyle5},\epsilon-\omega)\right)\right]\\\nonumber&
+\frac{i}{4d}\int\frac{d\epsilon{d}\omega}{(2\pi)^2}d\mathbf{r}_{\scriptscriptstyle2}...d\mathbf{r}_{\scriptscriptstyle6}\frac{\partial{n_P(\omega)}}{\partial\omega}\chi_{e,h}(\epsilon-\omega)F_i(\epsilon)\left[n_F(\epsilon-\omega)-n_F(\epsilon)\right]
\left[V^{R}(\mathbf{r}_{\scriptscriptstyle5},\mathbf{r}_{\scriptscriptstyle2},\omega)-V^{A}(\mathbf{r}_{\scriptscriptstyle5},\mathbf{r}_{\scriptscriptstyle2},\omega)\right]\\\nonumber
&\left[\left(g_{eq}^A(\mathbf{r}_{\scriptscriptstyle2},\mathbf{r}_{\scriptscriptstyle3},\epsilon)-g_{eq}^R(\mathbf{r}_{\scriptscriptstyle2},\mathbf{r}_{\scriptscriptstyle3},\epsilon)\right)v_{j}(\mathbf{r}_{\scriptscriptstyle3},\mathbf{r}_{\scriptscriptstyle4}){g}_{eq}^{A}(\mathbf{r}_{\scriptscriptstyle4},\mathbf{r}_{\scriptscriptstyle5},\epsilon)
-
g_{eq}^R(\mathbf{r}_{\scriptscriptstyle2},\mathbf{r}_{\scriptscriptstyle3},\epsilon)v_{j}(\mathbf{r}_{\scriptscriptstyle3},\mathbf{r}_{\scriptscriptstyle4})\left({g}_{eq}^{A}(\mathbf{r}_{\scriptscriptstyle4},\mathbf{r}_{\scriptscriptstyle5},\epsilon)-{g}_{eq}^{R}(\mathbf{r}_{\scriptscriptstyle4},\mathbf{r}_{\scriptscriptstyle5},\epsilon)\right)\right]\\\nonumber
&\left[{g}_{eq}^{A}(\mathbf{r}_{\scriptscriptstyle5},\mathbf{r}_{\scriptscriptstyle6},\epsilon-\omega)v_{j}(\mathbf{r}_{\scriptscriptstyle6},\mathbf{r}_{\scriptscriptstyle1})\left(g_{eq}^A(\mathbf{r}_{\scriptscriptstyle1},\mathbf{r}_{\scriptscriptstyle2},\epsilon-\omega)-g_{eq}^R(\mathbf{r}_{\scriptscriptstyle1},\mathbf{r}_{\scriptscriptstyle2},\epsilon-\omega)\right)
-
\left({g}_{eq}^{A}(\mathbf{r}_{\scriptscriptstyle5},\mathbf{r}_{\scriptscriptstyle6},\epsilon-\omega)-{g}_{eq}^{R}(\mathbf{r}_{\scriptscriptstyle5},\mathbf{r}_{\scriptscriptstyle6},\epsilon-\omega)\right)\right.\\\nonumber
&\left.v_{j}(\mathbf{r}_{\scriptscriptstyle6},\mathbf{r}_{\scriptscriptstyle1})g_{eq}^R(\mathbf{r}_{\scriptscriptstyle1},\mathbf{r}_{\scriptscriptstyle2},\epsilon-\omega)
\right].
\\\nonumber&
\end{align}
The two parts of the above equation originate from the two terms of the self-energy written in Eq.~\ref{eq:OR-Sigma2}. In the second part we have redefined the frequency $\epsilon+\omega\rightarrow\epsilon$. One can recognize that Eq.~\ref{eq:OR-Currents3} poses the structure of Eq.~\ref{eq:generalJ}, and as such satisfies the Onsager relation. Furthermore, we may reformulate the expressions for the electric current as a response to $\boldsymbol{\nabla}T$  and the heat current generated by $\mathbf{E}$  in a compact way:
\begin{subequations}\label{eq:OR-Js}
\begin{align}\label{eq:OR-Je}
&\mathbf{E}\mathbf{j}_e(\boldsymbol{\nabla}T)=-iT\int\frac{d\epsilon{d}\omega}{(2\pi)^2}d\mathbf{r}_{\scriptscriptstyle5}
[n_F(\epsilon-\omega)-n_F(\epsilon)]^{-1}
[V_{eq}^{R}(\mathbf{r}_{2},\mathbf{r}_{5},\omega)-V_{eq}^{A}(\mathbf{r}_{2},\mathbf{r}_{5},\omega)]\\\nonumber
&\times\left\{G_{\mathbf{E}}^{<}(\mathbf{r}_{2},\mathbf{r}_{5},\epsilon){G}_{\boldsymbol{\nabla}T}^{<}(\mathbf{r}_{2},\mathbf{r}_{5},\epsilon-\omega)
+G_{\boldsymbol{\nabla}T}^{<}(\mathbf{r}_{2},\mathbf{r}_{5},\epsilon)G_{\mathbf{E}}^{<}(\mathbf{r}_{2},\mathbf{r}_{5},\epsilon-\omega)\right].
\end{align}
\begin{align}\label{eq:OR-Jh}
&\boldsymbol{\nabla}T\mathbf{j}_h(\mathbf{E})=iT^2\int\frac{d\epsilon{d}\omega}{(2\pi)^2}d\mathbf{r}_{\scriptscriptstyle5}
[n_F(\epsilon-\omega)-n_F(\epsilon)]^{-1}
[V_{eq}^{R}(\mathbf{r}_{2},\mathbf{r}_{5},\omega)-V_{eq}^{A}(\mathbf{r}_{2},\mathbf{r}_{5},\omega)]\\\nonumber&\times
\left\{G_{\boldsymbol{\nabla}T}^{<}(\mathbf{r}_{2},\mathbf{r}_{5},\epsilon)
{G}_{\mathbf{E}}^{<}(\mathbf{r}_{2},\mathbf{r}_{5},\epsilon-\omega)+
G_{\mathbf{E}}^{<}(\mathbf{r}_{2},\mathbf{r}_{5},\epsilon)G_{\boldsymbol{\nabla}T}^{<}(\mathbf{r}_{2},\mathbf{r}_{5},\epsilon-\omega)\right].
\end{align}
\end{subequations}
The diagrammatic representation of these contributions to the current is presented in Fig.~\ref{fig:Je1}. It is worth nothing that  one may identify $A_{eq}P_{eq}$ appearing in the general proof of the Onsager relation (see Eq.~\ref{eq:OR-Struct2})  with the effective interaction. Then, the contributions to the thermoelectric currents  can be obtained from Eq.~\ref{eq:OR-Currents3} by substituting  $V^R-V^A$ with $A_{eq}P_{eq}$. Correspondingly, the compact form of the currents presented in Eq.~\ref{eq:OR-Js} can be reproduced. To explain what we have in mind,  let us look at the self-energy given in Fig.~\ref{fig:SelfEnergy2st}(b) and Eq.~\ref{eq:OR-Sigma2nd}. Following the same manipulations, the electric and heat currents generated by this self-energy can be written as:
\begin{subequations}
\begin{align}\label{eq:OR-JeMiddle}\nonumber
&\mathbf{E}\mathbf{j}_e(\boldsymbol{\nabla}T)=T\int\frac{d\epsilon{d}\omega{d}\omega'}{(2\pi)^3}d\mathbf{r}_{\scriptscriptstyle2}...d\mathbf{r}_{\scriptscriptstyle4}
\left[n_F(\epsilon-\omega-\omega')-n_F(\epsilon)\right]^{-1}\left[n_P(\omega)-n_P(-\omega')\right]
g_{eq}^{R}(\mathbf{r}_{\scriptscriptstyle1},\mathbf{r}_{\scriptscriptstyle2},\epsilon-\omega)g_{eq}^{A}(\mathbf{r}_{\scriptscriptstyle3},\mathbf{r}_{\scriptscriptstyle4},\epsilon-\omega')\\\nonumber
&\left\{(V_{eq}^{R}(\mathbf{r}_{\scriptscriptstyle1},\mathbf{r}_{\scriptscriptstyle3},\omega)-V_{eq}^{A}(\mathbf{r}_{\scriptscriptstyle1},\mathbf{r}_{\scriptscriptstyle3},\omega))G_{\boldsymbol{\nabla}T}^{<}(\mathbf{r}_{\scriptscriptstyle2},\mathbf{r}_{\scriptscriptstyle3},\epsilon-\omega-\omega')
({V}_{eq}^{R}(\mathbf{r}_{\scriptscriptstyle2},\mathbf{r}_{\scriptscriptstyle4},\omega')-{V}_{eq}^{A}(\mathbf{r}_{\scriptscriptstyle2},\mathbf{r}_{\scriptscriptstyle4},\omega'))
G_{\mathbf{E}}^{<}(\mathbf{r}_{\scriptscriptstyle4},\mathbf{r}_{\scriptscriptstyle1},\epsilon)\right.\\
&\left.+
(V^{R}(\mathbf{r}_{\scriptscriptstyle3},\mathbf{r}_{\scriptscriptstyle1},\omega')-V^{A}(\mathbf{r}_{\scriptscriptstyle3},\mathbf{r}_{\scriptscriptstyle1},\omega'))G_{\boldsymbol{\nabla}T}^{<}(\mathbf{r}_{\scriptscriptstyle4},\mathbf{r}_{\scriptscriptstyle1},\epsilon-\omega-\omega')
({V}^{R}(\mathbf{r}_{\scriptscriptstyle4},\mathbf{r}_{\scriptscriptstyle2},\omega)-{V}^{A}(\mathbf{r}_{\scriptscriptstyle4},\mathbf{r}_{\scriptscriptstyle2},\omega'))
G_{\mathbf{E}}^{<}(\mathbf{r}_{\scriptscriptstyle2},\mathbf{r}_{\scriptscriptstyle3},\epsilon)\right],
\end{align}
\begin{align}\label{eq:OR-JhMiddle}\nonumber
&\boldsymbol{\nabla}T\mathbf{j}_h(\mathbf{E})=-T^2\int\frac{d\epsilon{d}\omega{d}\omega'}{(2\pi)^3}d\mathbf{r}_{\scriptscriptstyle2}...d\mathbf{r}_{\scriptscriptstyle4}
\left[n_F(\epsilon-\omega-\omega')-n_F(\epsilon)\right]^{-1}
\left[n_P(\omega)-n_P(-\omega')\right]
g_{eq}^{R}(\mathbf{r}_{\scriptscriptstyle1},\mathbf{r}_{\scriptscriptstyle2},\epsilon-\omega)g_{eq}^{A}(\mathbf{r}_{\scriptscriptstyle3},\mathbf{r}_{\scriptscriptstyle4},\epsilon-\omega')\\\nonumber
&\left\{(V_{eq}^{R}(\mathbf{r}_{\scriptscriptstyle1},\mathbf{r}_{\scriptscriptstyle3},\omega)-V_{eq}^{A}(\mathbf{r}_{\scriptscriptstyle1},\mathbf{r}_{\scriptscriptstyle3},\omega))G_{\mathbf{E}}^{<}(\mathbf{r}_{\scriptscriptstyle2},\mathbf{r}_{\scriptscriptstyle3},\epsilon-\omega-\omega')
({V}_{eq}^{R}(\mathbf{r}_{\scriptscriptstyle2},\mathbf{r}_{\scriptscriptstyle4},\omega')-{V}_{eq}^{A}(\mathbf{r}_{\scriptscriptstyle2},\mathbf{r}_{\scriptscriptstyle4},\omega'))
G_{\boldsymbol{\nabla}T}^{<}(\mathbf{r}_{\scriptscriptstyle4},\mathbf{r}_{\scriptscriptstyle1},\epsilon)\right.\\
&\left.+
(V^{R}(\mathbf{r}_{\scriptscriptstyle3},\mathbf{r}_{\scriptscriptstyle1},\omega')-V^{A}(\mathbf{r}_{\scriptscriptstyle3},\mathbf{r}_{\scriptscriptstyle1},\omega'))G_{\mathbf{E}}^{<}(\mathbf{r}_{\scriptscriptstyle4},\mathbf{r}_{\scriptscriptstyle1},\epsilon-\omega-\omega')
({V}^{R}(\mathbf{r}_{\scriptscriptstyle4},\mathbf{r}_{\scriptscriptstyle2},\omega)-{V}^{A}(\mathbf{r}_{\scriptscriptstyle4},\mathbf{r}_{\scriptscriptstyle2},\omega'))
G_{\boldsymbol{\nabla}T}^{<}(\mathbf{r}_{\scriptscriptstyle2},\mathbf{r}_{\scriptscriptstyle3},\epsilon)\right],
\end{align}
\end{subequations}
\end{widetext}
The diagrammatic interpretation of these currents is presented in Fig.~\ref{fig:Je3}.

Let us return to the discussion of the self-energy presented in Eq.~\ref{eq:OR-Sigma2}. We shall examine the contributions to the thermoelectric currents in which the external perturbation enters $\Sigma_{\mathbf{F}}$ through the propagator of the interactions. These contributions, corresponding to the Coulomb drag terms, were analyzed in Appendix~\ref{App:Drag}. It is shown there that the electric current as a response to a temperature gradient can be written as:
\begin{align}\label{eq:OR-JeDrag}\nonumber
&j_{e}^{i}=\frac{1}{E}\int\frac{{d}\omega}{2\pi}d\mathbf{r}_{\scriptscriptstyle2}d\mathbf{r}_{\scriptscriptstyle3}d\mathbf{r}_{\scriptscriptstyle4}\left(\frac{\partial{n_P(\omega)}}{\partial\omega}\right)^{-1}V_{eq}^{A}(\mathbf{r}_{\scriptscriptstyle3},\mathbf{r}_{\scriptscriptstyle4},\omega)
\\\nonumber&\hspace{-2mm}\left[n_P(-\omega)\Pi_{\mathbf{E}}^{<}(\mathbf{r}_{\scriptscriptstyle4},\mathbf{r}_{\scriptscriptstyle1},\omega)+n_P(\omega)\Pi_{\mathbf{E}}^{>}(\mathbf{r}_{\scriptscriptstyle4},\mathbf{r}_{\scriptscriptstyle1},\omega)\right]\hspace{-1mm}V_{eq}^{R}(\mathbf{r}_{\scriptscriptstyle1},\mathbf{r}_{\scriptscriptstyle2},\omega)
\\
&\hspace{-2mm}\left[n_P(-\omega)\Pi_{\boldsymbol{\nabla}T}^{<}(\mathbf{r}_{\scriptscriptstyle2},\mathbf{r}_{\scriptscriptstyle3},\omega)+n_P(\omega)\Pi_{\boldsymbol{\nabla}T}^{>}(\mathbf{r}_{\scriptscriptstyle2},\mathbf{r}_{\scriptscriptstyle3},\omega)\right].
\end{align}
Similarly, the heat current generated by an electric field can be formulated is
\begin{align}\label{eq:OR-JhDrag}\nonumber
&j_{h}^{i}=-\frac{T}{\boldsymbol{\nabla}{T}}\int\frac{{d}\omega}{2\pi}d\mathbf{r}_{\scriptscriptstyle2}d\mathbf{r}_{\scriptscriptstyle3}d\mathbf{r}_{\scriptscriptstyle4}\left(\frac{\partial{n_P(\omega)}}{\partial\omega}\right)^{-1}V_{eq}^{A}(\mathbf{r}_{\scriptscriptstyle3},\mathbf{r}_{\scriptscriptstyle4},\omega)
\\&\hspace{-2mm}\left[n_P(-\omega)\Pi_{\boldsymbol{\nabla}T}^{<}(\mathbf{r}_{\scriptscriptstyle4},\mathbf{r}_{\scriptscriptstyle1},\omega)+n_P(\omega)\Pi_{\boldsymbol{\nabla}T}^{>}(\mathbf{r}_{\scriptscriptstyle4},\mathbf{r}_{\scriptscriptstyle1},\omega)\right]
\\\nonumber
&\hspace{-2mm}V_{eq}^{R}(\mathbf{r}_{\scriptscriptstyle1},\mathbf{r}_{\scriptscriptstyle2},\omega)\left[n_P(-\omega)\Pi_{\mathbf{E}}^{<}(\mathbf{r}_{\scriptscriptstyle2},\mathbf{r}_{\scriptscriptstyle3},\omega)+n_P(\omega)\Pi_{\mathbf{E}}^{>}(\mathbf{r}_{\scriptscriptstyle2},\mathbf{r}_{\scriptscriptstyle3},\omega)\right].
\end{align}
One may see that Eqs.~\ref{eq:OR-JeDrag} and~\ref{eq:OR-JhDrag} are connected through the Onsager relation under the condition of microscopic reversibility. However, it is not obvious how to recognize in the above expression the structure of Eq.~\ref{eq:generalJ} used in the general proof of the Onsager relation. To resolve this issue, one has to go to Eq.~\ref{eq:Drag-JE} in Appendix~\ref{App:Drag}. Then, instead of analyzing each polarization operator $\hat{\Pi}_{\mathbf{F}}$  separately, one should concentrate on combinations of the kind $[(1-n_F(\epsilon))n_F(\epsilon-\omega)\Pi_{\mathbf{F}}^{<}(\mathbf{q},\omega)+
n_F(\epsilon)(n_F(\epsilon-\omega)-1)\Pi_{\mathbf{F}}^{>}(\mathbf{q},\omega)]$. Following Sec.~\ref{sec:Onsager}, we shift the arguments of the Green's functions  inside $\Sigma_{\mathbf{F}}(\epsilon)$ in such a way that they all  contain $\epsilon$. The polarization operator will be written as:
\begin{align}\label{eq:OR-DragPi}\nonumber
\Pi^{<,>}(\omega)&=\frac{1}{2}\int\frac{d\Omega}{2\pi}\left[G^{<,>}(\epsilon-\Omega)G^{>,<}(\epsilon-\Omega-\omega)\right.\\
&\left.+
G^{<,>}(\epsilon+\Omega)G^{>,<}(\epsilon+\Omega-\omega)\right].
\end{align}
Now, the Coulomb drag term contains four distribution functions. To simplify the expression, we group these functions into two pairs, and use the identities in Eq.~\ref{eq:OR-identities2} for each of the pairs. Here, we pair the distributions functions in such a way that the difference between the two frequencies in each pair is $\Omega$. Namely, one pair contains $n_F(\epsilon)$ and $n_F(\epsilon-\Omega)$, while the other includes $n_F(\epsilon-\omega)$ and $n_F(\epsilon-\omega-\Omega)$. Following the steps described here, one can obtain an expression for the Coulomb drag term that has the same structure as Eq.~\ref{eq:generalJ}. Consequently, one may present the Coulomb drag contributions to the thermoelectric currents in a compact form similar to Eqs.~\ref{eq:OR-Je} and~\ref{eq:OR-Jh}.

%
%

\section{The quantum kinetic approach in the presence of superconducting fluctuations}\label{sec:ElectricCond-SCF}

In this work, we derived the currents as a response to an electric field and a temperature gradient for electrons interacting through the density channel, e.g., the Coulomb interaction. Here we show that following the same scheme one can find the expression for the currents in the presence of  superconducting fluctuations.

The difference between the fluctuations in the two channels reveals itself most clearly when the electric current or the response to an electric field are studied. The action for electrons interacting through the Cooper channel in the presence of an electric field is
\begin{widetext}
\begin{align}\label{eq:EC-SC-S}
&\mathcal{S}=\int{d\mathbf{r}dt}\left\{\phantom{\frac{1}{1}}\hspace{-2mm}\sum_{s}\psi_s^{\dag}(\mathbf{r},t)i\partial_t\psi_s(\mathbf{r},t)
-\sum_{s}\frac{(\boldsymbol{\nabla}\psi_s^{\dag}(\mathbf{r},t))(\boldsymbol{\nabla}\psi_s(\mathbf{r},t))}{2m}-\sum_{s}\psi_s^{\dag}(\mathbf{r},t)\left[e\mathbf{r}\cdot\mathbf{E}+V_{imp}-\mu\right]\psi_s(\mathbf{r},t)\right.\\\nonumber
&\left.-\left[\Delta(\mathbf{r},t)\psi_{\uparrow}^{\dag}(\mathbf{r},t)\psi_{\downarrow}^{\dag}(\mathbf{r},t)+h.c.\right]
-\frac{|\Delta(\mathbf{r},t)|^{2}}{\lambda}\right\}.
\end{align}
\end{widetext}
It follows from the action that the charge of the electrons, $-e|\psi(\mathbf{r})|^2$, is not conserved unless the charge carried by the interaction field, $2e|\Delta(\mathbf{r})|^2$, is also included. Correspondingly, the continuity equation for the current of the quasiparticles acquires a source term $-e\sum_s\partial_t|\psi_s(\mathbf{r},t)|^2+\boldsymbol{\nabla}\mathbf{j}_{e}^{qp}(\mathbf{r},t)
=-2ie[\Delta^{\dag}(\mathbf{r},t)\psi_{\downarrow}(\mathbf{r},t)\psi_{\uparrow}(\mathbf{r},t)-\Delta(\mathbf{r},t)\psi_{\uparrow}^{\dag}(\mathbf{r},t)\psi_{\downarrow}^{\dag}(\mathbf{r},t)]$. Hence, the expression for the electric current is the sum of the currents of the quasiparticles and interacting field:
\begin{align}\label{eq:EC-SC-Je}
&\mathbf{j}_e=\frac{ie}{2}\int{d\mathbf{r}'dt'}\left[2\hat{\underline{\mathbf{v}}}(\mathbf{r},t;\mathbf{r}',t')\hat{\underline{G}}(\mathbf{r}',t';\mathbf{r},t)\right.\\\nonumber
&\left.+\hspace{-1mm}2\hat{\underline{\boldsymbol{\mathcal{V}}}}(\mathbf{r},t;\mathbf{r}',t')\hat{\underline{L}}(\mathbf{r}',t';\mathbf{r},t)
\right]^{<}+h.c.
\end{align}
Here, $\hat{\underline{L}}(\mathbf{r},t;\mathbf{r}',t')$ is the gauge invariant propagator of the interaction in the Cooper channel, and we use the notation $\hat{\underline{\boldsymbol{\mathcal{V}}}}(\mathbf{r},t;\mathbf{r}',t')=-i(\mathbf{r-r}')\underline{\hat{\Pi}}_{sc}(\mathbf{r},t;\mathbf{r}',t')$ to emphasize that this term is analogous to the renormalized velocity given in Eq.~\ref{eq:EC-velocityQP}. [Since the propagator $\hat{L}$ does not have a dimension of an inverse energy, $\hat{\boldsymbol{\mathcal{V}}}$ does not have the dimension of velocity.] The factor $2$ in the contribution to the current from the Green's function of the quasiparticles is due to the sum over the two spin directions, while the factor of $2$ in the contribution of $\hat{L}$ is because the superconducting fluctuations carry a charge of $2e$.

Since $\Delta$ is a charged field, the kinetic equation for its propagator $\hat{L}(\mathbf{r},t;\mathbf{r}',t')$ resembles the kinetic equation of the quasiparticle Green's function, see Eq.~\ref{eq:EC-QKE-E}, rather than Eq.~\ref{eq:EC-DEV} for the neutral interaction field:
\begin{align}\label{eq:QKEelectricSimple}\nonumber
&-\lambda^{-1}\hat{\underline{L}}(\boldsymbol{\rho},\omega;imp)
=\delta(\boldsymbol{\rho})-\hspace{-1mm}\int\hspace{-1mm}{d\mathbf{r}_{\scriptscriptstyle1}}\hat{\underline{\Pi}}\left(\boldsymbol{\rho}-\mathbf{r}_{\scriptscriptstyle1},\omega;imp\right)\\
&\times\hspace{-1mm}\left\{1-e\mathbf{E}
\left[\mathbf{r}_{\scriptscriptstyle1}\frac{\overleftarrow{\partial}}{\partial\omega}-(\boldsymbol{\rho}-\mathbf{r}_{\scriptscriptstyle1})\frac{\overrightarrow{\partial}}{\partial\omega}\right]\right\}
\hat{\underline{L}}\left(\mathbf{r}_{\scriptscriptstyle1},\omega;imp\right).
\end{align}
Using Eqs.~\ref{eq:EC-SC-Je} and~\ref{eq:QKEelectricSimple}, we have reproduced the known results for the paraconductivity~\cite{Aslamazov1968,Maki1968,Varlamov} and for the corrections to the magneto-resistance in disordered superconducting films.~\cite{Galitski2001}

Let us turn to the response to the temperature gradient, which is analyzed with the help of the gravitational field.
Unlike the response to an electric field, the quantum kinetic equation and $\hat{L}$ in the presence of a gravitational field has the same form as Eq.~\ref{eq:QKET-DEVTransform} for $\hat{V}$:
\begin{align}\label{eq:EC-S-DEL}
&-\lambda^{-1}\hat{\underline{\underline{L}}}(\mathbf{r},t;\mathbf{r}',t')=\delta(\mathbf{r}-\mathbf{r}')\delta(t-t')\\\nonumber
&-\int{d\mathbf{r}_{\scriptscriptstyle1}dt_{\scriptscriptstyle1}}\hat{\underline{\underline{\Pi}}}_{sc}(\mathbf{r},t;\mathbf{r}_{\scriptscriptstyle1},t_{\scriptscriptstyle1})\hat{\underline{\underline{L}}}(\mathbf{r}_{\scriptscriptstyle1},t_{\scriptscriptstyle1};\mathbf{r}',t').
\end{align}
Here,  $\hat{\underline{\underline{L}}}$  is transformed according to Eq.~\ref{eq:QKET-GaugeTrans}. Note that just like for $\hat{\underline{\underline{V}}}$, the dependence of the propagator $\hat{\underline{\underline{L}}}$ on the gravitational field is only through its self-energy $\underline{\underline{\hat{\Pi}}}_{sc}$. In view of the  similarity of the kinetic equations for $\hat{\underline{\underline{V}}}$ and $\hat{\underline{\underline{L}}}$, it is natural that the heat current (both as a response to $\mathbf{E}$ and $\boldsymbol{\nabla}T$) does not acquire any additional terms in the presence of fluctuations in the Cooper channel and Eqs.~\ref{eq:HC-HeatCurrentT} and~\ref{eq:HC-HeatCurrentE2} are still valid. Finally, the expression for the electric current as a response to a temperature gradient includes two contributions; one is the electric current of the quasiparticles (described by the $\boldsymbol{\nabla}T$-dependent Green's function) and the other  is from the electric current carried by the interaction field (described by the $\boldsymbol{\nabla}T$-dependent propagator $\hat{\underline{\underline{L}}}$).

We applied this scheme for the calculation of the Nernst effect in disordered films above the superconducting transition.~\cite{KM2008}


\begin{thebibliography}{99}
\bibitem{comment} Notice that Eq.~\ref{eq:CurrentNonInt} can be rewriten using the commutation relation for the field operators as:
$\mathbf{j}_h(\tau)=\frac{1}{2}\sum_{\mathbf{p}}\frac{\partial\varepsilon_{\mathbf{p}}}{\partial\mathbf{p}}\{c_{\mathbf{p}}^{\dag}[c_{\mathbf{p}},H]-[c_{\mathbf{p}}^{\dag},H]c_{\mathbf{p}}\}$, where $H$ is the Hamiltonian.
\end{thebibliography}
\end{document}